\documentclass[twocolumn]{aastex631}

\usepackage{amsmath}
\usepackage{gensymb}
\usepackage{multirow}
\shorttitle{Enabling Kilonova Science with Roman}
\shortauthors{}
\graphicspath{{./}{figures/}}
\begin{document}

\title{Enabling Kilonova Science with Nancy Grace Roman Space Telescope}

\author[0000-0002-8977-1498]{Igor Andreoni}
\altaffiliation{Neil Gehrels Fellow}
\affil{Joint Space-Science Institute, University of Maryland, College Park, MD 20742, USA}
\affil{Department of Astronomy, University of Maryland, College Park, MD 20742, USA}
\affil{Astrophysics Science Division, NASA Goddard Space Flight Center, Mail Code 661, Greenbelt, MD 20771, USA}
\email{andreoni@umd.edu}

\author[0000-0002-8262-2924]{Michael W. Coughlin}
\affil{School of Physics and Astronomy, University of Minnesota, Minneapolis, Minnesota 55455, USA}

\author[0000-0002-9225-7756]{Alexander W. Criswell}
\affil{School of Physics and Astronomy, University of Minnesota, Minneapolis, Minnesota 55455, USA}

\author[0000-0002-8255-5127]{Mattia Bulla}
\affil{Department of Physics and Earth Science, University of Ferrara, via Saragat 1, 44122 Ferrara, Italy}
\affil{INFN, Sezione di Ferrara, via Saragat 1, I-44122 Ferrara, Italy}
\affil{INAF, Osservatorio Astronomico d’Abruzzo, via Mentore Maggini snc, 64100 Teramo, Italy}

\author[0009-0008-9546-2035]{Andrew Toivonen}
\affil{School of Physics and Astronomy, University of Minnesota, Minneapolis, Minnesota 55455, USA}

\author[0000-0001-9898-5597]{Leo P. Singer}
\affil{Astroparticle Physics Laboratory, NASA Goddard Space Flight Center, Code 661, Greenbelt, MD 20771, USA}

\author[0000-0002-6011-0530]{Antonella Palmese}
\affil{McWilliams Center for Cosmology, Department of Physics, Carnegie Mellon University, Pittsburgh, PA 15213, USA}

\author[0000-0002-2942-3379]{E.~Burns}
\affiliation{Department of Physics and Astronomy, Louisiana State University, Baton Rouge, LA 70803 USA}

\author{Suvi Gezari}
\affiliation{Space Telescope Science Institute, 3700 San Martin Drive, Baltimore, MD 21218, USA}
\affiliation{Department of Physics and Astronomy, Johns Hopkins University, 3400 N. Charles St., Baltimore, MD 21218, USA}

\author{Mansi M. Kasliwal}
\affiliation{Division of Physics, Mathematics and Astronomy, California Institute of Technology, Pasadena, CA 91125, USA}

\author[0000-0002-9108-5059]{R. Weizmann Kiendrebeogo}
\affiliation{Laboratoire de Physique et de Chimie de l’Environnement, Université Joseph KI-ZERBO, Ouagadougou, Burkina Faso}
\affiliation{Artemis, Observatoire de la Côte d’Azur, Université Côte d’Azur, Boulevard de l'Observatoire, 06304 Nice, France}
\affiliation{School of Physics and Astronomy, University of Minnesota, Minneapolis, Minnesota 55455, USA}

\author{Ashish Mahabal}
\affiliation{Division of Physics, Mathematics and Astronomy, California Institute of Technology, Pasadena, CA 91125, USA}

\author{Takashi J. Moriya}
\affil{National Astronomical Observatory of Japan, National Institutes of Natural Sciences, 2-21-1 Osawa, Mitaka, Tokyo 181-8588, Japan}
\affil{School of Physics and Astronomy, Faculty of Science, Monash University, Clayton, Victoria 3800, Australia}

\author{Armin Rest}
\affil{Space Telescope Science Institute, Baltimore, MD 21218, USA}
\affil{Department of Physics and Astronomy, The Johns Hopkins University, Baltimore, MD 21218, USA}

\author{Dan Scolnic}
\affil{Department of Physics, Duke University, Durham, NC, 27708, USA}

\author{Robert A. Simcoe}
\affil{MIT Kavli Institute for Astrophysics and Space Research, 77 Massachusetts Avenue, Cambridge, 02139, Massachusetts, USA}

\author[0000-0001-9226-4043]{Jamie Soon}
\affil{Research School of Astronomy and Astrophysics, Australian National University, Cotter Rd, Weston Creek, ACT 2611, Australia}

\author[0000-0003-2434-0387]{Robert Stein}
\affiliation{Division of Physics, Mathematics and Astronomy, California Institute of Technology, Pasadena, CA 91125, USA}

\author[0000-0001-9304-6718]{Tony Travouillon}
\affil{Advanced Instrumentation and Technology Centre, Research School of Astronomy and Astrophysics, Australian National University, Mt Stromlo Observatory, Cotter Road, Weston Creek, Australia}

\begin{abstract}
 Binary neutron star mergers and neutron star--black hole mergers are multi-messenger sources that can be detected in gravitational waves and in electromagnetic radiation.
 The low electron fraction of neutron star merger ejecta favors the production of heavy elements such as lanthanides and actinides via rapid neutron capture ($r$-process).  The decay of these unstable nuclei powers an infrared-bright transient called a ``kilonova". The discovery of a population of kilonovae will allow us to determine if neutron star mergers are the dominant sites for $r$-process element nucleosynthesis, constrain the equation of state of nuclear matter, and make independent measurements of the Hubble constant. 
The Nancy Grace Roman Space Telescope (Roman) will have a unique combination of depth, near-infrared sensitivity, and wide field of view. These characteristics will enable Roman's discovery of GW counterparts that will be missed by optical telescopes, such as kilonova that are associated with large distances, high lanthanide fractions, high binary mass-ratios, large dust extinction in the line of sight, or that are observed from equatorial viewing angles. In preparation for Roman's launch and operations, our analysis suggests to (i) make available a rapid ($\sim 1$ week) Target of Opportunity mode for GW follow-up; (ii) include observations of the High Latitude Time-Domain survey footprint in at least two filters (preferably the F158 and F213 filters) with a cadence of $\lesssim 8$ days; (iii) operate in synergy with Rubin Observatory. Following these recommendations, we expect that 1--6 kilonovae can be identified by Roman via target of opportunity observations of well localized ($A < 10$\,deg$^2$, 90\% C.I.) neutron star mergers during 1.5 years of the LIGO-Virgo-KAGRA fifth (or $\sim$4--21 in during the sixth) observing run. A sample of 5--40 serendipitously discovered kilonovae can be collected in a 5-year high latitude survey.

\end{abstract}


\section{Introduction} \label{sec:intro}

The first binary neutron star (BNS) merger discovered in gravitational waves (GWs), named GW170817, was accompanied by a short-duration gamma-ray burst \citep[GRB; e.g.,][]{Abbott2017gw_grb, Goldstein2017}, an optical/infrared transient \citep[e.g.,][]{Coulter17,Arcavi2017GW,Tanvir17,Lipunov2017,Soares-Santos2017,Valenti:2017ngx}, and a radio \citep[e.g.,][]{Alexander2017, Hallinan:2017woc} and X-ray \citep[e.g.,][]{Margutti:2017cjl, Troja2017} afterglow.  The multi-messenger discovery has led to hundreds of studies addressing, for example, astrophysics of energetic phenomena \citep[][]{Gottlieb2018, Mooley2018Nat, Ghirlanda2019Sci, Nativi2021a, Nativi2021b, Salafia2019, Salafia2021, Mooley2022} and cosmology \citep{AbbottNSdiscovery,AbbottH0, Guidorzi2017, Hjorth17, Hotokezaka2019NatAs, CoDi2019,CoAn2020, DiCo2020, Wang2021H0, Bulla2022Univ, Palmese2023H0arXiv}. Combining GW and electromagnetic (EM) information provided us with unique insights on the equation of state of neutron stars and extreme nuclear physics \citep{BaJu2017, MaMe2017, CoDi2018, CoDi2018b, CoDi2019b, AnEe2018, MoWe2018,RaPe2018,Lai2019,DiCo2020, Nicholl2021}. 

The optical and near infrared (nIR) transient counterpart to BNS and neutron star--black hole (NSBH) mergers, called a ``kilonova'' or ``macronova'' \citep[see][for a review]{Metzger:2019zeh}, is powered by the radioactive decay of heavy elements synthesized in the neutron-rich ejecta via rapid neutron capture (``$r$-process"). 
Observations of the GW170817 kilonova were used to measure the heavy-element content of the ejecta  \citep{ChBe2017,Coulter17, Cowperthwaite17, Kasen17, Kasliwal17, Kilpatrick2017, PiDa2017, RoFe2017, Smartt17, Rosswog2018, WaHa2019,KaKa2019}. 
The {\it Spitzer} space observatory has provided us with the most constraining nIR data in the late (nebular) phase, suggesting that elements in the second (120 $\leq$ A $\leq$ 140) and third (180 $\leq$ A $\leq$ 200) $r$-process abundance peak of the solar heavy element distribution were synthesized \citep{Kasliwal2022}. 
Recent reviews of the GW and EM observations of GW170817 can be found in  \cite{Nakar2020PhR} and \cite{Margutti21}.

The GW and EM observations of GW170817 enabled extremely broad science, however we have barely peered into the potential of multi-messenger astronomy to answer profound questions on the origin of heavy elements in the Universe, cosmology, and fundamental physics. A population of kilonovae must be found to meet this purpose. 
Despite numerous optical and infrared surveys that have been active during the last decade, kilonovae have largely remained elusive because of their intrinsic rarity, rapid evolution, and low luminosity compared to other extragalactic transients such as supernovae. Localization skymaps that are issued with GW triggers are typically coarse, thus wide-field capabilities are required for discovery of an EM counterpart. Moreover, kilonovae are expected to be bright in the nIR and could be too faint to be seen in the optical \citep[e.g.,][]{Tanaka:2013ixa, Kasen17}. New sensitive, wide-field observatories are therefore needed to enable GW multi-messenger sciences in the years to come.

In this paper, we discuss the unique capabilities of {\it Nancy Grace Roman Space Telescope} (hereafter ``Roman") for the discovery and characterization of kilonovae. This work was prompted by the Call for Input into Roman's Core Community Surveys issued on April 11, 2023. We consider two avenues for kilonova discovery: target of opportunity observations triggered by GW events detected in real time, and ``serendipitous'' identification of kilonovae in survey data. We also discuss future work that will greatly benefit multi-messenger observations with Roman. Cosmological parameters from \cite{Planck2020cosmo} and the AB magnitude system are adopted throughout the paper.

\section{Roman unique capabilities for kilonova science}

The Wide Field Instrument (WFI\footnote{\label{note:WFI} \url{https://roman.gsfc.nasa.gov/science/WFI_technical.html}}) is a 288 Mpx optical/nIR camera with an active field of view of 0.281 deg$^2$, excluding detector gaps. The large field of view is achieved by employing 18 detectors in a $6 \times 3 $ array (Fig.\,\ref{fig:skymap}).
WFI is equipped with 8 science filters with overlapping band passes spanning 0.48--2.3 $\mu$m. Roman WFI observations will comprise Core Community Surveys (CCSs) and General Astrophysics (GA) surveys. The CCSs will include a High Latitude Wide Area survey, a Galactic Bulge Time Domain survey, and a High Latitude Time Domain (HLTD) survey. The 19\,deg$^2$ HLTD survey will lead to the discovery of thousands of extragalactic transient sources thanks to the enormous volume that it will probe. The imaging sensitivity is reported in Tab.\,\ref{tab:sensitivity} as on the WFI technical overview webpage\textsuperscript{\ref{note:WFI}} for 55\,s and 1\,hr of exposure time, which we use here as reference for shallow and deep survey modes.

\begin{table*}
    \centering
    \begin{tabular}{ccccccccc}
    \hline\hline
Exp. &	F062&	F087&	F106&	F129&	F158&	F184&	F213&	F146 \\
\hline
& \multicolumn{8}{c}{Central wavelength (width) [$\mu m$]} \\
\cline{2-9}
& 0.620	(0.280)	& 0.869	(0.217) & 1.060 (0.265) & 1.293	(0.323) & 1.577	(0.394) & 1.842	(0.317) & 2.125 (0.35) & 1.464 (1.030)\\
\cline{2-9}
& \multicolumn{8}{c}{Limiting magnitude} \\
\cline{2-9}
1\,hr &	28.5&	28.2&	28.1&	28.0&	28.0&	27.4&	26.2&	28.4 \\
55\,s &	25.5&	25.1&	25.1&	25.0&	24.9&	24.4&	23.7&	25.9 \\
\hline
    \end{tabular}
    \caption{ Central wavelength (width) of the Roman WFI bandpasses, expressed in microns. The last two rows show limiting magnitudes (AB) for $5\sigma$ point source detection per filter assuming 1\,hr and 55\,s of Roman exposure time, as described in the WFI technical overview webpage\textsuperscript{\ref{note:WFI}}. In the 1\,hr case, the signal-to-noise calculations assume coaddition of four 900\,s exposures, while in the 55\,s case it assumes a single exposure.}
    \label{tab:sensitivity}
\end{table*}

Roman will achieve a unique combination of i) wide field of view, ii) depth, and iii) near-infrared sensitivity. These characteristics make it an excellent discovery instrument for EM counterparts to neutron star mergers both during GW follow-up via target of opportunity (ToO) observations as well as via un-triggered searches in the CCS/HLTD footprint.

There have been a variety of previous analyses that describe Roman's potential for identifying kilonovae, predominantly focused on serendipitous searches and targeted follow-up analyses. Focusing on serendipitous observations, \cite{Scolnic2018} found that Roman would identify $\sim$\,16 kilonovae similar to GW170817 in a 45\,sq. deg. deep drilling field assuming a 5 day cadence over a 2 year survey. \cite{Chase2022} evaluated Roman's detectability of a grid of kilonova models over time and redshift, showing that Roman may detect kilonovae out to $z \sim 1$. \cite{Hounsell2018} do not focus on kilonovae, but instead estimate the effect of Roman cadence on identification of supernovae. \cite{Ma2023} show how Roman could detect $\sim$\,3 lensed kilonovae in follow-up of Cosmic Explorer observations of lensed binary neutron stars. 

In this work, we focused our analysis on the definition of the observing strategy. We employed up-to-date simulated skymaps and compact merger rates to determine the number of kilonovae that could be observed with Roman.  We also used a new grid of kilonova light curves, obtained with 3D Monte Carlo radiative transfer code \citep{Bulla2019, Bulla2023MNRAS, Anand2023arXivChemDist}, to probe the detectability of kilonovae observed at different viewing angles and changing physical parameters such as the electron fraction of the dynamical ejecta and total ejecta masses.

\subsection{\bf Motivation for ToO observations}

\begin{figure}[t]
    \centering
    \includegraphics[width=\columnwidth]{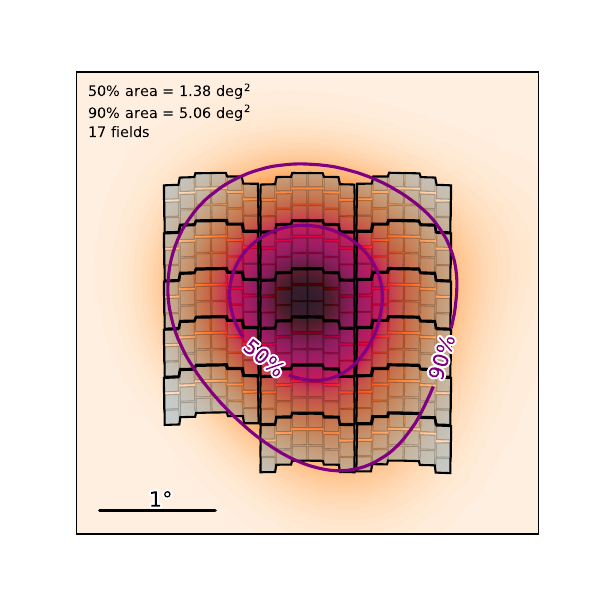}
    \caption{An example LVK skymap for a well-localized event (area $A \sim 5$\,deg$^2$; 90\% C.I.) tiled by Roman. Darker regions indicate higher localization probability. Purple contours indicate the 50\% and 90\% confidence intervals of the probability map. Black contours mark the Roman field of view, inside which individual detectors (grey patches) are separated by small gaps. }
    \label{fig:skymap}
\end{figure}

Follow-up of compact binary coalescence candidates from interferometric gravitational-wave detectors such as LIGO \citep{aLIGO}, Virgo \citep{adVirgo}, and KAGRA \citep{AkEA2019}, together referred to as LVK hereafter, remain the most promising pathway to the identification of kilonovae beyond GW170817. ToO observations with Roman will consist of tiling the localization probability skymaps that are issued by LVK when a trigger occurs (Fig.\,\ref{fig:skymap}). Complementary GW and EM observations of neutron star mergers enable combined constraints on the neutron star environment, merger products, and the original binary system. These opportunities will only continue to grow with the improved reach of the gravitational-wave detectors and the sensitivity of facilities such as Roman. Predictions for the sky localizations produced in forthcoming gravitational-wave runs indicate that follow-up by wide field-of-view surveys will be necessary \citep{petrov_data-driven_2022a}. While comprehensive and deep coverage can be expected in the optical from the ground, Roman will be one of the only near-infrared to infrared instruments capable of surveying the $\sim$few to tens of square degrees that will be required. While sky localizations continue to improve, the vast majority of sources will be near threshold, and therefore have localizations that require tiled coverage by wide-field facilities.

According to the Roman requirements document from November 05, 2020, Baseline and Threshold Technical Requirements \#8, ``Roman shall have the capability to respond to ToO requests to point at an object in the sky within the Field of Regard of the observatory within two weeks of notification to the Science Operation Center". We therefore assume that it will be possible to activate ToO observations with Roman. As demonstrated below, a time lag of two weeks between a BNS or NSBH merger and Roman observations is likely too slow a response to meet our science goals, so we encourage looking into making it possible to activate ToOs with a faster turnaround.  

\subsection{\bf Motivation for serendipitous searches}

Follow-up observations of gravitational-wave candidates are not the only way to find kilonovae. Serendipitous observations of the sky are ``free'' in the sense that they do not require re-pointing of the system and open the discovery space given the much larger set of images to mine. The survey will yield candidate kilonovae, in addition to more common phenomena such as supernovae, and their identification becomes a technical and follow-up challenge. However, this method is highly sensitive to the survey cadence of the system, given that simply ``detecting'' an object is not enough, and confirmation of candidate nature must be done in real time due to the requirement of follow-up observations. Their identification requires many coordinated elements: real-time algorithms filtering and identifying candidates, photometric and spectroscopic follow-up, and model comparisons to understand their physics. Therefore, careful considerations of both cadence and filter selection will have an important effect on the success of fast transient science cases in general.

\section{Strategies}\label{sec:strategies}

Kilonova discovery with Roman will happen via ToO observations, triggered by the detection of GW signals by LVK, and during the extragalactic time-domain survey.  Only JWST will be capable of acquiring spectra of distant kilonova candidates found by Roman, but ToO observations with JWST will have to be initiated parsimoniously. Adopting a sensible observing strategy (during ToOs and CCSs) can significantly increase the discovery rate and enable the unambiguous identification of a population of kilonovae.

The intrinsic rate of BNS and NSBH mergers is an important factor to estimate the expected number of kilonova detections. The rates derived from GW observations are $R_{\textrm{BNS}} = 210^{+240}_{-120}$\,Gpc$^{-3}$\,y$^{-1}$ and $R_{\textrm{NSBH}} = 8.6^{+9.7}_{-5.0}$\,Gpc$^{-3}$\,y$^{-1}$ \citep{theligo-virgo-kagracollaboration_observing_2023, coughlin_ligo/virgo/kagra_2022, singer_lpsinger/observing-scenarios-simulations:_2022}. 
A kilonova is expected to originate from NSBH mergers when the disruption of the neutron star occurs outside the innermost stable circular orbit (ISCO) of the black hole, which depends on the binary mass ratio, black hole spin, and neutron star equation of state \citep[see e.g.,][]{Foucart:2012vn, Kawaguchi2016}. The detection rates for BNS and NSBH mergers depend on the sensitivity of the LVK detectors \citep{Abbott2020LRRprospects}. Roman is expected to launch in 2027, so its operations will likely overlap with the fifth and sixth LVK observing runs \citep[O5 and O6;][]{Abbott2020LRRprospects}. In \S\ref{sec:metrics} we discuss the number of well-localized GW events that we estimated from simulations.

Transient discoveries usually occur via image subtraction between a new ``science" image and an archival ``template" or ``reference" image. Even if a new source is bright and isolated enough to be identified without image subtraction \citep[for example using the \texttt{Source Extractor} software;][]{Bertin1996}, it is necessary to compare the science image with deep images where the source is absent to establish its transient nature.
Although it is possible that Roman archival images are available in the sky region where a GW follow-up ToO is initiated, it is unlikely. Multiple epochs must therefore be acquired to enable image differencing to reveal the presence of the transient, especially when it is embedded in a luminous host. A deep template when the kilonova has disappeared ($\gtrsim$ 1--2 months, depending on the distance) can then be acquired to enable photometry of the kilonova (again, via image subtraction) free of contamination from the host and from the transient itself.

In this section, minimal and preferred strategies are suggested for kilonova discovery via ToO observations and during the HLTD survey. Our analysis is aided by a set of kilonova models obtained with the 3D Monte Carlo radiative transfer code \textsc{possis} \citep{Bulla2019, Bulla2023MNRAS}. The most recent grid of \textsc{possis} models, employed in this work, is also presented in detail in \cite{Anand2023arXivChemDist}. Out of the 1024 models within the grid, we employ  as a baseline model the one with parameters that are closest to those inferred by \cite{Anand2023arXivChemDist} for the GW170817 kilonova. We refer to it as a ``polar" model, labelling it GW170817$_{\rm pol}$, because of the relatively small viewing angle ($\theta = 26$\,deg) between the observer and the rotation axis of the binary. We then employ a model GW170817$_{\rm eq}$ with the same parameters as GW170817$_{\rm pol}$, but where the viewing angle is equatorial ($\theta = 90$\,deg). Finally, we employed other models in which physical parameters are varied such as the averaged electron fraction $Y_e$ of the dynamical ejecta and the total ejecta mass. A description of the model parameters, together with the adopted values for the different models, is presented in Tab.\,\ref{tab:models}.

\begin{table}[t]
    \centering
    \begin{tabular}{l c c c c}
    \hline\hline
   Model & $M_{\rm dyn}$ & $Y_{\rm e, dyn}$ &  $M_{\rm wind}$  & $\theta$ \\
    & $(M_{\odot})$ &  &  $(M_{\odot})$ & (deg) \\
    \hline
   GW170817$_{\rm pol}$ & 0.001 & 0.20 & 0.050 &  26 \\
   GW170817$_{\rm eq}$ & 0.001 & 0.20 & 0.050 &  90 \\
   low $Y_{e}$, $M_{ej}$ & 0.001 & 0.15 & 0.010 &  45 \\
   low $Y_{\rm e}$ & 0.001 & 0.15 & 0.050 &  26 \\
   high $M_{\rm ej}$ & 0.020 & 0.20 & 0.090 & 26 \\
   low $M_{\rm ej}$ & 0.001 & 0.20 & 0.010 & 26 \\
    \hline
    \end{tabular}
    \caption{Models used for the analysis, from a grid obtained with the 3D Monte Carlo radiative trasnfer code \textsc{possis} \citep{Bulla2019, Bulla2023MNRAS} and presented in \cite{Anand2023arXivChemDist}. The grid is constructed with ejecta modelled as in \cite{Bulla2023MNRAS} and varying five free parameters: the mass ($M_{\rm dyn}$) and average electron fraction ($Y_{\rm e, dyn}$) of the dynamical ejecta, assuming an average velocity $v_{\rm dyn} = 0.15c$, and the mass ($M_{\rm wind}$) of the post-merger disk-wind ejecta assuming an average velocity $v_{\rm wind} = 0.03c$.  The model closest to the best-fit to GW170817 \citep[as obtained in][]{Anand2023arXivChemDist} is here indicated as GW170817$_{\rm pol}$ because of the almost-polar viewing angle. The model GW170817$_{\rm eq}$ has the same parameters as GW170817$_{\rm pol}$ but the viewing angle is equatorial ($\theta = 90$\,deg). In the other models, we varied the electron fraction $Y_e$ of the dynamical ejecta and the ejecta mass.}
    \label{tab:models}
\end{table}

\subsection{Minimal Strategies}

The minimal strategies are designed to enable the identification of a likely EM counterpart to a BNS or NSBH merger. We require at least two detections in at least two filters for the following reasons:
\begin{itemize}
    \item at least two detections are needed to confirm that the source is astrophysical (i.e., not a detector artefact, cosmic ray, or near-Earth moving object)
    \item at least two detections in one filter will allow us to perform initial image subtraction to find varying sources in the images, even if the transient has not completely faded away yet; timely identification of EM counterpart candidates is extremely important for multi-wavelength follow-up with narrow-field telescopes. They can also help us determine the rising or fading timescale of the transient, which will aid photometric classification
    \item detections in at least two filters at similar epochs will place (minimal) constraints on the spectral energy distribution (SED) of the source, and therefore on its temperature and bolometric luminosity
    \item at least two detections in at least two filters at different epochs, will help model the source thanks to the measurement of its color and color evolution (in other words, the evolution of the SED)
\end{itemize}

\begin{figure}[t]
    \centering
    \includegraphics[width=\columnwidth]{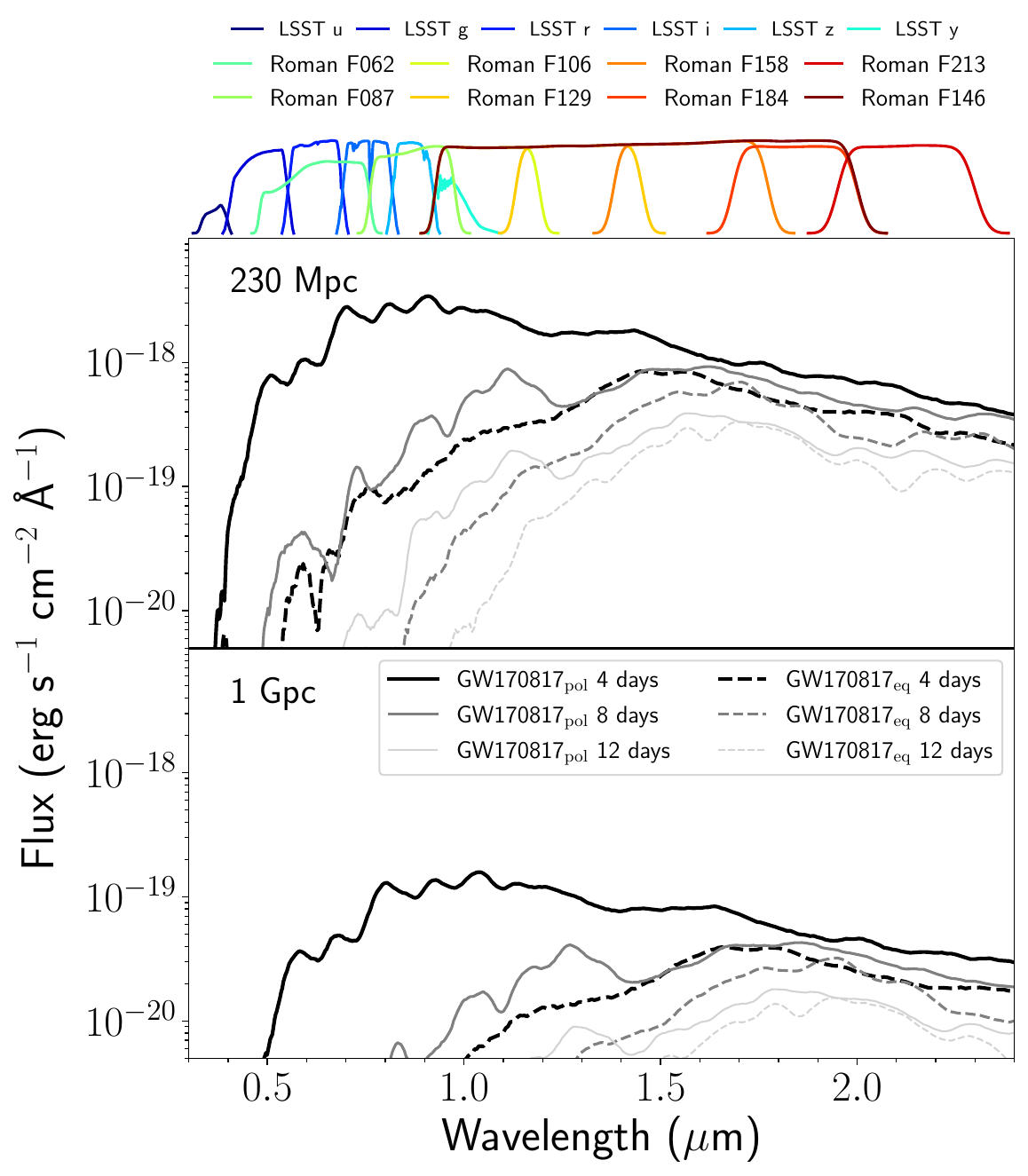}
    \caption{Kilonova spectra computed with \textsc{possis} \citep{Bulla2019,Bulla2023MNRAS} at 230 Mpc (top) and 1 Gpc (bottom). Spectra are shown at three different epochs (4, 8 and 12\,d after the merger) for the model to GW170817 kilonova (GW170817$_{\rm pol}$, solid lines) and for the same model viewed from an equatorial viewing angle (GW170817$_{\rm eq}$, dashed lines), see Tab.~\ref{tab:models} for more details. Rubin LSST and Roman passbands are shown at the top.}
    \label{fig:spec}
\end{figure}

\begin{figure*}
    \centering
    \includegraphics[width=\textwidth]{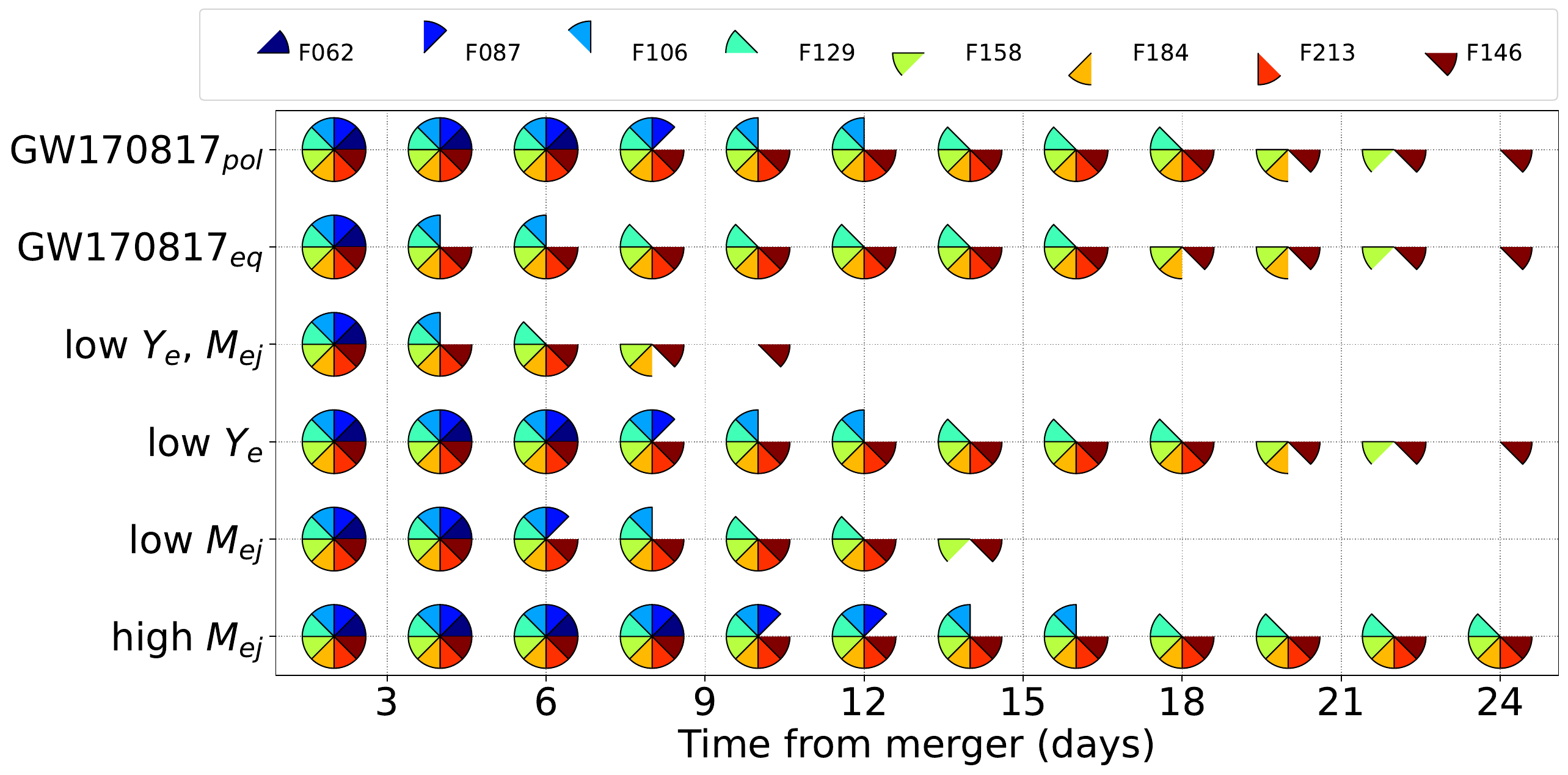}
    \includegraphics[width=\textwidth]{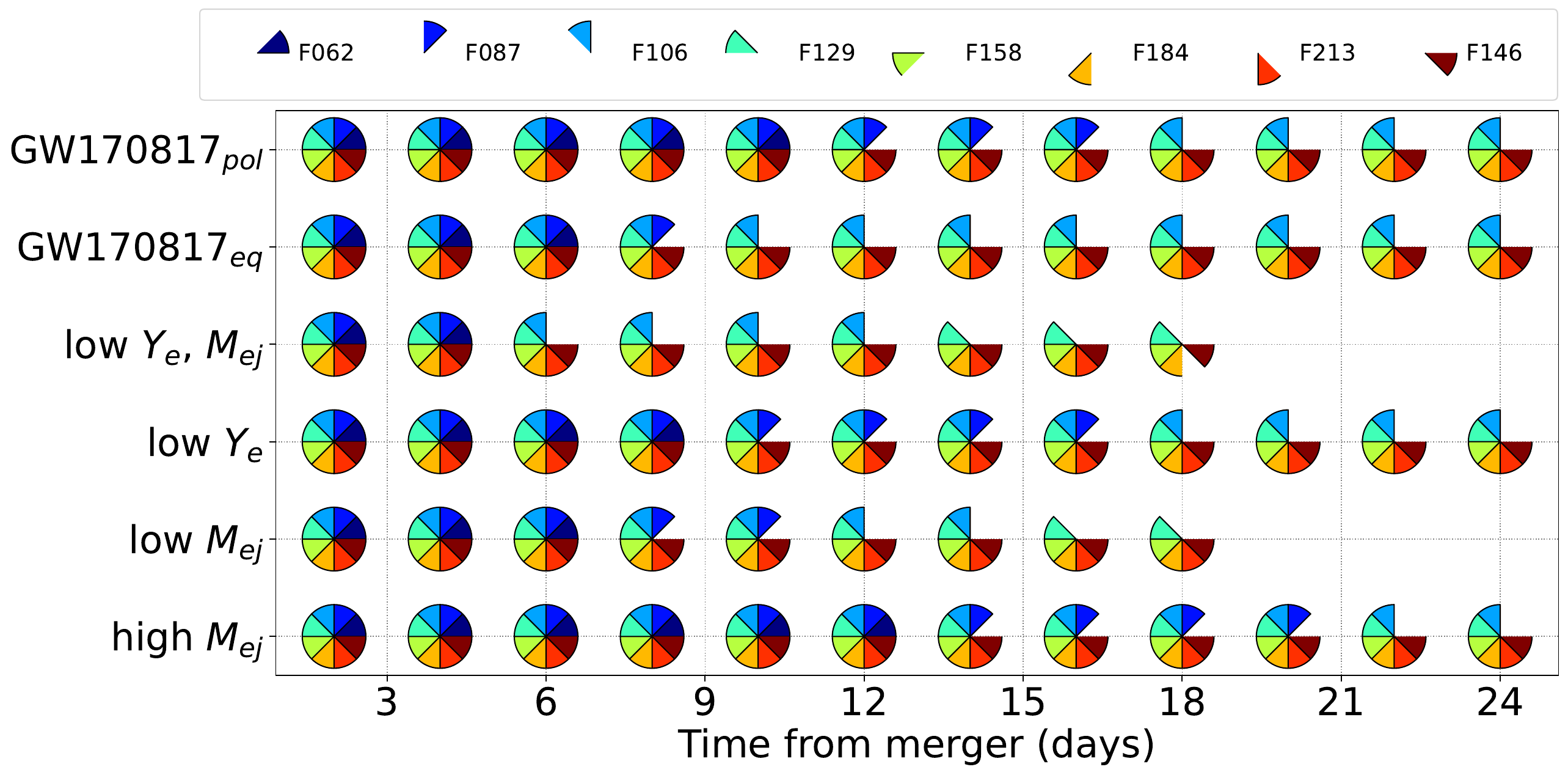}
    \caption{Per-filter detection at interval of two days for kilonova models described in Tab.\,\ref{tab:models} at a distance of 230\,Mpc for an exposure time of 55s (top) and 1hr (bottom).}
    \label{fig:gantt_near}
\end{figure*}

\begin{figure*}
    \centering
    \includegraphics[width=\textwidth]{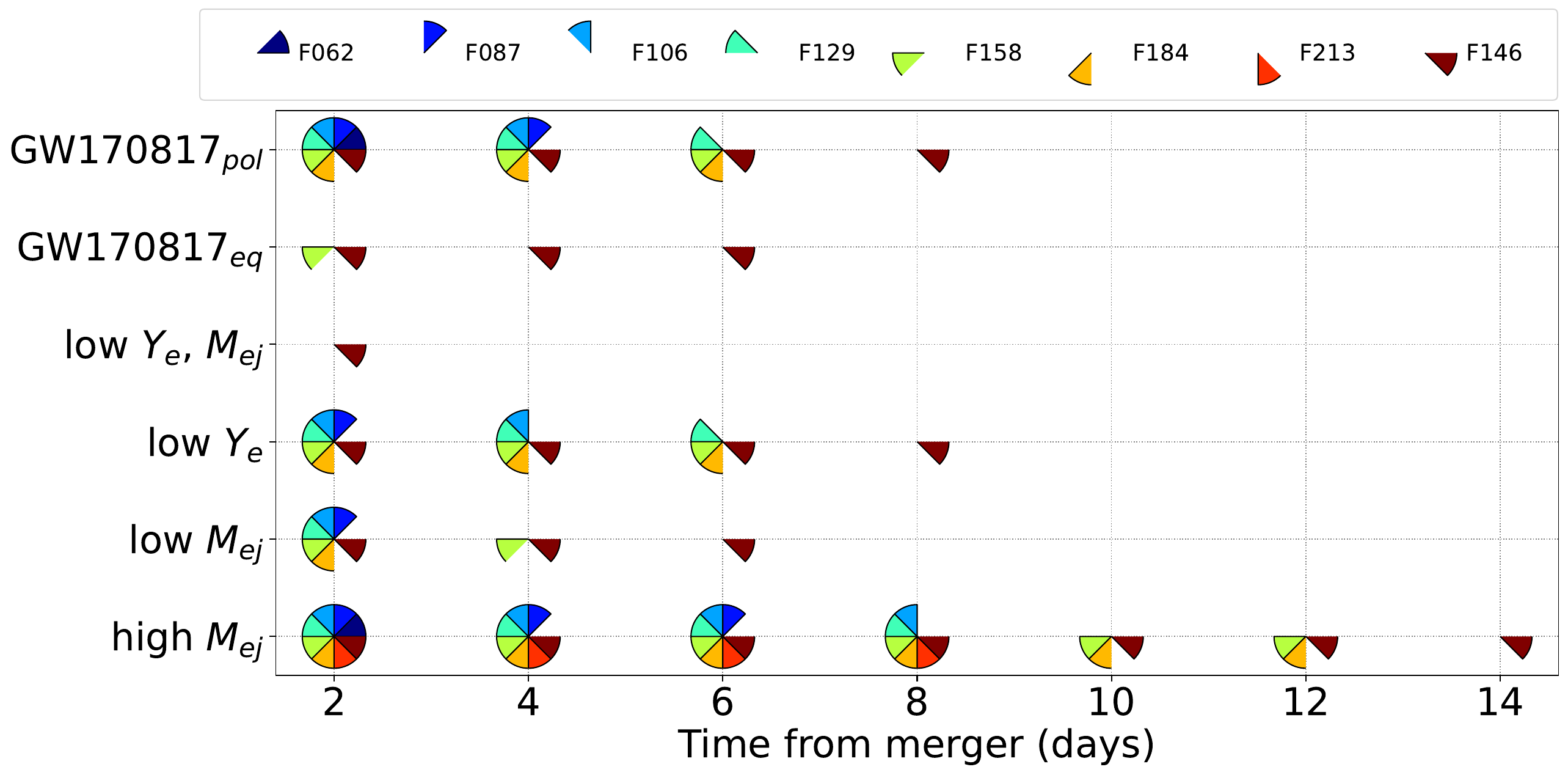}
    \includegraphics[width=\textwidth]{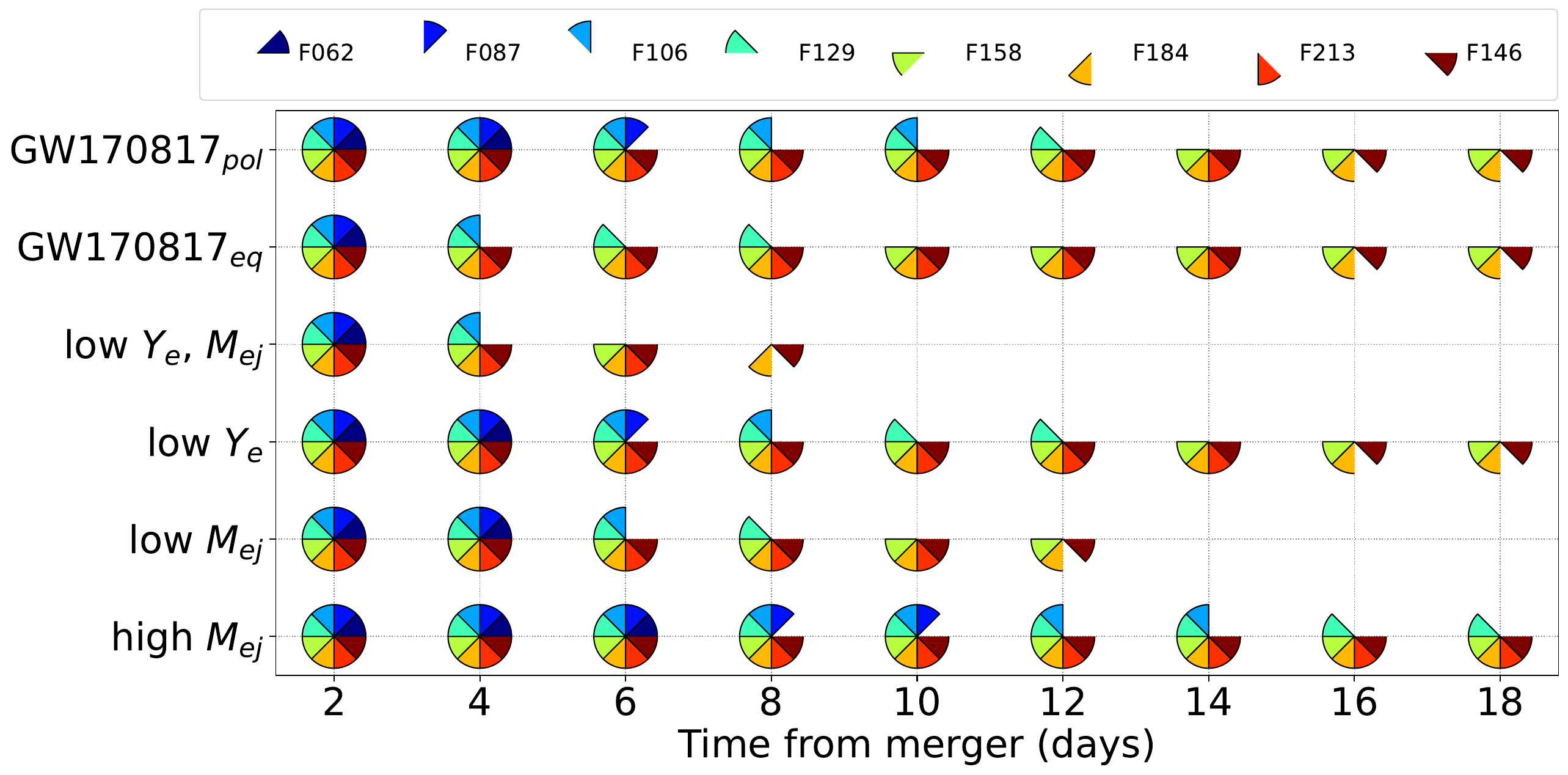}
    \caption{Per-filter detection at interval of two days for kilonova models described in Tab.\,\ref{tab:models} at a distance of 1\,Gpc for an exposure time of 55s (top) and 1hr (bottom).}
    \label{fig:gantt_far}
\end{figure*}

\begin{figure*}
    \centering
    \includegraphics[width=\textwidth]{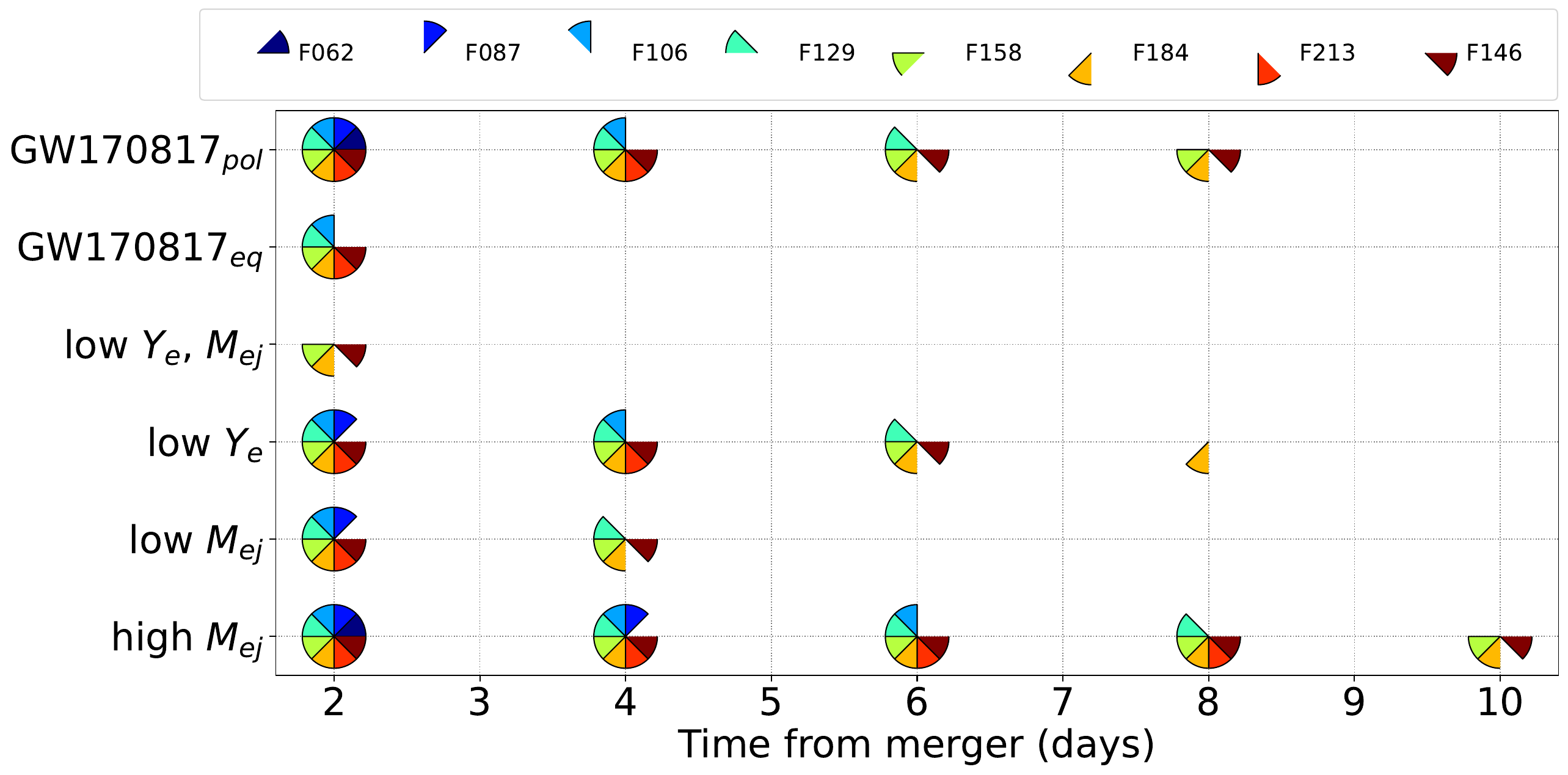}
    \caption{Per-filter detection at interval of two days for kilonova models described in Tab.\,\ref{tab:models} at a luminosity distance of 3\,Gpc for an exposure time of 1hr. An exposure time of a few minutes is insufficient to detect most types of kilonovae at such large distances.}
    \label{fig:gantt_veryfar}
\end{figure*}

Broadband photometry using the F146 bandpass (central wavelength 1.464\,$\mu m$, width 1.030 $\mu m$; see Tab.\,\ref{tab:sensitivity} and Fig.\,\ref{fig:spec}) may appear to be advantageous, but we discourage it. The main benefit of employing a broadband filter is that sources would be observable for a couple of days longer than using other filters (Fig.\,\ref{fig:gantt_near}--\ref{fig:gantt_veryfar}).  F146 would also be less sensitive to different SED shapes and is more likely to return source detection with high signal-to-noise ratio (S/N). On the other hand, by the time the difference in detection limit between F146 and other filters becomes noticeable (2--3 weeks after merger), dozens of candidate counterparts may be identified by Roman, most of which will likely be too faint for any telescope (aside from JWST and maybe a few others) to follow them up. It is unrealistic to expect that JWST will be used for classification of more than a couple of transients per cycle via disruptive ToO requests. Unlike broadband photometry, multi-band photometry in other filters will enable the selection of several EM counterpart candidates, for which deep multi-wavelength follow-up can be initiated. 

\subsubsection{Target of Opportunity}
\label{sec:minimal-ToO}

\begin{figure*}
    \centering
    \includegraphics[width=0.32\textwidth]{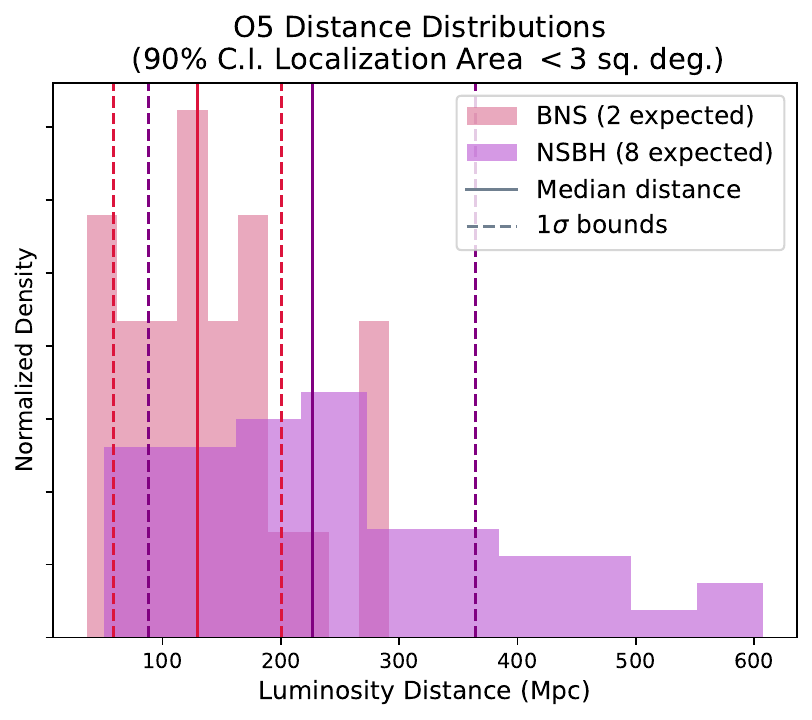}
    \includegraphics[width=0.32\textwidth]{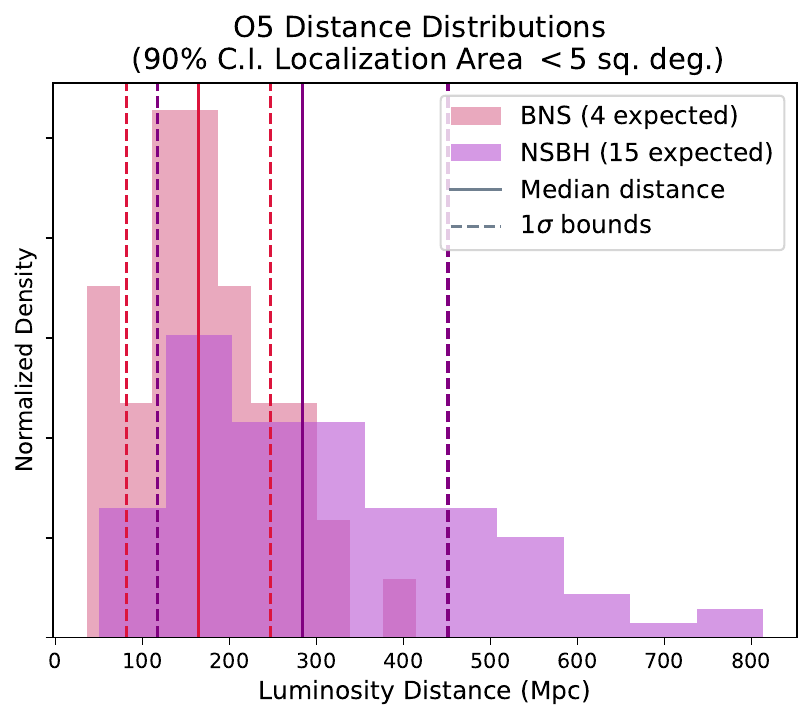}
    \includegraphics[width=0.32\textwidth]{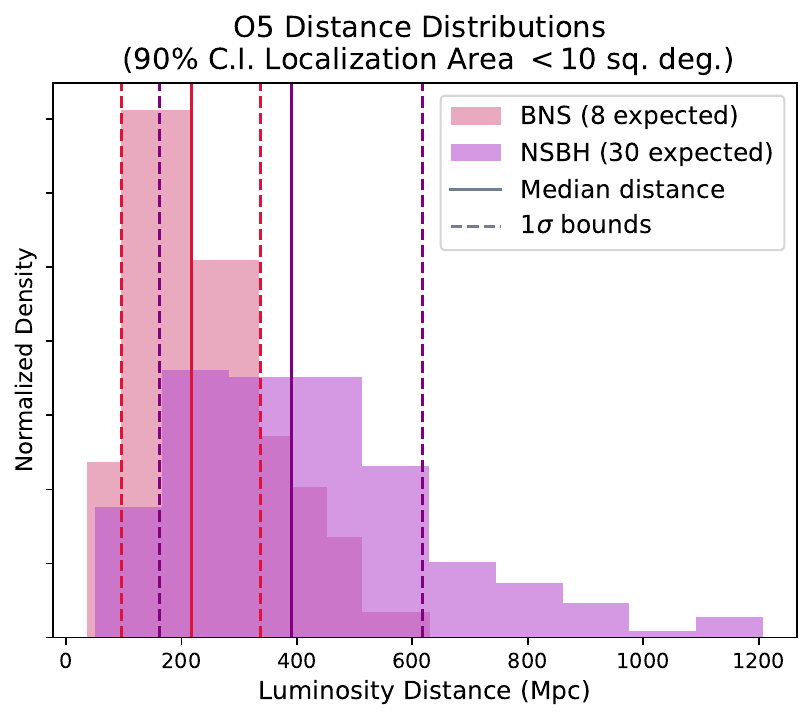}
    
    \includegraphics[width=0.32\textwidth]{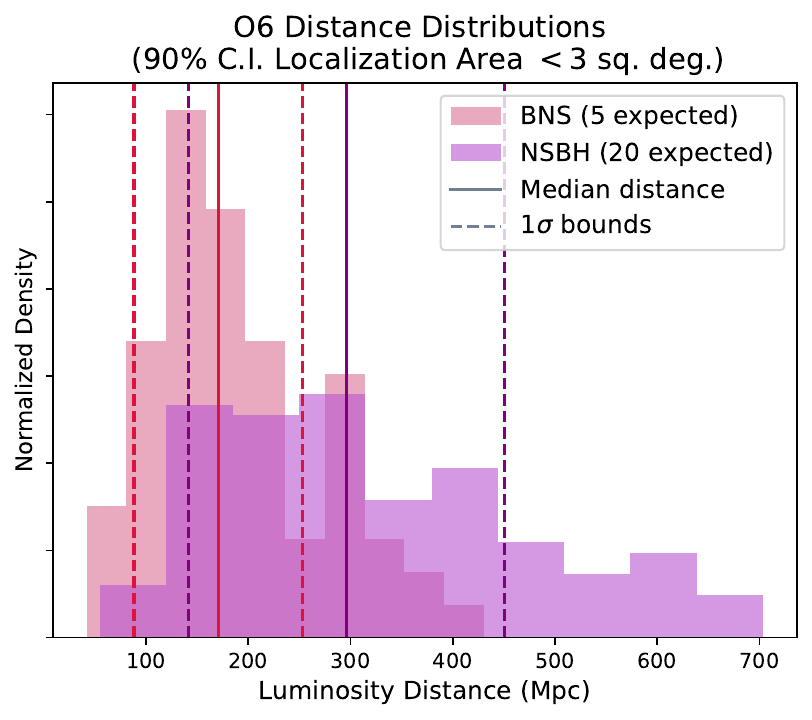}
    \includegraphics[width=0.32\textwidth]{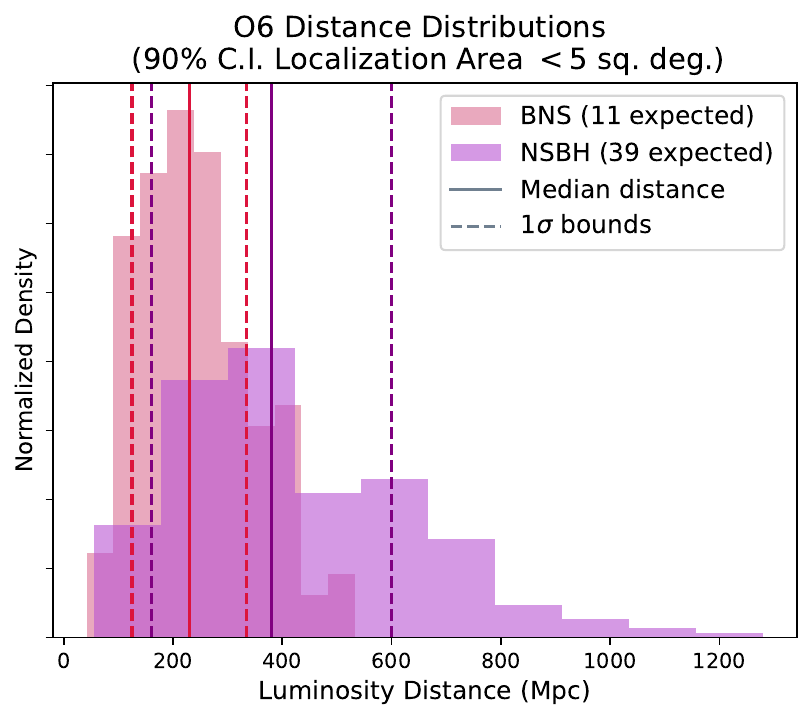}
    \includegraphics[width=0.32\textwidth]{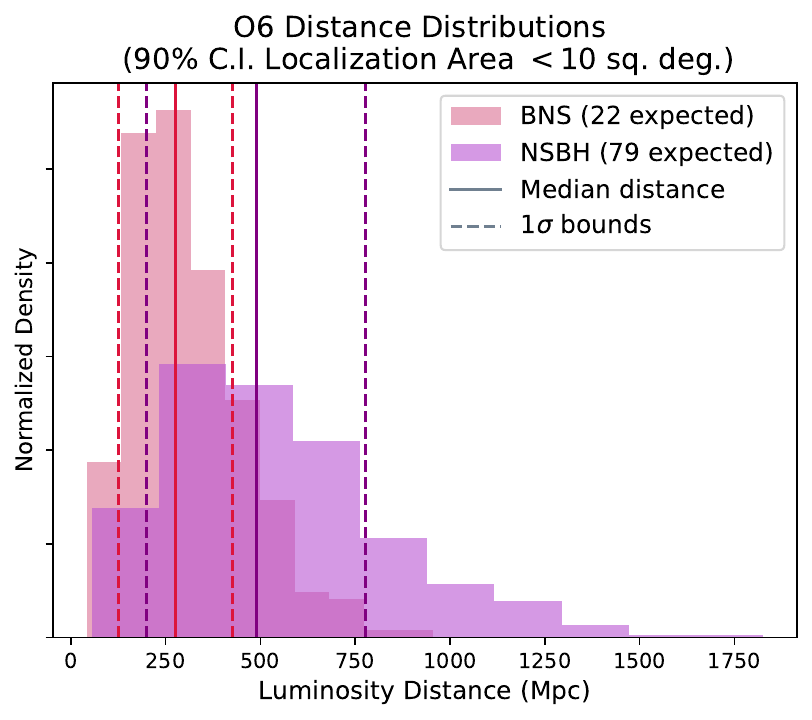}
    
    \caption{Luminosity distance distributions for well-localized BNS and NSBH mergers (shown in pink and purple, respectively) expected to be found during the LVK O5 (top) and O6 (bottom) runs as determined from the simulations by \citet{petrov_data-driven_2022a}; see \S\ref{sec:metrics}. The plots show such distributions for events localized better than 3\,deg$^2$ (left), 5\,deg$^2$ (center), and 10\,deg$^2$ (right).  Median distances are indicated with solid lines and the $1\sigma$ standard deviations are indicated with dashed lines.}
    \label{fig:distributions}
\end{figure*}

% Fig.\,\ref{fig:gantt_near} represents detectability of different kilonova models at a distance of 230\,Mpc, which is the median of the distance distribution (Fig.\,\ref{fig:distributions}) in O6 for BNS events better localized than 5\,deg$^2$ (used as reference).
Fig.\,\ref{fig:gantt_near} represents detectability of different kilonova models at a distance of 230\,Mpc, which is the median distance of BNS events better localized than 5\,deg$^2$ in the \citet{petrov_data-driven_2022a} simulation\footnote{See Fig.\,\ref{fig:distributions} for the distribution of event distances for well-localized events in the \citet{petrov_data-driven_2022a} simulations and \S\ref{sec:metrics} for further details regarding our treatment of the \citet{petrov_data-driven_2022a} simulations.}  of O6 (used as reference). 
A kilonova similar to GW170817 would be detectable for up to $\sim 3$ weeks from the merger in at least 2 filters using 55\,s exposures. A kilonova with lower electron fraction and ejecta mass would be observable for only $\sim 8$ days with 55\,s exposures and $\sim 18$ days with 1\,hr exposures. Population studies using follow-up observations of short GRBs showed that typical kilonovae may be less luminous than GW170817. For example, \cite{Ascenzi2019MNRAS} found an $H$ filter peak magnitude in the range of $[-16.2$, $-13.1]$, with GW170817 having peaked at $H \sim -15$\,mag. Observational constraints on the presence of optical counterparts to neutron star merger candidates during the third LVK observing run indicated that some kilonovae must be fainter than GW170817, possibly having $M_{ej} < 0.05$\,$M_{\odot}$ or a wide half-opening angle of the lanthanide-rich ejecta component \citep{Kasliwal2020}. 
\cite{Setzer2023} also reached the conclusion that GW170817 was brighter than the typical kilonova in the simulated kilonova population that they obtained.

Consequently, a 2-week response time as per the Roman requirements document may not be enough to enable kilonova detection over multiple epochs at (and beyond) median O6 distances for typical kilonovae. To exploit the unique potential of Roman to discover distant and neutron-rich kilonovae, a rapid ($\lesssim 7$\,days) ToO mode is needed. Even if a distant or faint counterpart is detected at later times ($\gtrsim 14$\,days), its identification with Roman will have to wait at least a few more days, when the field is imaged again and image subtraction is performed. By then, it will likely be too late for multi-wavelength follow-up in the nIR, optical and at higher energies to be performed, and there might not be enough information for unambiguous photometric classification. We therefore recommend that a faster ($\lesssim 7$\,days) ToO mode is made possible for Roman.

\begin{figure*}[t]
    \centering
    \includegraphics[width=0.49\textwidth]{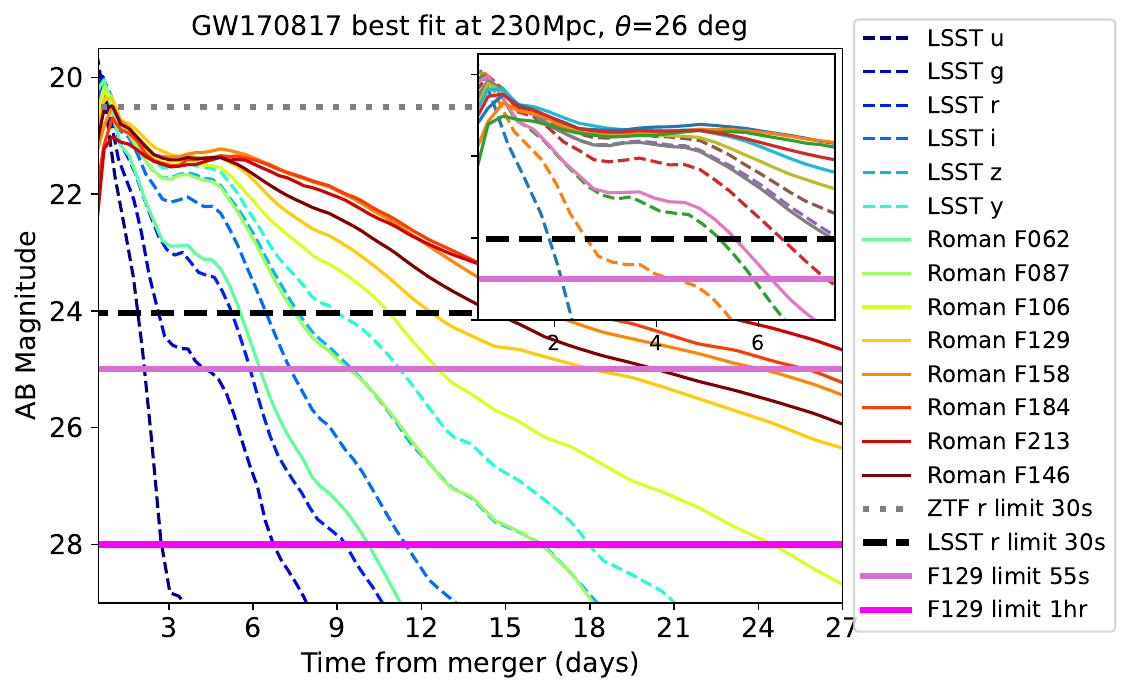}
    \includegraphics[width=0.49\textwidth]{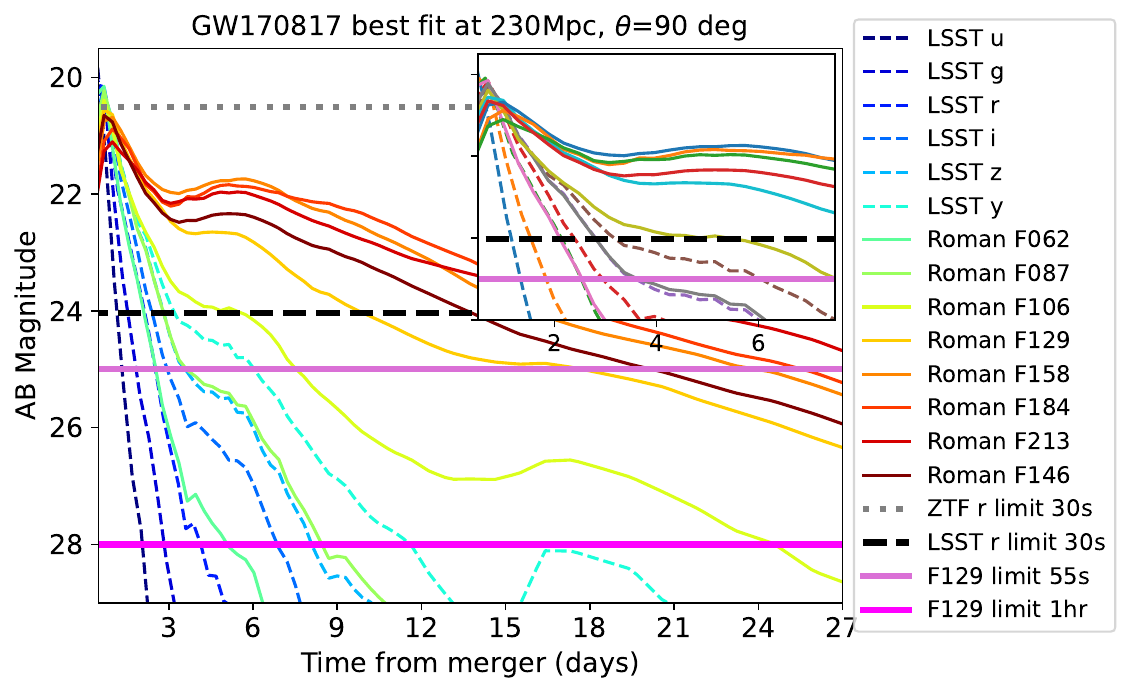}
    \includegraphics[width=0.49\textwidth]{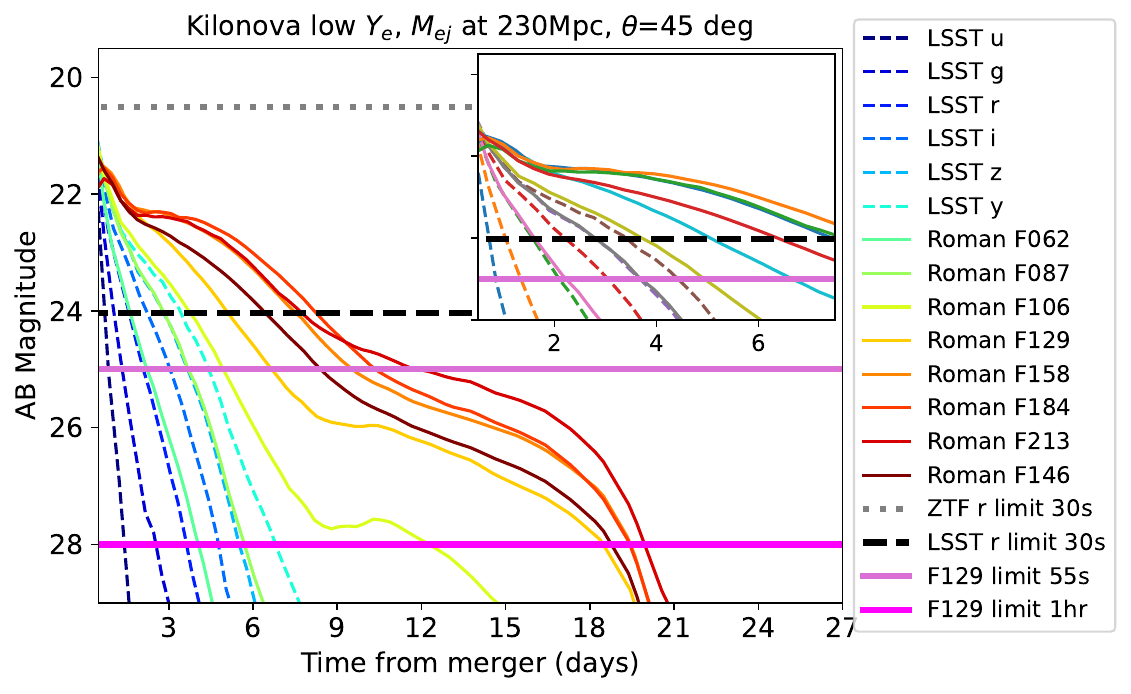}
    \includegraphics[width=0.49\textwidth]{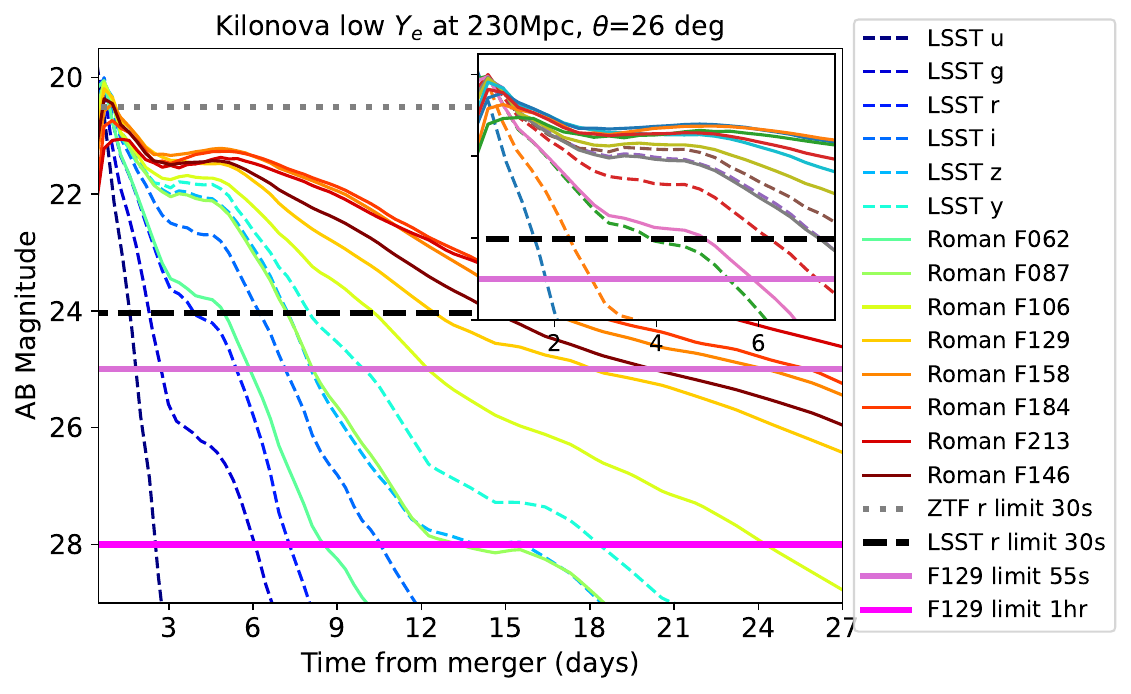}
    \includegraphics[width=0.49\textwidth]{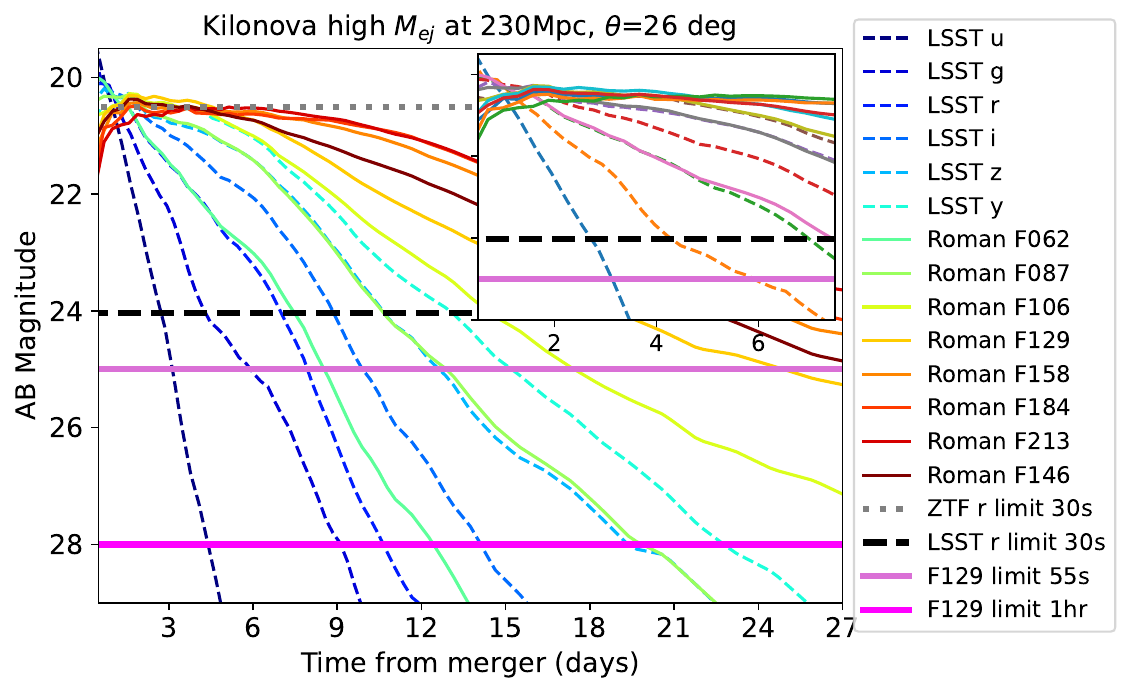}
    \includegraphics[width=0.49\textwidth]{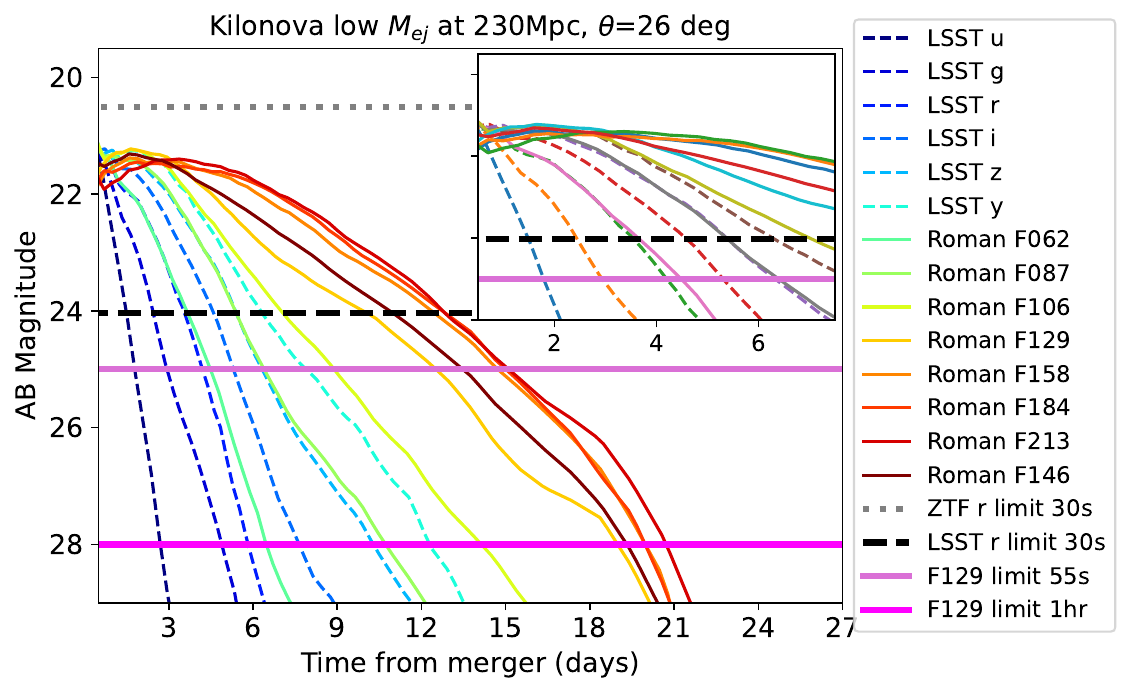}
    \caption{The luminosity, evolution timescale, and color of kilonova light curves depend on the inclination of the merging binary as well as on the mass and composition of the ejecta. In these plots, kilonovae are placed at a luminosity distance of 230\,Mpc, which is the median distance expected for BNS detection in O6 for events localized better then $A < 5$\,deg$^2$ (Fig\ref{fig:distributions}). Top: a kilonova similar to GW170817 viewed from a quasi-polar angle \citep[left, best fit to GW170817;][]{Anand2023arXivChemDist, Bulla2023MNRAS} and viewed from a quasi-equatorial angle (right), in the Rubin/LSST and Roman photometric bands. Then kilonovae are shown with most parameters unchanged from the best fit to GW170817 but with lower ejecta mass, electron fraction $Y_e$, and 45\,deg viewing angle (center left), low (center right) electron fraction, high (bottom left) and low (bottom right) ejecta masses. See Tab.\,\ref{tab:models} for details about the models. Horizontal lines represent source detection limit for the Zwicky Transient Facility \citep[ZTF;][]{Graham2019PASP, Bellm2019PASP} optical survey, Rubin, and Roman. The light curve is brighter and longer-lived in the nIR than in the optical in all cases after the first week from the merger. The advantage of observing in the nIR instead of in optical bands is particularly evident for kilonovae observed at equatorial viewing angles and with low electron fraction in their ejecta.}
    \label{fig:lightcurves_near}
\end{figure*}

For GW follow-up observations, we recommend the following minimal strategy:

\begin{itemize}
    \item Follow-up observations with Roman should be initiated for well-localized ($A < 10$\,deg$^2$) BNS and NSBH events likely to harbor an EM counterpart \citep[i.e. have remnant ejecta, according to the \texttt{HasRemnant} parameter estimated by the GW analysis;][]{foucart_remnant_2018,theligo-virgo-kagracollaboration_observing_2023}. We expect $\sim 3$ events to meet these criteria in 1.5 years of O5 and $\sim 9$ in 1.5 years of O6 and lay within the Roman field of regard (Tab.\,\ref{tab:counts}--\ref{tab:exclusion}). 

    \item Two epochs shall be obtained by tiling the whole 90\% area of interest (Fig.\,\ref{fig:skymap}) in F158 and F213 filters. The first epoch, if obtained in the first week, can have $\gtrsim 1$\,min of exposure time. The second epoch, acquired $\sim 4-7$ days later, will require $\sim 1$\,hr of exposure time because the kilonova will have faded significantly (Fig.\,\ref{fig:lightcurves_near}). 
    
    \item Late-time ($> 1$ month) templates shall be acquired with $\sim 1$\,hr exposures only on those pointings of the tiling grid where confirmed or most probable counterparts are found
\end{itemize}

\begin{figure*}[t]
    \centering
    \includegraphics[width=0.49\textwidth]{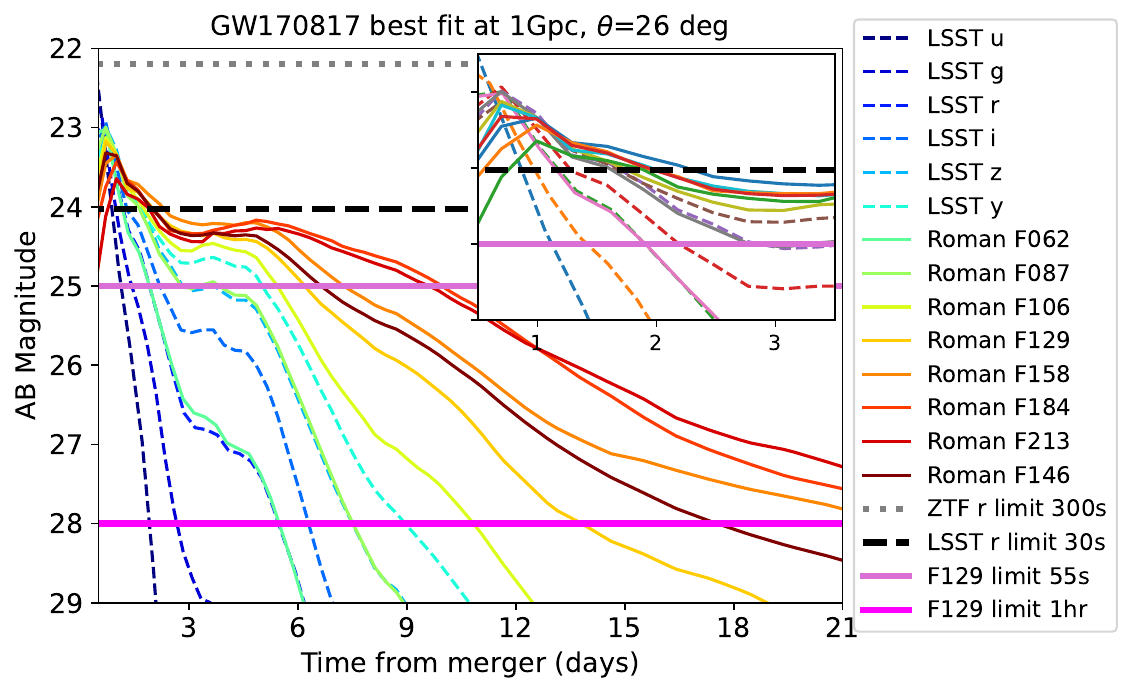}
    \includegraphics[width=0.49\textwidth]{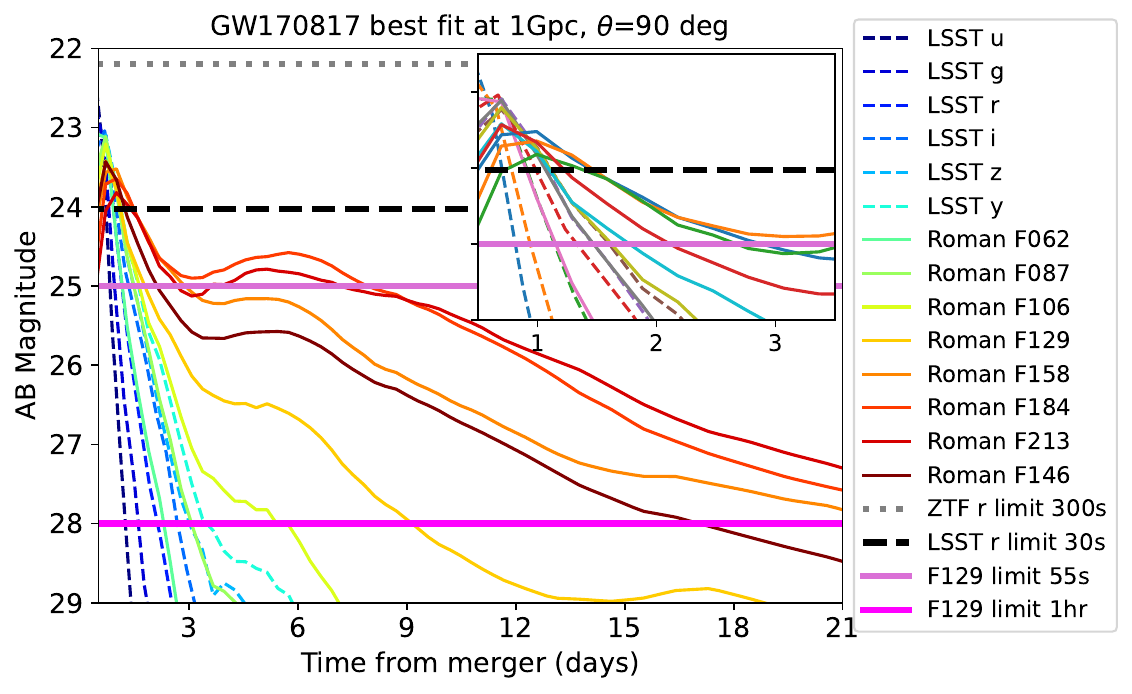}
    \includegraphics[width=0.49\textwidth]{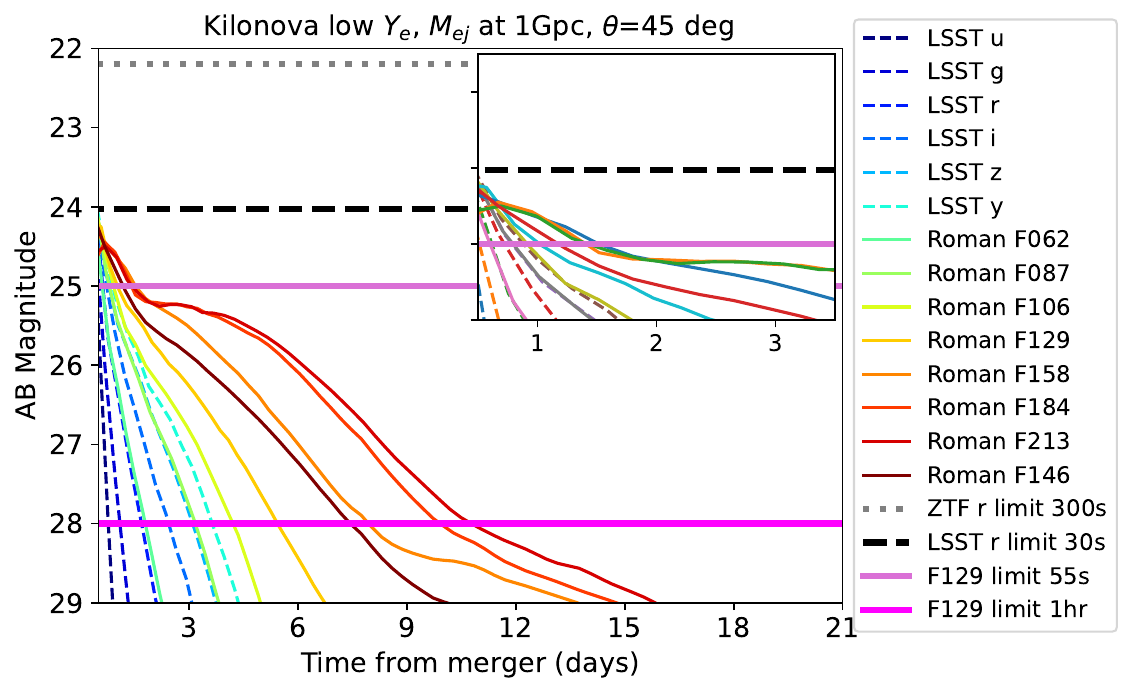}
    \includegraphics[width=0.49\textwidth]{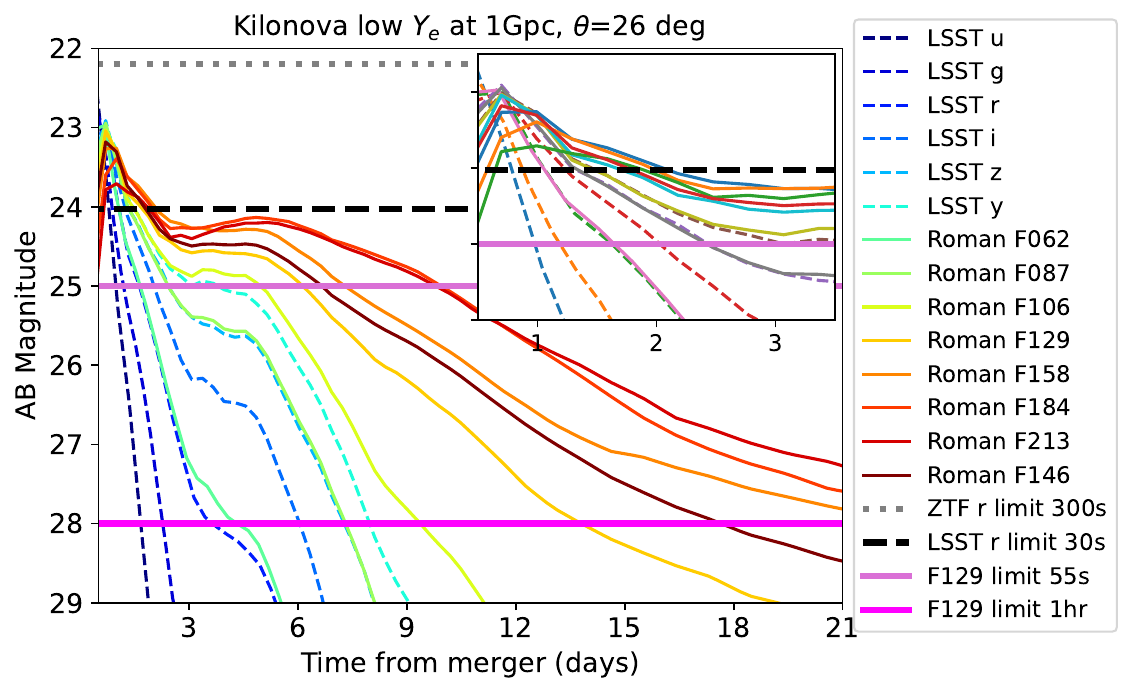}
    \includegraphics[width=0.49\textwidth]{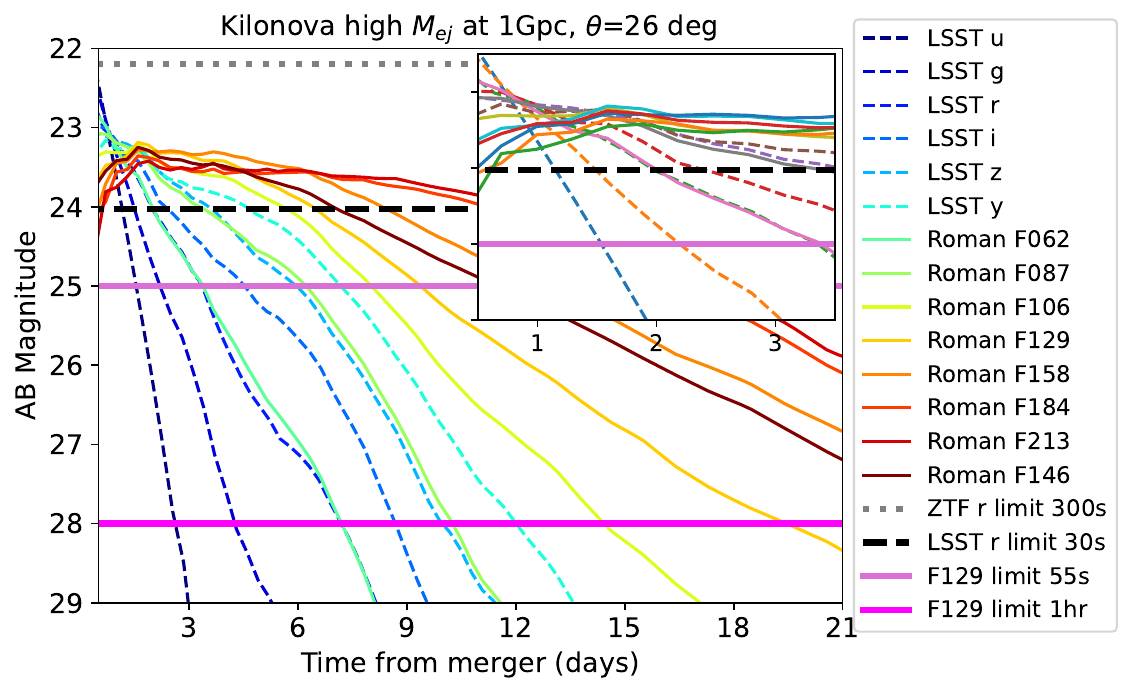}
    \includegraphics[width=0.49\textwidth]{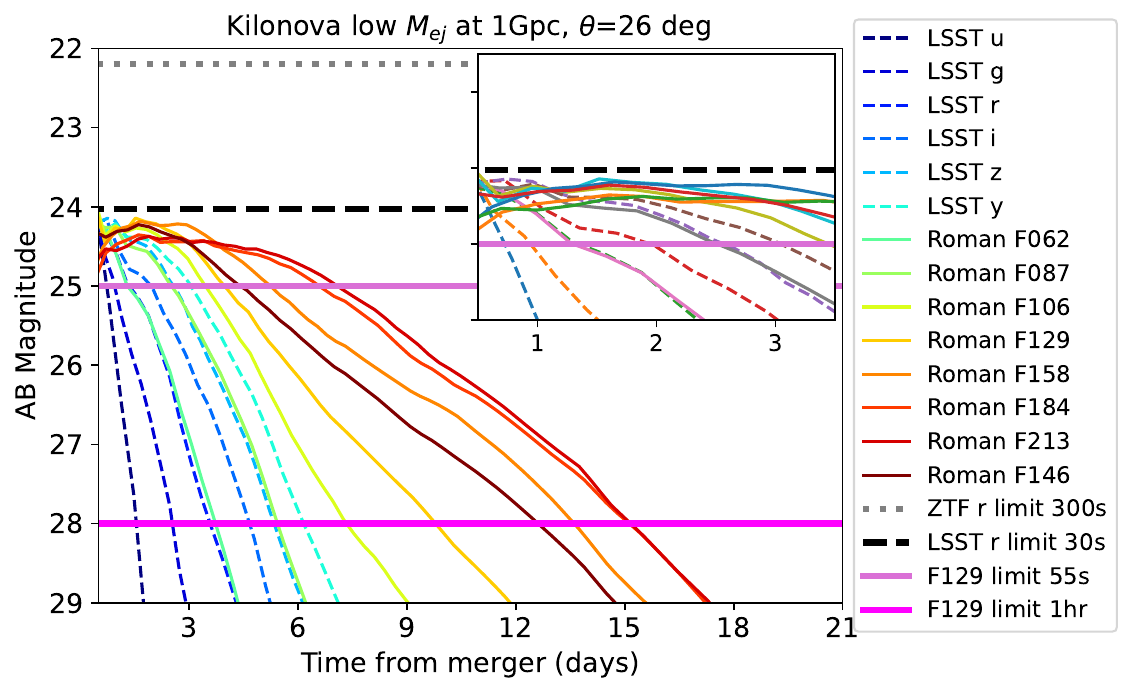}
    \caption{Same as Fig.\,\ref{fig:lightcurves_near}, but where the kilonovae are placed at a luminosity distance of 1\,Gpc.}
    \label{fig:lightcurves_far}
\end{figure*}

\begin{figure*}[t]
    \centering
    \includegraphics[width=0.49\textwidth]{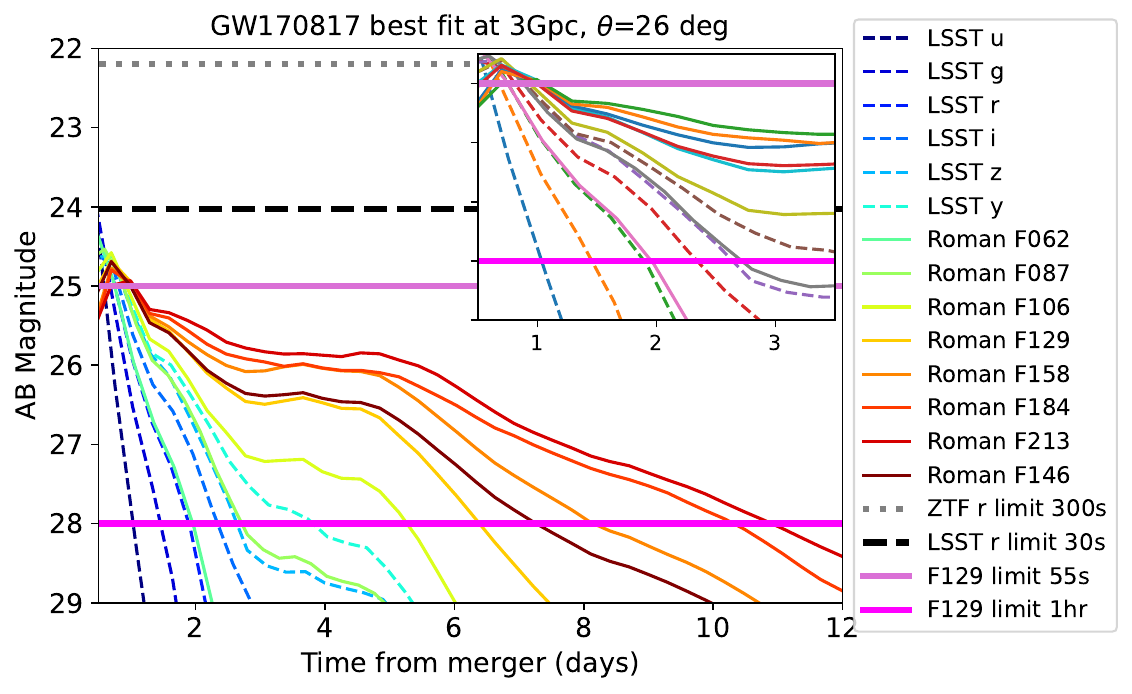}
    \includegraphics[width=0.49\textwidth]{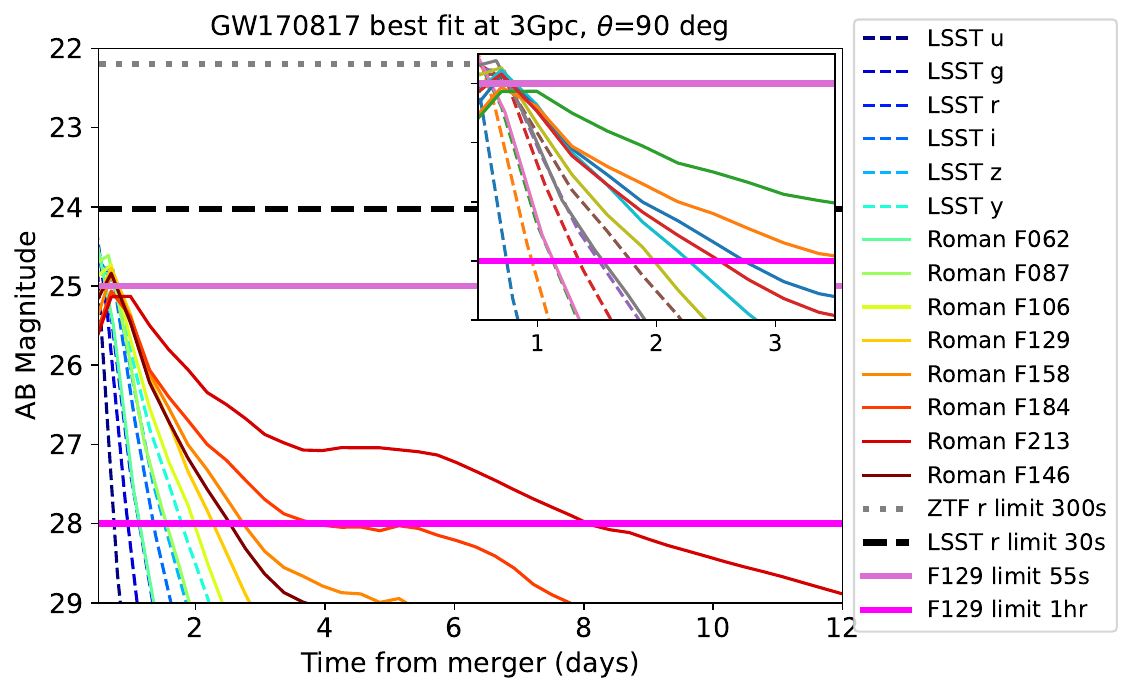}
    \includegraphics[width=0.49\textwidth]{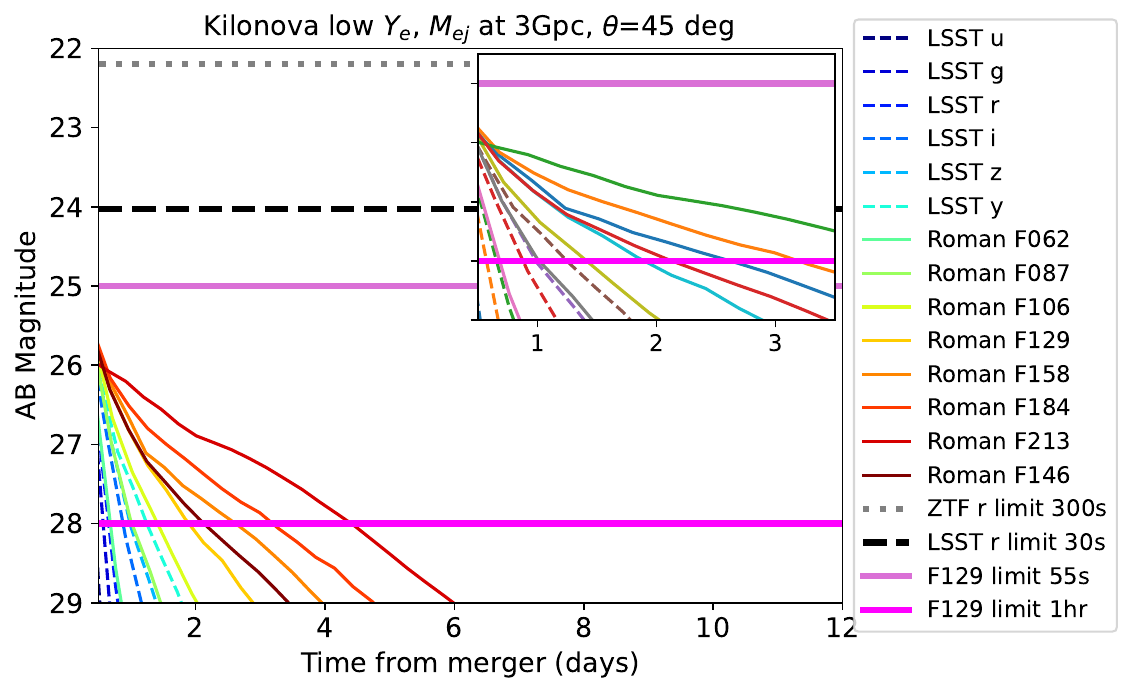}
    \includegraphics[width=0.49\textwidth]{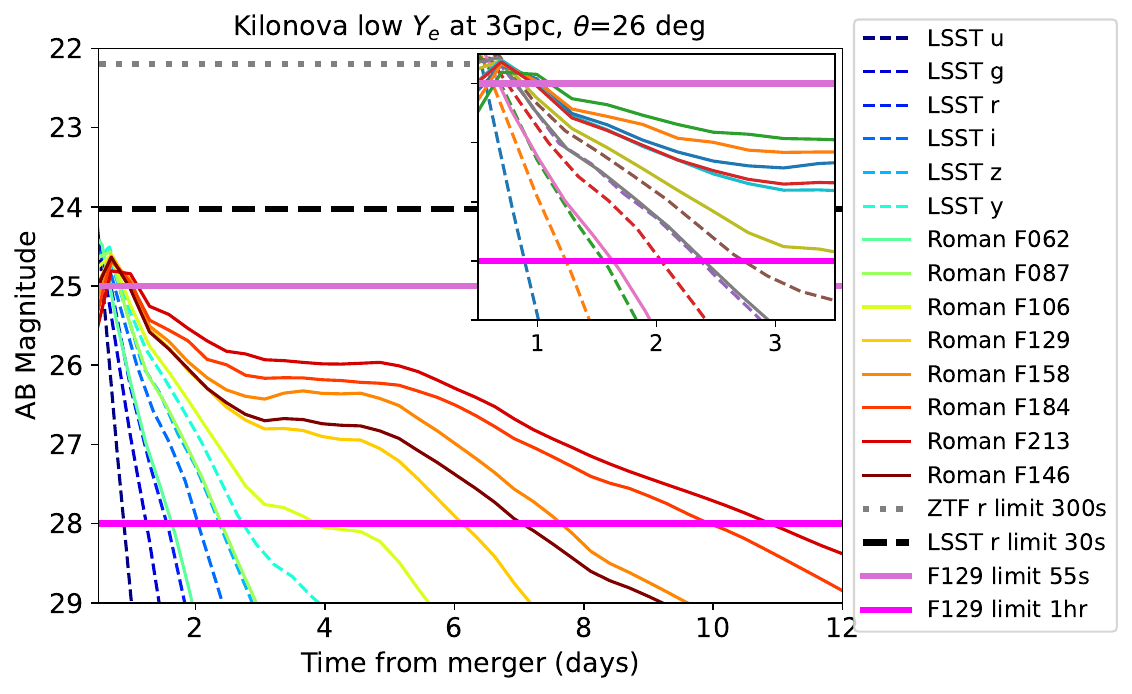}
    \includegraphics[width=0.49\textwidth]{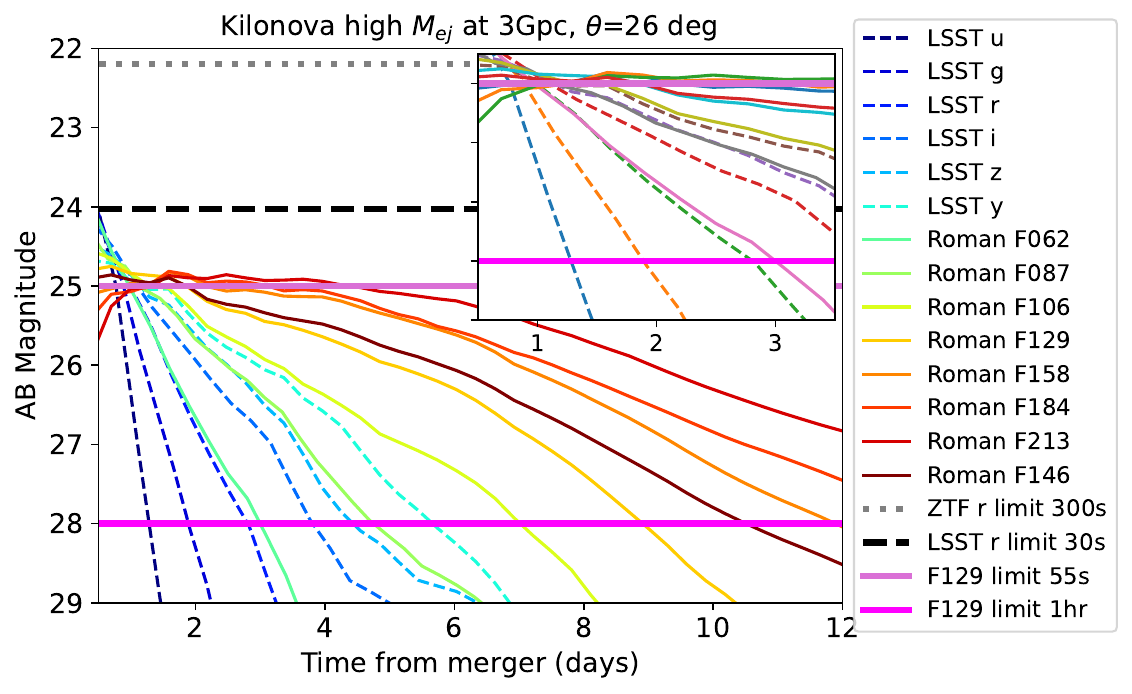}
    \includegraphics[width=0.49\textwidth]{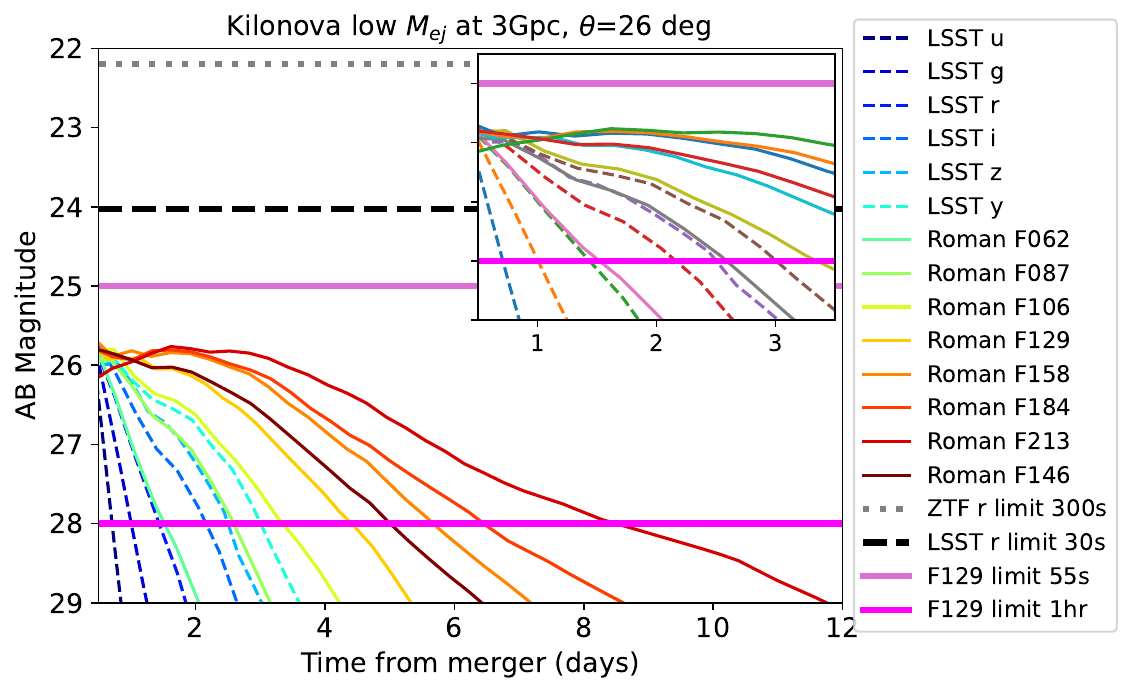}
    \caption{Same as Fig.\,\ref{fig:lightcurves_near}, but where the kilonovae are placed at a luminosity distance of 3\,Gpc.}
    \label{fig:lightcurves_veryfar}
\end{figure*}

The F158 and F213 filters were chosen for three reasons. The first is to take full advantage of the nIR capabilities of Roman, which can enable discovery of kilonovae missed by optical telescopes (Fig.\,\ref{fig:lightcurves_near}--\ref{fig:lightcurves_veryfar}), or provide excellent photometry in complementary regions of the spectrum. The second is because kilonovae are expected to be longer lived and brighter in the nIR than in the optical (Fig.\,\ref{fig:spec}--\ref{fig:lightcurves_veryfar}). The third is that ``skipping'' one of Roman's consecutive bands (in this case F184, which is placed between F158 and F213, see Fig.\,\ref{fig:spec}) will enable more robust modeling of the SED.

\subsubsection{Serendipitous}
\label{sec:minimal-serendip}

Roman has the opportunity to discover kilonovae serendipitously during the HLTD survey. Here, we consider a baseline footprint of 19\,deg$^2$, which is small in comparison to ground-based surveys such as the Zwicky Transient Facility, Pan-STARRS, ATLAS, ASAS-SN, or Rubin. For example, the Rubin Legacy Survey of Space and Time (LSST) survey footprint will cover $\sim 18,000$\,deg$^2$ \citep{Ivezic2019}.  However, the volume probed by Roman will be enormous because of the depth it can reach, thus hundreds of transients will be detected. As during GW follow-up, it will be impossible to spectroscopically classify all the transients found by Roman, the vast majority of which will be fainter than 23\,mag AB. In fact, 8m-class optical telescopes on the ground can typically characterize sources as faint as $\sim 22$\,mag in about 1\,hr of wall clock time with low-resolution spectroscopy. Photometric classification -- or at least pre-selection of kilonova candidates -- will be the only way to identify a population of these elusive sources.   

Dedicated kilonova searches have been carried out with wide-field instruments \citep[e.g.,][]{Doctor2017, Bianco2019, AnKo2020, Andreoni2021ztfrest, McBrien2021MNRAS}. Aided by observations of GW170817 kilonova and the availability of kilonova model grids \citep[e.g.,][]{Kasen17, Bulla2019}, these searches have provided tools and know-how to discriminate between kilonova candidates and the most common classes of extragalactic transients. In the infrared, more work needs to be done to achieve robust photometric classification of kilonova candidates. As explained above, multiple detections in at least two filters are the absolute minimum requirement to be able to monitor the evolution of luminosity and color of the transients. A systematic cadence also makes it possible to establish if a transient is fast (i.e., it appears and fades away within a few days), as expected from kilonovae, or if it is evolving slowly like supernovae and tidal disruption events.

For the minimal HLTD survey, we envision observations that are:
\begin{itemize}
    \item regularly cadenced in time; same-night observations of the same field in different filters are strongly encouraged. Same-night observations of the same field in the same filter have a much smaller impact in the nIR than in the optical (see for example Fig.\,\ref{fig:lightcurves_near}), thus they are not considered in this work
    \item cadenced such that a kilonova can be detected at least twice in at least two filters (preferably F158 and F213)
    \item similarly to the case of ToO observations, we deem it more useful to carry out the HLTD survey in nIR filters and avoid the broadband filter F146. This will enable more robust photometric classification, SED modeling, and will provide a unique complement to synergetic observations in the optical and at other wavelengths
    \item including deep ($\sim$1 \,hr) observations in at least two filters 
\end{itemize}

These recommendations are derived from the metrics described in section \S\ref{sec:metrics}.

\subsection{Preferred Strategies}

Preferred strategies will increase the scientific return of the minimal strategies by providing better-sampled light curves in 3 nIR bands.

\subsubsection{Target of Opportunity}

The preferred ToO strategies for GW follow-up are based on the minimal strategy described in \S\ref{sec:minimal-ToO}. 
We recommend the following preferred strategy:

\begin{itemize}
    \item Roman follow-up should be initiated with the preferred strategy for well-localized ($A < 10$\,deg$^2$) BNS and NSBH events likely to harbor an EM counterpart. The minimal strategy should be applied for the follow-up of less well-localized GW sources (for example, 10\,deg$^2 < A < 15$\,deg$^2$) to increase the total sample of identified kilonovae. The ongoing fourth LVK observing run (O4), which is due to continue until the end of 2024, will give us a better handle on the event rates to expect in O5 and O6.

    \item Three epochs shall be obtained by tiling the whole 90\% area of interest (Fig.\,\ref{fig:skymap}) in F158, F184, and F213 filters. This will provide data for excellent SED modeling of the kilonova as well as the other candidates, enabling robust photometric classification and monitoring of the temperature and luminosity evolution. This is even more valuable if observations are conducted in synergy with optical observatories. Like for the minimal strategy, the first epoch, if obtained early, can have $\gtrsim 1$\,minute of exposure time. The second and third epochs, acquired $\sim 4$ and $\sim 8$ days later, will require $\sim 1$\,hr of exposure time.

    \item If a counterpart is identified, a fourth epoch should be obtained in F158, F184, and F213 filters with $> 1$\,hr of exposure time $\gtrsim 3$ weeks from the merger. This epoch will be important to estimate the amount of $r$-process elements from the second and third abundance peaks \citep[e.g.,][]{Hotokezaka2020, Hotokezaka2021, Kasliwal2022}, which is one of the science goals that Roman is uniquely capable of achieving.  
    
    \item Late-time ($> 1$ month) templates shall be acquired with $\gtrsim 1$\,hr of exposure time in the whole region tiled during the follow-up if a counterpart is found only after the second epoch. These will not only be used to obtain accurate photometry of the EM counterpart (free of contamination from the host and from the kilonova's own flux), but they will also enable serendipitous fast-transient science in the large volume observed during the follow-up. 
\end{itemize}

\subsubsection{Serendipitous}

In the same spirit, for the preferred strategies for serendipitous kilonova discovery we recommend HLTD observations that are:

\begin{itemize}
    \item regularly cadenced in time and with multi-band observations occurring on the same day
    \item cadenced such that a kilonova can be detected at least twice in at least three filters (excluding F146), or at least three times in at least two filters
    \item including deep ($\sim$1 \,hr) observations in at least two filters 
    \item carried across a larger footprint, possibly with area $A > 40$\,deg$^2$ (see \S\ref{sec:metrics}), to enable the identification of a sample of elusive kilonovae with low ejecta mass and/or with high neutron content in the ejecta (i.e., with low electron fraction) 
\end{itemize}

These recommendations are derived from the metrics described in section \S\ref{sec:metrics}.

\section{Metrics and Figures of Merit}
\label{sec:metrics}
%(iv) the impact of different observational strategy choices on the science investigation, expressed via appropriate metrics or figures of merit that the CCS definition committees can readily utilize. 

\subsection{Target of Opportunity}

The number of kilonovae expected in the volume that Roman will probe is set by the instrument limiting magnitude combined with the neutron star merger rate (i.e., number of events per unit of time and volume), the kilonova luminosity function (i.e., the distribution of peak luminosity), or the distribution of physical quantities such as the ejecta mass, velocity, and composition. For ToO observations, the number of expected LVK triggers, their localization, and the expected distance distribution are also important factors.  Some constraints on the kilonova luminosity function were derived from the study of kilonova candidates found during follow-up of short GRBs \citep[e.g.,][]{Gompertz2018, Ascenzi2019MNRAS, Rossi2020} and the non-detection of counterparts --- barring GW170817 --- during the follow-up of GW events in the second and third LVK observing run \citep{Kasliwal2020}. However, the kilonova luminosity function remains highly uncertain, so we present results obtained for a number of kilonova models as described at the beginning of \S\ref{sec:strategies} and in Tab.\,\ref{tab:models}. 

The number of detected counterparts via ToO observations will depend on the number and nature of well-localized triggers that will be issued by the LVK network. We use the data-driven observing scenario simulations of \citet{petrov_data-driven_2022a} to determine the expected number and distance distribution of ToO triggers from BNS and NSBH events in O5 and O6. We assume a fiducial duration of 1.5 years for both O5 and O6, and use the current best estimates at time of writing for the BNS (NSBH) merger rate of $210^{+240}_{-120}\ (8.6^{+9.7}_{-5.0})\ \mathrm{Gpc}^{-3}\mathrm{year}^{-1}$ \citep{theligo-virgo-kagracollaboration_observing_2023,coughlin_ligo/virgo/kagra_2022,singer_lpsinger/observing-scenarios-simulations:_2022}. For O5 and O6, we determine the expected number of ToO triggers for each of 3 localization thresholds by selecting only those events from the \citet{petrov_data-driven_2022a} simulations with 90\% confidence interval (C.I.) GW sky localization areas $\leq$ 3, 5, and 10 square degrees.  The simulated occurrence rate for such events is then scaled to the astrophysical rate, combining the assumed log-normal astrophysical rate error with Poisson counting statistics but neglecting Monte Carlo uncertainty, as described in (e.g.) \citet{petrov_data-driven_2022a,theligo-virgo-kagracollaboration_observing_2023,kiendrebeogo_updated_2023}.\footnote{ A detailed explanation of these statistical considerations and Python code for the calculation as performed for \citet{theligo-virgo-kagracollaboration_observing_2023} can be found in \citet{singer_lpsinger/observing-scenarios-simulations:_2022}.} Assuming a field of view of 0.281 deg$^2$, it would take 11, 18, and 36 pointings to cover those sky areas with Roman. Analysis for larger sky areas can be carried out in the future, depending on updated rates and specific constraints on Roman ToO observations. 

The expected total number of GW events for each threshold are presented in Tab.~\ref{tab:counts}, and the corresponding expected number of Roman ToO triggers --- accounting for Roman's field of regard and assuming isotropy of GW events \citep{essick_anisotropy_2023} --- are presented in Tab.\,\ref{tab:exclusion}. The corresponding distributions of event distances are shown in Fig.\,\ref{fig:distributions}, with summary statistics presented in Tab.\,\ref{tab:distances}. We neglect the impact of erroneous ToO triggers due to false-alarm GW alerts from terrestrial sources, as such false-alarm alerts are unlikely to meet stringent localization requirements. Additionally, it is worth noting that at most 10-20\% of NSBH mergers are expected to yield an electromagnetically-bright counterpart \citep{biscoveanu_population_2023,broekgaarden_impact_2021,drozda_black_2022,fragione_black-hole-neutron-star_2021,roman-garza_the_2021}. As such, the values quoted here for NSBH merger ToO triggers are likely optimistic in terms of prospects for detection of NSBH EM counterparts. These numbers should be nonetheless representative of the expected ToO trigger rate; even the absence of an EM counterpart is scientifically interesting (e.g., \citealt{coughlin_implications_2020a}) and a sufficiently well-localized merger will likely prompt ToO observations regardless of the inferred NSBH masses.

We stress that, even during ToO observations, the mere detection of a transient is unlikely going to be enough to identify a robust kilonova candidate at large distances. Deep multi-band photometry is required.

\subsection{Serendipitous}

Expected numbers of sources detectable in a 19\,deg$^2$ survey footprint over a range of absolute magnitudes are presented in Tab.\,\ref{tab:rates}. This represents the potential of detecting kilonovae reaching a certain luminosity when the observations are performed.

In order to quantify the impact of different observing strategies for serendipitous kilonova identification, we developed a \texttt{multi\_detect} metric, which returns the number of kilonovae detectable at least N times in at least M filters. We injected 500,000 kilonovae for each model, uniformly distributed in comoving volume between 100 Mpc and 7\,Gpc ($z\sim 1$). We then assumed a set of regular time gaps between observations and shifted the phase of the kilonovae to simulate realistic recovery rates. The time gaps that we explored are of 2, 4, 8, and 16 days between returns to the same field with the same filter(s). The results of our analysis are summarized in Tab.\,\ref{tab:serendip1_noF146_2filt}--\ref{tab:serendip5_F158-and-F213_3det_2filt}. We assumed a footprint of 19\,deg$^2$ in all tables but in Tab.\,\ref{tab:serendip5_F158-and-F213_3det_2filt}, where the area was doubled. The number of kilonova detections scale linearly with the area, so it is straightforward to calculate the number of events expected if wider or smaller footprints are considered. 

Based on Tab.\,\ref{tab:serendip1_noF146_2filt}--\ref{tab:serendip5_F158-and-F213_3det_2filt}, we can draw the following conclusions:

\begin{itemize}
    \item Employing 55\,s exposures for the survey can yield $\sim 1$ kilonova per year only for observations performed in all filters with a cadence of $1-4$ days. This can be hardly considered efficient. If $\sim 1$\,minute exposures will be employed in the HLTD survey, the survey area should be increased to $> 40$\,deg$^2$ to recover $\sim 1$ kilonova similar to GW170817, or kilonovae with higher ejecta masses.
    
    \item Deep multi-band observations ($\sim 1$\,hr of exposure time) can yield up to $\sim 31$ kilonovae similar to GW170817 per year, detected in at least two filters (Tab.\,\ref{tab:serendip1_noF146_2filt}) in the optical and nIR, assuming a survey footprint of 19\,deg$^2$. If only the F158 and F213 filters are employed, up to $\sim 11$ ($\sim 7$) kilonovae similar to GW170817 can be recovered per year at least twice (three times) in both filters, as shown in Tab.\ref{tab:serendip2_F158-and-F213_2det_2filt} (Tab.\,\ref{tab:serendip5_F158-and-F213_3det_2filt}). If the observations are spaced by at most 4 days, we can expect to detect up to $\sim$8 ($\sim$4) kilonovae similar to GW170817 in the F158 and F213 filters. If the observations are spaced by at most 8 days, we can expect to detect up to $\sim$6 ($\sim$4) kilonovae similar to GW170817 in the F158 and F213 filters.
    
\end{itemize}

We recommend that the HLTD survey includes observations of the footprint in at least two filters (preferably F158+F213) with a $\lesssim 8$ days time gap between returns to the same field. Observations in the two filters should occur on the same day to enable color and color evolution to be measured. Such a strategy would greatly complement Rubin observations, if a synergy between the two observatories is put in place (see also \S\ref{sec:synergies}).  

Future studies should address whether requiring at least 2 detections in at least 3 filters (Tab.\,\ref{tab:serendip3_noF146_3filt}), as opposed to requiring at least 3 detections in at least 2 filters (Tab.\,\ref{tab:serendip4_noF146_3det}--\ref{tab:serendip5_F158-and-F213_3det_2filt}), better enables photometric classification and separation of kilonova candidates from other nIR transients.

\begin{table}[]
    \centering
    \begin{tabular}{lccccc}
    \hline\hline
    \multirow{2}{*}{Area}  & \multicolumn{4}{c}{Estimated Number of GW Events} \\ 
        \cline{2-5}  
 & O5 BNS & O5 NSBH & O6 BNS & O6 NSBH \\ 
\hline 
$\leq3$ deg$^2$  & $2.2^{+2.7}_{-1.2}$ & $0.3^{+0.4}_{-0.2}$ & $5.5^{+6.8}_{-3.1}$ & $0.8^{+1.1}_{-0.6}$\\ 
$\leq5$ deg$^2$ & $3.6^{+4.5}_{-2.0}$ & $0.6^{+0.8}_{-0.4}$ & $10.7^{+13.3}_{-6.0}$ & $1.6^{+2.1}_{-1.1}$\\ 
$\leq10$ deg$^2$ & $6.7^{+8.3}_{-3.7}$ & $1.1^{+1.5}_{-0.8}$ & $19.6^{+24.3}_{-11.0}$ & $3.0^{+4.0}_{-2.1}$\\ 
    \hline
    \end{tabular}
    \caption{Estimated number of GW events, broken down across merger type and LVK observing run, for three different thresholds of GW sky localization area. Quoted values are the median and 90\% C.I. ToO trigger count, assuming log-normal astrophysical merger rate uncertainty and Poisson counting uncertainty; see \S\ref{sec:metrics} for details.}
    \label{tab:counts}
\end{table}

\begin{table}[]
    \centering
    \begin{tabular}{lccccc}
    \hline\hline
    \multirow{2}{*}{Area}  & \multicolumn{4}{c}{Estimated Number of ToO Triggers} \\ 
        \cline{2-5}  
 & O5 BNS & O5 NSBH & O6 BNS & O6 NSBH \\ 
\hline 
$\leq3$ deg$^2$  & $0.9^{+1.1}_{-0.5}$ & $0.1^{+0.2}_{-0.1}$ & $2.3^{+2.8}_{-1.3}$ & $0.3^{+0.5}_{-0.2}$\\ 
$\leq5$ deg$^2$ & $1.5^{+1.9}_{-0.8}$ & $0.2^{+0.3}_{-0.2}$ & $4.4^{+5.5}_{-2.5}$ & $0.7^{+0.9}_{-0.5}$\\ 
$\leq10$ deg$^2$ & $2.8^{+3.4}_{-1.5}$ & $0.5^{+0.6}_{-0.3}$ & $8.1^{+10.0}_{-4.5}$ & $1.2^{+1.6}_{-0.9}$\\ 
    \hline
    \end{tabular}
    \caption{Estimated number of GW ToO triggers that will fall within the Roman field of regard.}
    \label{tab:exclusion}
\end{table}

\begin{table}[]
    \centering
    \begin{tabular}{lccccc}
    \hline\hline
    \multirow{2}{*}{Area}  & \multicolumn{4}{c}{Luminosity Distance (Mpc)} \\ 
        \cline{2-5}  
 & O5 BNS & O5 NSBH & O6 BNS & O6 NSBH \\ 
\hline 
$\leq3$ deg$^2$  & $130^{+71}_{-71}$ & $226^{+138}_{-138}$ & $171^{+82}_{-82}$ & $296^{+154}_{-154}$\\ 
$\leq5$ deg$^2$ & $164^{+83}_{-83}$ & $284^{+167}_{-167}$ & $230^{+105}_{-105}$ & $380^{+220}_{-220}$\\ 
$\leq10$ deg$^2$ & $200^{+113}_{-113}$ & $374^{+226}_{-226}$ & $274^{+142}_{-142}$ & $473^{+282}_{-282}$\\ 
    \hline
    \end{tabular}
    \caption{Median and $1\sigma$ bounds of the luminosity distance distributions shown in Fig.~\ref{fig:distributions}.}
    \label{tab:distances}
\end{table}

\begin{table}[]
    \centering
    \begin{tabular}{lccccc}
    \hline\hline
Lim & M=$-14$ & M=$-15$ & M=$-16$ & M=$-17$ \\
\hline
25.0  & $0.07^{+0.08}_{-0.04}$ & $0.24^{+0.27}_{-0.13}$ & $0.74^{+0.84}_{-0.42}$ & $2.15^{+2.45}_{-1.23}$ \\
28.0  & $2.15^{+2.45}_{-1.23}$ & $5.73^{+6.54}_{-3.27}$ & $13.89^{+15.88}_{-7.94}$ & $30.48^{+34.84}_{-17.42}$ \\
    \hline
    \end{tabular}
    \caption{Number of kilonovae expected to be there every year in the HLTD survey area, in any filter, with absolute magnitudes ranging from $M = -14$\,mag to $M = -17$\,mag given magnitude limits of 25\,mag and 28\,mag, which correspond to approximate magnitude limits for 55s and 1hr of Roman integration time. A BNS rate of $210^{+240}_{-120}$\,Gpc$^{-3}$\,y$^{-1}$ is assumed \citep{theligo-virgo-kagracollaboration_observing_2023,coughlin_ligo/virgo/kagra_2022,singer_lpsinger/observing-scenarios-simulations:_2022} and an area of 19\,deg$^2$ for the Roman HLTD survey.}
    \label{tab:rates}
\end{table}

\section{Synergies} \label{sec:synergies}

To maximize multi-wavelength monitoring at high cadence, synergy with multi-wavelength time-domain surveys should be organized. A particularly productive synergy can be put in place between Roman and Rubin Observatory, which will provide cadenced observations with neighboring passbands from 300\,nm to 2000\,nm. We recommend that Rubin Deep Drilling Fields (DDF\footnote{\url{https://www.lsst.org/scientists/survey-design/ddf}}, which include COSMOS, ELAIS S1, XMM-LSS, and Extended Chandra Deep Field-South) are considered among the preferred regions of the sky where the HLTD survey will occur. A synergy between Roman and Rubin will also be valuable during ToO observations \citep[see for example][regarding GW follow-up with Rubin]{Andreoni2022RubinToO}, as they will likely be the only facilities that can detect distant or intrinsically faint kilonovae. Cadenced observations in both optical and nIR bands will enable accurate modeling of multi-component kilonovae such as GW170817 \citep[e.g.,][]{Cowperthwaite:2017dyu, Kasen17, Kasliwal17, Smartt17, Villar17}. 

While kilonovae could be too faint to be detected in the optical (even by Rubin, which can observe $> 2$ magnitudes deeper than most wide-field telescopes at comparable exposure times, see Fig.\,\ref{fig:lightcurves_near}--\ref{fig:lightcurves_veryfar}), Roman will be able to successfully identify kilonovae even when the ejecta are particularly neutron rich or have low mass, and when the viewing angle is more equatorial. We note that extremely low ejecta mass, even down to orders of magnitude below what considered in the ``low $M_{ej}$'' case considered here, are expected for high mass mergers such as GW190425 (\citealt{Camilletti2022, Foley2020}; although with large uncertainties depending on various factors including mass ratio and neutron star equation of state), one of the two high confidence binary neutron star mergers detected so far in GWs. Such low ejecta masses would result in a kilonova that is orders of magnitude fainter than GW170817, and potentially fainter than the models considered here, which would lower the probability of detection from ground-based optical searches, including Rubin. Given the broad neutron star mass distribution found by LVK \citep{LVKpop2023}, and the fact that the more massive events are also louder (in GWs) than their less massive counterparts at the same distance, it is also possible that the next binary neutron star detections will include a significant fraction of high mass binaries like GW190425. These will potentially have very low ejecta mass, making Roman a unique telescope to detect electromagnetic counterparts to LVK GW sources. 

This effort will be consistent with other synergies that are currently being organized. For example, systematic radio monitoring of the Rubin DDF with the upcoming Deep Synoptic Array \footnote{\url{https://www.deepsynoptic.org/}} (DSA-2000). These synergies will naturally result in rich data sets to mine for kilonovae and will be crucial to diagnose the nature of transients of different classes, especially those too faint (intrinsically or because of their distance) for spectroscopic follow-up observations. These will constitute the overwhelming majority of transients discovered with Roman and Rubin. 

\section{Summary and Future Work}
\label{sec:conclusion}

We have discussed here how the unique capabilities of Roman can benefit multi-messenger astronomy and kilonova astrophysics to aid the definition of the Core Community Surveys.

Roman is uniquely placed to enable GW multi-messenger science. However, this may be achievable if Roman observing programs include ToO observations with a response time more rapid than the 2 weeks indicated in the Roman requirements document. We expect that 1--6 ToO observations can be performed with Roman to follow-up well localized ($A < 10$\,deg$^2$, 90\% C.I.) BNS or NSBH mergers during 1.5 years of LVK O5, or $\sim$4--21 during 1.5 years of O6.

For the HLTD survey, we recommend that it includes observations of the footprint in at least two filters (preferably the F158 and F213) to be acquired on the same day, and repeated regularly with a cadence of $\lesssim 8$ days. Assuming a footprint of 19\,deg$^2$ and 1\,hr exposure times, we expect this strategy to yield $\sim$ 1--8 kilonovae similar to GW170817 per year and build a rich sample of 5--40 serendipitously discovered kilonovae by the end of the 5-year survey. As of July 2023, no confirmed kilonovae have been found in wide-field survey data independently of gamma-ray burst or GW triggers, thus such a sample of kilonovae discovered by Roman will be invaluable to understand their diversity and the possible dependence on redshift. 

A choice of the time-domain survey footprint that maximizes synergetic observations with Rubin Observatory is strongly recommended (see \S\ref{sec:synergies}).

Real-time data processing will be a necessary ingredient to the success of multi-messenger projects with Roman. It will enable the identification of kilonovae while they are still bright and the activation of multi-wavelength follow-up. This is important for both ToO observations and serendipitous kilonova searches. 

Future work may include studying in greater detail kilonova photometric classification in the infrared, which can improve the minimum requirements for the survey design. The ability to use Roman (or Roman + Rubin) light curves to discriminate between kilonovae and other types of extragalaxtic transients such as supernovae will be crucial. The creation of a metrics analysis framework (MAF), similar to the one developed\footnote{\url{https://github.com/lsst/rubin_sim}} for Rubin LSST, can greatly help evaluate what the best CCS strategies can be for a broad range of science cases.

\begin{acknowledgments}
 We thank the anonymous referee for the useful feedback provided during the review process. MWC and AT acknowledge support from the National Science Foundation with grant numbers PHY-2010970 and OAC-2117997.
\end{acknowledgments}

\software{astropy \citep{2013AA...558A..33A,2018AJ....156..123A}, matplotlib \citep{Hunter:2007}}

\begin{table*}[]
    \centering
    \begin{tabular}{cccccccccc}
\hline\hline
Model & KN/y & KN/y & gap 2d & KN/y & gap 4d & KN/y & gap 8d & KN/y & gap 16d\\
 & gap 1d & gap 2d & /gap 1d & gap 4d & /gap 1d & gap 8d & /gap 1d & gap 16d & /gap 1d\\
\hline
\multicolumn{10}{c}{Exposure: 55s } \\
\hline
GW170817$_{pol}$ & $0.7^{+0.8}_{-0.4}$ & $0.6^{+0.7}_{-0.4}$ & 0.9 & $0.6^{+0.6}_{-0.3}$ & 0.8 & $0.4^{+0.4}_{-0.2}$ & 0.6 & $0.2^{+0.2}_{-0.1}$ & 0.3\\
GW170817$_{eq}$ & $0.2^{+0.2}_{-0.1}$ & $0.2^{+0.2}_{-0.1}$ & 0.9 & $0.1^{+0.2}_{-0.1}$ & 0.8 & $0.1^{+0.1}_{-0.1}$ & 0.7 & $0.1^{+0.1}_{-0.0}$ & 0.3\\
low $Y_{e}$, $M_{ej}$ & $0.1^{+0.1}_{-0.0}$ & $0.1^{+0.1}_{-0.0}$ & 0.9 & $0.1^{+0.1}_{-0.0}$ & 0.8 & $< 0.1$ & 0.4 & $< 0.1$ & 0.2\\
low $Y_{e}$ & $0.5^{+0.5}_{-0.3}$ & $0.4^{+0.5}_{-0.2}$ & 0.9 & $0.4^{+0.4}_{-0.2}$ & 0.9 & $0.3^{+0.3}_{-0.2}$ & 0.6 & $0.1^{+0.2}_{-0.1}$ & 0.3\\
low $M_{ej}$ & $0.6^{+0.7}_{-0.3}$ & $0.5^{+0.5}_{-0.3}$ & 0.8 & $0.3^{+0.4}_{-0.2}$ & 0.5 & $0.2^{+0.2}_{-0.1}$ & 0.3 & $0.1^{+0.1}_{-0.1}$ & 0.2\\
high $M_{ej}$ & $1.3^{+1.5}_{-0.7}$ & $1.3^{+1.5}_{-0.7}$ & 1.0 & $1.3^{+1.4}_{-0.7}$ & 1.0 & $0.9^{+1.0}_{-0.5}$ & 0.7 & $0.5^{+0.6}_{-0.3}$ & 0.4\\
\hline
\multicolumn{10}{c}{Exposure: 1hr } \\
\hline
GW170817$_{pol}$ & $14.7^{+16.8}_{-8.4}$ & $14.7^{+16.8}_{-8.4}$ & 1.1 & $12.8^{+14.6}_{-7.3}$ & 0.9 & $5.7^{+6.5}_{-3.3}$ & 0.4 & $4.2^{+4.9}_{-2.4}$ & 0.3\\
GW170817$_{eq}$ & $2.5^{+2.9}_{-1.5}$ & $1.8^{+2.1}_{-1.1}$ & 0.7 & $1.3^{+1.5}_{-0.8}$ & 0.5 & $1.3^{+1.4}_{-0.7}$ & 0.5 & $0.9^{+1.0}_{-0.5}$ & 0.3\\
low $Y_{e}$, $M_{ej}$ & $1.5^{+1.7}_{-0.8}$ & $1.4^{+1.6}_{-0.8}$ & 0.8 & $1.0^{+1.2}_{-0.6}$ & 0.6 & $0.7^{+0.8}_{-0.4}$ & 0.4 & $0.3^{+0.4}_{-0.2}$ & 0.2\\
low $Y_{e}$ & $12.6^{+14.4}_{-7.2}$ & $11.1^{+12.7}_{-6.4}$ & 0.9 & $9.4^{+10.8}_{-5.4}$ & 0.7 & $5.7^{+6.5}_{-3.2}$ & 0.4 & $3.1^{+3.5}_{-1.7}$ & 0.2\\
low $M_{ej}$ & $13.5^{+15.4}_{-7.7}$ & $10.4^{+11.9}_{-5.9}$ & 0.8 & $7.1^{+8.1}_{-4.1}$ & 0.5 & $3.8^{+4.4}_{-2.2}$ & 0.3 & $2.0^{+2.3}_{-1.2}$ & 0.2\\
high $M_{ej}$ & $16.7^{+19.1}_{-9.6}$ & $16.7^{+19.1}_{-9.6}$ & 1.0 & $15.8^{+18.0}_{-9.0}$ & 0.9 & $11.5^{+13.2}_{-6.6}$ & 0.7 & $6.4^{+7.4}_{-3.7}$ & 0.4\\
\hline
    \end{tabular}
    \caption{For each model, we computed the expected kilonova (KN) detections in the HLTD survey per year using a \texttt{multi\_detect} metric in which we required at least N=2 detections in at least M=2 filters. The broadband filter F146 was excluded from the calculation. The area considered was $A=19$\,deg$^2$ and the kilonova rate was assumed to be $R_{\textrm{BNS}} = 210^{+240}_{-120}$\,Gpc$^{-3}$\,y$^{-1}$. Rarer kilonovae from NSBH mergers could also be found. The first column shows the model name (see Tab.\,\ref{tab:models}); the second, fourth, sixth, and eighth columns present the expected number of kilonovae detected within the survey footprint assuming regular time gaps between observations of 2, 4, 8, and 16 days. The third, fifth, seventh, and ninth columns indicate the ratio between the number of kilonova recovered with 2, 4, 8, 16 days and a gap of 1 day. The total number of kilonovae expected to happen in the volume of interest (i.e., within a luminosity distance of $\sim 7$\,Gpc) is $16.7^{+19.1}_{-9.6}$\,y$^{-1}$.}
    \label{tab:serendip1_noF146_2filt}
\end{table*}

\begin{table*}[]
    \centering
    \begin{tabular}{cccccccccc}
\hline\hline
Model & KN/y & KN/y & gap 2d & KN/y & gap 4d & KN/y & gap 8d & KN/y & gap 16d\\
 & gap 1d & gap 2d & /gap 1d & gap 4d & /gap 1d & gap 8d & /gap 1d & gap 16d & /gap 1d\\
\hline
\multicolumn{10}{c}{Exposure: 55s } \\
\hline
GW170817$_{pol}$ & $0.1^{+0.1}_{-0.1}$ & $0.1^{+0.1}_{-0.1}$ & 1.0 & $0.1^{+0.1}_{-0.1}$ & 1.0 & $0.1^{+0.1}_{-0.0}$ & 0.8 & $< 0.1$ & 0.3\\
GW170817$_{eq}$ & $0.0^{+0.1}_{-0.0}$ & $< 0.1$ & 0.9 & $< 0.1$ & 0.9 & $< 0.1$ & 0.6 & $< 0.1$ & 0.4\\
low $Y_{e}$, $M_{ej}$ & $< 0.1$ & $< 0.1$ & 1.0 & $< 0.1$ & 0.8 & $< 0.1$ & 0.4 & $< 0.1$ & 0.2\\
low $Y_{e}$ & $0.1^{+0.1}_{-0.1}$ & $0.1^{+0.1}_{-0.1}$ & 1.0 & $0.1^{+0.1}_{-0.1}$ & 1.0 & $0.1^{+0.1}_{-0.0}$ & 0.8 & $< 0.1$ & 0.4\\
low $M_{ej}$ & $0.1^{+0.1}_{-0.1}$ & $0.1^{+0.1}_{-0.1}$ & 1.0 & $0.1^{+0.1}_{-0.0}$ & 0.8 & $0.0^{+0.1}_{-0.0}$ & 0.5 & $< 0.1$ & 0.3\\
high $M_{ej}$ & $0.3^{+0.3}_{-0.2}$ & $0.3^{+0.3}_{-0.2}$ & 1.0 & $0.3^{+0.3}_{-0.2}$ & 0.9 & $0.3^{+0.3}_{-0.1}$ & 0.9 & $0.2^{+0.2}_{-0.1}$ & 0.6\\
\hline
\multicolumn{10}{c}{Exposure: 1hr } \\
\hline
GW170817$_{pol}$ & $5.1^{+5.9}_{-2.9}$ & $3.8^{+4.4}_{-2.2}$ & 0.7 & $3.8^{+4.4}_{-2.2}$ & 0.7 & $2.7^{+3.1}_{-1.5}$ & 0.5 & $1.8^{+2.1}_{-1.0}$ & 0.3\\
GW170817$_{eq}$ & $1.3^{+1.5}_{-0.7}$ & $1.3^{+1.5}_{-0.7}$ & 1.0 & $1.3^{+1.5}_{-0.7}$ & 1.0 & $1.1^{+1.3}_{-0.6}$ & 0.9 & $0.7^{+0.8}_{-0.4}$ & 0.6\\
low $Y_{e}$, $M_{ej}$ & $1.0^{+1.2}_{-0.6}$ & $1.0^{+1.1}_{-0.6}$ & 0.9 & $0.8^{+0.9}_{-0.4}$ & 0.7 & $0.4^{+0.5}_{-0.2}$ & 0.4 & $0.2^{+0.3}_{-0.1}$ & 0.2\\
low $Y_{e}$ & $3.7^{+4.2}_{-2.1}$ & $3.3^{+3.8}_{-1.9}$ & 0.9 & $3.2^{+3.6}_{-1.8}$ & 0.9 & $2.3^{+2.6}_{-1.3}$ & 0.6 & $1.4^{+1.5}_{-0.8}$ & 0.4\\
low $M_{ej}$ & $3.2^{+3.6}_{-1.8}$ & $3.1^{+3.5}_{-1.8}$ & 1.0 & $2.2^{+2.5}_{-1.3}$ & 0.7 & $1.5^{+1.7}_{-0.8}$ & 0.5 & $0.7^{+0.8}_{-0.4}$ & 0.2\\
high $M_{ej}$ & $16.7^{+19.0}_{-9.5}$ & $14.7^{+16.9}_{-8.4}$ & 0.9 & $12.8^{+14.6}_{-7.3}$ & 0.8 & $8.2^{+9.4}_{-4.7}$ & 0.5 & $4.7^{+5.4}_{-2.7}$ & 0.3\\
\hline
    \end{tabular}
    \caption{Same as Tab.\,\ref{tab:serendip1_noF146_2filt}, but employing only the F158 and F213 filters. At least two detections in each filter are required (N=2, M=2).}
    \label{tab:serendip2_F158-and-F213_2det_2filt}
\end{table*}

\begin{table*}[]
    \centering
    \begin{tabular}{cccccccccc}
\hline\hline
Model & KN/y & KN/y & gap 2d & KN/y & gap 4d & KN/y & gap 8d & KN/y & gap 16d\\
 & gap 1d & gap 2d & /gap 1d & gap 4d & /gap 1d & gap 8d & /gap 1d & gap 16d & /gap 1d\\
\hline
\multicolumn{10}{c}{Exposure: 55s } \\
\hline
GW170817$_{pol}$ & $0.6^{+0.7}_{-0.3}$ & $0.6^{+0.6}_{-0.3}$ & 1.0 & $0.4^{+0.5}_{-0.2}$ & 0.7 & $0.2^{+0.2}_{-0.1}$ & 0.4 & $0.1^{+0.1}_{-0.0}$ & 0.2\\
GW170817$_{eq}$ & $0.1^{+0.1}_{-0.1}$ & $0.1^{+0.1}_{-0.0}$ & 0.7 & $0.1^{+0.1}_{-0.0}$ & 0.5 & $< 0.1$ & 0.4 & $< 0.1$ & 0.2\\
low $Y_{e}$, $M_{ej}$ & $0.1^{+0.1}_{-0.0}$ & $0.0^{+0.1}_{-0.0}$ & 0.8 & $< 0.1$ & 0.5 & $< 0.1$ & 0.3 & $< 0.1$ & 0.1\\
low $Y_{e}$ & $0.4^{+0.5}_{-0.2}$ & $0.4^{+0.4}_{-0.2}$ & 0.9 & $0.3^{+0.4}_{-0.2}$ & 0.8 & $0.2^{+0.2}_{-0.1}$ & 0.5 & $0.1^{+0.1}_{-0.1}$ & 0.2\\
low $M_{ej}$ & $0.5^{+0.5}_{-0.3}$ & $0.4^{+0.4}_{-0.2}$ & 0.8 & $0.2^{+0.3}_{-0.1}$ & 0.5 & $0.1^{+0.1}_{-0.1}$ & 0.3 & $0.1^{+0.1}_{-0.0}$ & 0.1\\
high $M_{ej}$ & $1.3^{+1.5}_{-0.7}$ & $1.3^{+1.5}_{-0.7}$ & 1.0 & $1.2^{+1.3}_{-0.7}$ & 0.9 & $0.8^{+0.9}_{-0.5}$ & 0.6 & $0.5^{+0.5}_{-0.3}$ & 0.4\\
\hline
\multicolumn{10}{c}{Exposure: 1hr } \\
\hline
GW170817$_{pol}$ & $12.8^{+14.6}_{-7.3}$ & $8.4^{+9.6}_{-4.8}$ & 0.8 & $6.3^{+7.2}_{-3.6}$ & 0.6 & $3.7^{+4.2}_{-2.1}$ & 0.4 & $1.4^{+1.6}_{-0.8}$ & 0.1\\
GW170817$_{eq}$ & $1.8^{+2.0}_{-1.0}$ & $1.4^{+1.6}_{-0.8}$ & 0.7 & $1.3^{+1.5}_{-0.8}$ & 0.7 & $1.2^{+1.4}_{-0.7}$ & 0.6 & $0.7^{+0.9}_{-0.4}$ & 0.4\\
low $Y_{e}$, $M_{ej}$ & $1.3^{+1.5}_{-0.7}$ & $1.1^{+1.3}_{-0.6}$ & 0.9 & $0.8^{+1.0}_{-0.5}$ & 0.7 & $0.5^{+0.6}_{-0.3}$ & 0.4 & $0.2^{+0.3}_{-0.1}$ & 0.2\\
low $Y_{e}$ & $7.8^{+8.9}_{-4.4}$ & $6.8^{+7.8}_{-3.9}$ & 0.9 & $5.5^{+6.3}_{-3.1}$ & 0.7 & $3.3^{+3.8}_{-1.9}$ & 0.4 & $2.0^{+2.3}_{-1.1}$ & 0.3\\
low $M_{ej}$ & $8.7^{+9.9}_{-5.0}$ & $6.4^{+7.3}_{-3.6}$ & 0.7 & $4.0^{+4.5}_{-2.3}$ & 0.4 & $2.4^{+2.7}_{-1.3}$ & 0.3 & $1.2^{+1.3}_{-0.7}$ & 0.1\\
high $M_{ej}$ & $16.7^{+19.1}_{-9.6}$ & $16.0^{+18.3}_{-9.2}$ & 1.0 & $13.3^{+15.2}_{-7.6}$ & 0.8 & $8.7^{+9.9}_{-5.0}$ & 0.5 & $5.0^{+5.7}_{-2.8}$ & 0.3\\
\hline
    \end{tabular}
    \caption{Same as Tab.\,\ref{tab:serendip1_noF146_2filt}, but at least two detections in at least three filters are required (N=2, M=3).}
    \label{tab:serendip3_noF146_3filt}
\end{table*}

\begin{table*}[]
    \centering
    \begin{tabular}{cccccccccc}
\hline\hline
Model & KN/y & KN/y & gap 2d & KN/y & gap 4d & KN/y & gap 8d & KN/y & gap 16d\\
 & gap 1d & gap 2d & /gap 1d & gap 4d & /gap 1d & gap 8d & /gap 1d & gap 16d & /gap 1d\\
\hline
\multicolumn{10}{c}{Exposure: 55s } \\
\hline
GW170817$_{pol}$ & $0.5^{+0.6}_{-0.3}$ & $0.5^{+0.6}_{-0.3}$ & 1.0 & $0.4^{+0.4}_{-0.2}$ & 0.7 & $0.3^{+0.3}_{-0.2}$ & 0.5 & $< 0.1$ & 0.0\\
GW170817$_{eq}$ & $0.1^{+0.2}_{-0.1}$ & $0.1^{+0.2}_{-0.1}$ & 1.0 & $0.1^{+0.2}_{-0.1}$ & 0.9 & $0.1^{+0.1}_{-0.1}$ & 0.6 & $< 0.1$ & 0.0\\
low $Y_{e}$, $M_{ej}$ & $0.1^{+0.1}_{-0.0}$ & $0.1^{+0.1}_{-0.0}$ & 0.9 & $< 0.1$ & 0.6 & $< 0.1$ & 0.3 & $< 0.1$ & 0.0\\
low $Y_{e}$ & $0.4^{+0.4}_{-0.2}$ & $0.4^{+0.4}_{-0.2}$ & 1.0 & $0.3^{+0.4}_{-0.2}$ & 0.8 & $0.2^{+0.3}_{-0.1}$ & 0.6 & $< 0.1$ & 0.0\\
low $M_{ej}$ & $0.3^{+0.4}_{-0.2}$ & $0.3^{+0.3}_{-0.1}$ & 0.8 & $0.2^{+0.2}_{-0.1}$ & 0.6 & $0.1^{+0.1}_{-0.1}$ & 0.3 & $< 0.1$ & 0.0\\
high $M_{ej}$ & $1.3^{+1.5}_{-0.7}$ & $1.3^{+1.5}_{-0.7}$ & 1.0 & $1.1^{+1.3}_{-0.7}$ & 0.9 & $0.8^{+0.9}_{-0.5}$ & 0.6 & $< 0.1$ & 0.0\\
\hline
\multicolumn{10}{c}{Exposure: 1hr } \\
\hline
GW170817$_{pol}$ & $12.7^{+14.5}_{-7.2}$ & $12.7^{+14.5}_{-7.2}$ & 1.0 & $8.5^{+9.7}_{-4.8}$ & 0.7 & $5.0^{+5.7}_{-2.8}$ & 0.4 & $< 0.1$ & 0.0\\
GW170817$_{eq}$ & $1.3^{+1.5}_{-0.7}$ & $1.3^{+1.5}_{-0.7}$ & 1.0 & $1.3^{+1.5}_{-0.7}$ & 1.0 & $1.1^{+1.3}_{-0.6}$ & 0.9 & $< 0.1$ & 0.0\\
low $Y_{e}$, $M_{ej}$ & $1.3^{+1.5}_{-0.7}$ & $1.1^{+1.3}_{-0.6}$ & 0.9 & $0.8^{+0.9}_{-0.4}$ & 0.6 & $0.5^{+0.5}_{-0.3}$ & 0.4 & $< 0.1$ & 0.0\\
low $Y_{e}$ & $9.0^{+10.3}_{-5.1}$ & $9.0^{+10.3}_{-5.1}$ & 1.0 & $7.0^{+7.9}_{-4.0}$ & 0.8 & $3.9^{+4.5}_{-2.2}$ & 0.4 & $< 0.1$ & 0.0\\
low $M_{ej}$ & $7.5^{+8.6}_{-4.3}$ & $5.8^{+6.6}_{-3.3}$ & 0.8 & $3.7^{+4.2}_{-2.1}$ & 0.5 & $2.3^{+2.6}_{-1.3}$ & 0.3 & $< 0.1$ & 0.0\\
high $M_{ej}$ & $16.7^{+19.1}_{-9.6}$ & $16.7^{+19.1}_{-9.6}$ & 1.0 & $14.3^{+16.3}_{-8.2}$ & 0.9 & $9.6^{+10.9}_{-5.5}$ & 0.6 & $< 0.1$ & 0.0\\
\hline
    \end{tabular}
    \caption{Same as Tab.\,\ref{tab:serendip1_noF146_2filt}, but at least three detections in at least two filters are required (N=3, M=2). %The footprint area was doubled to A=38\,deg$^2$.
    }
    \label{tab:serendip4_noF146_3det}
\end{table*}

\begin{table*}[]
    \centering
    \begin{tabular}{cccccccccc}
\hline\hline
Model & KN/y & KN/y & gap 2d & KN/y & gap 4d & KN/y & gap 8d & KN/y & gap 16d\\
 & gap 1d & gap 2d & /gap 1d & gap 4d & /gap 1d & gap 8d & /gap 1d & gap 16d & /gap 1d\\
\hline
\multicolumn{10}{c}{Exposure: 55s } \\
\hline
GW170817$_{pol}$ & $0.2^{+0.2}_{-0.1}$ & $0.2^{+0.2}_{-0.1}$ & 1.0 & $0.2^{+0.2}_{-0.1}$ & 0.9 & $0.1^{+0.2}_{-0.1}$ & 0.8 & $< 0.1$ & 0.0\\
GW170817$_{eq}$ & $0.1^{+0.1}_{-0.0}$ & $0.1^{+0.1}_{-0.0}$ & 1.0 & $0.1^{+0.1}_{-0.0}$ & 0.9 & $0.0^{+0.1}_{-0.0}$ & 0.7 & $< 0.1$ & 0.0\\
low $Y_{e}$, $M_{ej}$ & $< 0.1$ & $< 0.1$ & 0.7 & $< 0.1$ & 0.5 & $< 0.1$ & 0.3 & $< 0.1$ & 0.0\\
low $Y_{e}$ & $0.2^{+0.2}_{-0.1}$ & $0.2^{+0.2}_{-0.1}$ & 1.0 & $0.2^{+0.2}_{-0.1}$ & 1.0 & $0.1^{+0.1}_{-0.1}$ & 0.7 & $< 0.1$ & 0.0\\
low $M_{ej}$ & $0.1^{+0.2}_{-0.1}$ & $0.1^{+0.1}_{-0.1}$ & 1.0 & $0.1^{+0.1}_{-0.1}$ & 0.9 & $0.1^{+0.1}_{-0.0}$ & 0.5 & $< 0.1$ & 0.0\\
high $M_{ej}$ & $0.6^{+0.7}_{-0.3}$ & $0.6^{+0.7}_{-0.3}$ & 1.0 & $0.5^{+0.6}_{-0.3}$ & 0.9 & $0.5^{+0.6}_{-0.3}$ & 0.8 & $< 0.1$ & 0.0\\
\hline
\multicolumn{10}{c}{Exposure: 1hr } \\
\hline
GW170817$_{pol}$ & $6.3^{+7.2}_{-3.6}$ & $6.3^{+7.2}_{-3.6}$ & 1.0 & $3.5^{+4.0}_{-2.0}$ & 0.6 & $4.4^{+5.0}_{-2.5}$ & 0.7 & $< 0.1$ & 0.0\\
GW170817$_{eq}$ & $2.6^{+2.9}_{-1.5}$ & $2.6^{+2.9}_{-1.5}$ & 1.0 & $2.5^{+2.9}_{-1.4}$ & 1.0 & $1.9^{+2.1}_{-1.1}$ & 0.7 & $< 0.1$ & 0.0\\
low $Y_{e}$, $M_{ej}$ & $1.8^{+2.0}_{-1.0}$ & $1.5^{+1.8}_{-0.9}$ & 0.9 & $1.1^{+1.3}_{-0.7}$ & 0.7 & $0.7^{+0.8}_{-0.4}$ & 0.4 & $< 0.1$ & 0.0\\
low $Y_{e}$ & $6.3^{+7.2}_{-3.6}$ & $6.3^{+7.2}_{-3.6}$ & 1.0 & $5.1^{+5.8}_{-2.9}$ & 0.8 & $3.8^{+4.4}_{-2.2}$ & 0.6 & $< 0.1$ & 0.0\\
low $M_{ej}$ & $5.8^{+6.7}_{-3.3}$ & $4.1^{+4.7}_{-2.3}$ & 0.7 & $3.3^{+3.8}_{-1.9}$ & 0.5 & $2.2^{+2.6}_{-1.3}$ & 0.4 & $< 0.1$ & 0.0\\
high $M_{ej}$ & $25.3^{+29.0}_{-14.5}$ & $25.3^{+29.0}_{-14.5}$ & 1.0 & $20.7^{+23.7}_{-11.9}$ & 0.8 & $13.0^{+14.9}_{-7.4}$ & 0.5 & $< 0.1$ & 0.0\\
\hline
    \end{tabular}
    \caption{Same as Tab.\,\ref{tab:serendip1_noF146_2filt}, but employing only the F158 and F213 filters and requiring at least three detections in each filter (N=3, M=2). The footprint area was also doubled ($A = 38$\,deg$^2$) to provide values also for the 55\,s survey scenario. The total number of kilonovae expected to happen in the volume of interest (i.e., within a luminosity distance of $\sim 7$\,Gpc) is $33.5^{+38.3}_{-19.1}$\,y$^{-1}$.}
    \label{tab:serendip5_F158-and-F213_3det_2filt}
\end{table*}

\bibliographystyle{aasjournal}
\bibliography{references, references2, references3, references4, Roman_MMA}

\begin{thebibliography}{}
\expandafter\ifx\csname natexlab\endcsname\relax\def\natexlab#1{#1}\fi
\providecommand{\url}[1]{\href{#1}{#1}}
\providecommand{\dodoi}[1]{doi:~\href{http://doi.org/#1}{\nolinkurl{#1}}}
\providecommand{\doeprint}[1]{\href{http://ascl.net/#1}{\nolinkurl{http://ascl.net/#1}}}
\providecommand{\doarXiv}[1]{\href{https://arxiv.org/abs/#1}{\nolinkurl{https://arxiv.org/abs/#1}}}

\bibitem[{{Aasi et al}(2015)}]{aLIGO}
{Aasi et al}. 2015, Classical and Quantum Gravity, 32, 074001

\bibitem[{{Abbott} {et~al.}(2017{\natexlab{a}}){Abbott}, {Abbott}, {Abbott},
  {Acernese}, {Ackley}, {Adams}, {Adams}, {Addesso}, {Adhikari}, {Adya},
  {Affeldt}, {Afrough}, {Agarwal}, {Agathos}, {Agatsuma}, {Aggarwal}, {Aguiar},
  {Aiello}, {Ain}, {et~al.}}]{Abbott2017gw_grb}
{Abbott}, B.~P., {Abbott}, R., {Abbott}, T.~D., {et~al.} 2017{\natexlab{a}},
  \apjl, 848, L13, \dodoi{10.3847/2041-8213/aa920c}

\bibitem[{{Abbott} {et~al.}(2017{\natexlab{b}}){Abbott}, {Abbott}, {Abbott},
  {Acernese}, {Ackley}, {Adams}, {Adams}, {Addesso}, {Adhikari}, {Adya}, \&
  et~al.}]{AbbottNSdiscovery}
---. 2017{\natexlab{b}}, \apjl, 848, L12, \dodoi{10.3847/2041-8213/aa91c9}

\bibitem[{{Abbott} {et~al.}(2017{\natexlab{c}}){Abbott}, {Abbott}, {Abbott},
  {Acernese}, {Ackley}, {Adams}, {Adams}, {Addesso}, {Adhikari}, {Adya}, \&
  et~al.}]{AbbottH0}
---. 2017{\natexlab{c}}, \nat, 551, 85, \dodoi{10.1038/nature24471}

\bibitem[{{Abbott} {et~al.}(2020){Abbott}, {Abbott}, {Abbott}, {Abraham},
  {Acernese}, {Ackley}, {Adams}, {Adya}, {Affeldt}, {Agathos}, {Agatsuma},
  {Aggarwal}, {Aguiar}, {Aiello}, {Ain}, {Ajith}, {Akutsu}, {Allen}, {Allocca},
  {Aloy}, {Altin}, {Amato}, {Ananyeva}, {Anderson}, {Anderson}, {Ando},
  {Angelova}, {Antier}, {Appert}, {Arai}, {Arai}, {Arai}, {Araki}, {Araya},
  {Araya}, {Areeda}, {Ar{\`e}ne}, {Aritomi}, {Arnaud}, {Arun}, {Ascenzi},
  {Ashton}, {Aso}, {Aston}, {Astone}, {Aubin}, {Aufmuth}, {Aultoneal},
  {Austin}, {Avendano}, {Avila-Alvarez}, {Babak}, {Bacon}, {Badaracco},
  {Bader}, {Bae}, {Bae}, {Baiotti}, {Bajpai}, {Baker}, {Baldaccini},
  {Ballardin}, {Ballmer}, {Banagiri}, {Barayoga}, {Barclay}, {Barish},
  {Barker}, {Barkett}, {Barnum}, {Barone}, {Barr}, {Barsotti}, {Barsuglia},
  {Barta}, {Bartlett}, {Barton}, {Bartos}, {Bassiri}, {Basti}, {Bawaj},
  {Bayley}, {Bazzan}, {B{\'e}csy}, {Bejger}, {Belahcene}, {Bell}, {Beniwal},
  {Berger}, {Bergmann}, {Bernuzzi}, {Bero}, {Berry}, {Bersanetti}, {Bertolini},
  {Betzwieser}, {Bhandare}, {Bidler}, {Bilenko}, {Bilgili}, {Billingsley},
  {Birch}, {Birney}, {Birnholtz}, {Biscans}, {Biscoveanu}, {Bisht}, {Bitossi},
  {Bizouard}, {Blackburn}, {Blair}, {Blair}, {Blair}, {Bloemen}, {Bode},
  {Boer}, {Boetzel}, {Bogaert}, {Bondu}, {Bonilla}, {Bonnand}, {Booker},
  {Boom}, {Booth}, {Bork}, {Boschi}, {Bose}, {Bossie}, {Bossilkov}, {Bosveld},
  {Bouffanais}, {Bozzi}, {Bradaschia}, {Brady}, {Bramley}, {Branchesi}, {Brau},
  {Briant}, {Briggs}, {Brighenti}, {Brillet}, {Brinkmann}, {Brisson},
  {Brockill}, {Brooks}, {Brown}, {Brown}, {Brunett}, {Buikema}, {Bulik},
  {Bulten}, {Buonanno}, {Buskulic}, {Buy}, {Byer}, {Cabero}, {Cadonati},
  {Cagnoli}, {Cahillane}, {Bustillo}, {Callister}, {Calloni}, {Camp},
  {Campbell}, {Canepa}, {Cannon}, {Cannon}, {Cao}, {Cao}, {Capocasa},
  {Carbognani}, {Caride}, {Carney}, {Carullo}, {Casanueva Diaz}, {Casentini},
  {Caudill}, {Cavagli{\`a}}, {Cavalier}, {Cavalieri}, {Cella},
  {Cerd{\'a}-Dur{\'a}n}, {Cerretani}, {Cesarini}, {Chaibi}, {Chakravarti},
  {Chamberlin}, {Chan}, {Chan}, {Chao}, {Charlton}, {Chase},
  {Chassande-Mottin}, {Chatterjee}, {Chaturvedi}, {Chatziioannou},
  {Cheeseboro}, {Chen}, {Chen}, {Chen}, {Chen}, {Chen}, {Chen}, {Cheng},
  {Cheong}, {Chia}, {Chincarini}, {Chiummo}, {Cho}, {Cho}, {Cho},
  {Christensen}, {Chu}, {Chu}, {Chu}, {Chua}, {Chung}, {Chung}, {Ciani},
  {Ciobanu}, {Ciolfi}, {Cipriano}, {Cirone}, {Clara}, {Clark}, {Clearwater},
  {Cleva}, {Cocchieri}, {Coccia}, {Cohadon}, {Cohen}, {Colgan}, {Colleoni},
  {Collette}, {Collins}, {Cominsky}, {Constancio}, {Conti}, {Cooper}, {Corban},
  {Corbitt}, {Cordero-Carri{\'o}n}, {Corley}, {Cornish}, {Corsi}, {Cortese},
  {Costa}, {Cotesta}, {Coughlin}, {Coughlin}, {Coulon}, {Countryman},
  {Couvares}, {Covas}, {Cowan}, {Coward}, {Cowart}, {Coyne}, {Coyne},
  {Creighton}, {Creighton}, {Cripe}, {Croquette}, {Crowder}, {Cullen},
  {Cumming}, {Cunningham}, {Cuoco}, {Dal Canton}, {Travasso}, {Traylor},
  {Tringali}, {Trovato}, {Trozzo}, {Trudeau}, {Tsang}, {Tsang}, {Tse}, {Tso},
  {Tsubono}, {Tsuchida}, {Tsukada}, {Tsuna}, {Tsuzuki}, {Tuyenbayev},
  {Uchikata}, {Uchiyama}, {Ueda}, {Uehara}, {Ueno}, {Ueshima}, {Ugolini},
  {Unnikrishnan}, {Uraguchi}, {Urban}, {Ushiba}, {Usman}, {Vahlbruch},
  {Vajente}, {Valdes}, {van Bakel}, {van Beuzekom}, {van den Brand}, {van den
  Broeck}, {Vander-Hyde}, {van der Schaaf}, {van Heijningen}, {van Putten},
  {van Veggel}, {Vardaro}, {Varma}, {Vass}, {Vas{\'u}th}, {Vecchio},
  {Vedovato}, {Veitch}, {Veitch}, {Venkateswara}, {Venugopalan}, {Verkindt},
  {Vetrano}, {Vicer{\'e}}, {Viets}, {Vine}, {Vinet}, {Vitale}, {Vivanco}, {Vo},
  {Vocca}, {Vorvick}, {Vyatchanin}, {Wade}, {Wade}, {Wade}, {Walet}, {Walker},
  {Wallace}, {Walsh}, {Wang}, {Wang}, {Wang}, {Wang}, {Wang}, {Wang}, {Ward},
  {Warden}, {Warner}, {Was}, {Watchi}, {Weaver}, {Wei}, {Weinert}, {Weinstein},
  {Weiss}, {Wellmann}, {Wen}, {Wessel}, {We{\ss}els}, {Westhouse}, {Wette},
  {Whelan}, {Whiting}, {Whittle}, {Wilken}, {Williams}, {Williamson}, {Willis},
  {Willke}, {Wimmer}, {Winkler}, {Wipf}, {Wittel}, {Woan}, {Woehler},
  {Wofford}, {Worden}, {Wright}, {Wu}, {Wu}, {Wu}, {Wu}, {Wysocki}, {Xiao},
  {Xu}, {Yamada}, {Yamamoto}, {Yamamoto}, {Yamamoto}, {Yamamoto}, {Yancey},
  {Yang}, {Yap}, {Yazback}, {Yeeles}, {Yokogawa}, {Yokoyama}, {Yokozawa},
  {Yoshioka}, {Yu}, {Yu}, {Yuen}, {Yuzurihara}, {Yvert}, {Zadro{\.z}ny},
  {Zanolin}, {Zeidler}, {Zelenova}, {Zendri}, {Zevin}, {Zhang}, {Zhang},
  {Zhang}, {Zhao}, {Zhao}, {Zhou}, {Zhou}, {Zhu}, {Zhu}, {Zimmerman}, {Zucker},
  {Zweizig}, {Kagra Collaboration}, \& {VIRGO
  Collaboration}}]{Abbott2020LRRprospects}
---. 2020, Living Reviews in Relativity, 23, 3,
  \dodoi{10.1007/s41114-020-00026-9}

\bibitem[{{Abbott} {et~al.}(2023){Abbott}, {Abbott}, {Acernese}, {Ackley},
  {Adams}, {Adhikari}, {Adhikari}, {Adya}, {Affeldt}, {Agarwal}, {Agathos},
  {Agatsuma}, {Aggarwal}, {Aguiar}, {Aiello}, {Ain}, {Ajith}, {Akutsu}, {de
  Alarc{\'o}n}, {Akcay}, {Albanesi}, {Allocca}, {Altin}, {Amato}, {Anand},
  {Anand}, {Ananyeva}, {Anderson}, {Anderson}, {Ando}, {Andrade}, {Andres},
  {Andri{\'c}}, {Angelova}, {Ansoldi}, {Antelis}, {Antier}, {Antonini},
  {Appert}, {Arai}, {Arai}, {Arai}, {Araki}, {Araya}, {Araya}, {Areeda},
  {Ar{\`e}ne}, {Aritomi}, {Arnaud}, {Arogeti}, {Aronson}, {Arun}, {Asada},
  {Asali}, {Ashton}, {Aso}, {Assiduo}, {Aston}, {Astone}, {Aubin}, {Austin},
  {Babak}, {Badaracco}, {Bader}, {Badger}, {Bae}, {Bae}, {Baer}, {Bagnasco},
  {Bai}, {Baiotti}, {Baird}, {Bajpai}, {Ball}, {Ballardin}, {Ballmer},
  {Balsamo}, {Baltus}, {Banagiri}, {Bankar}, {Barayoga}, {Barbieri}, {Barish},
  {Barker}, {Barneo}, {Barone}, {Barr}, {Barsotti}, {Barsuglia}, {Barta},
  {Bartlett}, {Barton}, {Bartos}, {Bassiri}, {Basti}, {Bawaj}, {Bayley},
  {Baylor}, {Bazzan}, {B{\'e}csy}, {Bedakihale}, {Bejger}, {Belahcene},
  {Benedetto}, {Beniwal}, {Bennett}, {Bentley}, {Benyaala}, {Bergamin},
  {Berger}, {Bernuzzi}, {Berry}, {Bersanetti}, {Bertolini}, {LIGO Scientific
  Collaboration}, {VIRGO Collaboration}, \& {KAGRA Collaboration}}]{LVKpop2023}
{Abbott}, R., {Abbott}, T.~D., {Acernese}, F., {et~al.} 2023, Physical Review
  X, 13, 011048, \dodoi{10.1103/PhysRevX.13.011048}

\bibitem[{{Acernese et al}(2015)}]{adVirgo}
{Acernese et al}. 2015, Classical and Quantum Gravity, 32, 024001

\bibitem[{{Alexander} {et~al.}(2017){Alexander}, {Berger}, {Fong}, {Williams},
  {Guidorzi}, {Margutti}, {Metzger}, {Annis}, {Blanchard}, {Brout}, {Brown},
  {Chen}, {Chornock}, {Cowperthwaite}, {Drout}, {Eftekhari}, {Frieman}, {Holz},
  {Nicholl}, {Rest}, {Sako}, {Soares-Santos}, \& {Villar}}]{Alexander2017}
{Alexander}, K.~D., {Berger}, E., {Fong}, W., {et~al.} 2017, \apjl, 848, L21,
  \dodoi{10.3847/2041-8213/aa905d}

\bibitem[{{Anand} {et~al.}(2023){Anand}, {Pang}, {Bulla}, {Coughlin},
  {Dietrich}, {Healy}, {Hussenot-Desenonges}, {Jegou du Laz}, {Kasliwal},
  {Kunert}, {Markin}, {Mooley}, {Nedora}, \&
  {Neuweiler}}]{Anand2023arXivChemDist}
{Anand}, S., {Pang}, P. T.~H., {Bulla}, M., {et~al.} 2023, arXiv e-prints,
  arXiv:2307.11080, \dodoi{10.48550/arXiv.2307.11080}

\bibitem[{{Andreoni} {et~al.}(2020){Andreoni}, {Kool}, {Sagu{\'e}s Carracedo},
  {Kasliwal}, {Bulla}, {Ahumada}, {Coughlin}, {Anand}, {Sollerman}, {Goobar},
  {Kaplan}, {Loveridge}, {Karambelkar}, {Cooke}, {Bagdasaryan}, {Bellm},
  {Cenko}, {Cook}, {De}, {Dekany}, {Delacroix}, {Drake}, {Duev}, {Fremling},
  {Golkhou}, {Graham}, {Hale}, {Kulkarni}, {Kupfer}, {Laher}, {Mahabal},
  {Masci}, {Rusholme}, {Smith}, {Tzanidakis}, {Van Sistine}, \&
  {Yao}}]{AnKo2020}
{Andreoni}, I., {Kool}, E.~C., {Sagu{\'e}s Carracedo}, A., {et~al.} 2020, \apj,
  904, 155, \dodoi{10.3847/1538-4357/abbf4c}

\bibitem[{{Andreoni} {et~al.}(2021){Andreoni}, {Coughlin}, {Kool}, {Kasliwal},
  {Kumar}, {Bhalerao}, {Carracedo}, {Ho}, {Pang}, {Saraogi}, {Sharma},
  {Shenoy}, {Burns}, {Ahumada}, {Anand}, {Singer}, {Perley}, {De}, {Fremling},
  {Bellm}, {Bulla}, {Crellin-Quick}, {Dietrich}, {Drake}, {Duev}, {Goobar},
  {Graham}, {Kaplan}, {Kulkarni}, {Laher}, {Mahabal}, {Shupe}, {Sollerman},
  {Walters}, \& {Yao}}]{Andreoni2021ztfrest}
{Andreoni}, I., {Coughlin}, M.~W., {Kool}, E.~C., {et~al.} 2021, \apj, 918, 63,
  \dodoi{10.3847/1538-4357/ac0bc7}

\bibitem[{{Andreoni} {et~al.}(2022){Andreoni}, {Margutti}, {Salafia},
  {Parazin}, {Villar}, {Coughlin}, {Yoachim}, {Mortensen}, {Brethauer},
  {Smartt}, {Kasliwal}, {Alexander}, {Anand}, {Berger}, {Bernardini}, {Bianco},
  {Blanchard}, {Bloom}, {Brocato}, {Bulla}, {Cartier}, {Cenko}, {Chornock},
  {Copperwheat}, {Corsi}, {D'Ammando}, {D'Avanzo}, {H{\'e}l{\`e}ne Datrier},
  {Foley}, {Ghirlanda}, {Goobar}, {Grindlay}, {Hajela}, {Holz}, {Karambelkar},
  {Kool}, {Lamb}, {Laskar}, {Levan}, {Maguire}, {May}, {Melandri},
  {Milisavljevic}, {Miller}, {Nicholl}, {Nissanke}, {Palmese}, {Piranomonte},
  {Rest}, {Sagu{\'e}s-Carracedo}, {Siellez}, {Singer}, {Smith}, {Steeghs}, \&
  {Tanvir}}]{Andreoni2022RubinToO}
{Andreoni}, I., {Margutti}, R., {Salafia}, O.~S., {et~al.} 2022, \apjs, 260,
  18, \dodoi{10.3847/1538-4365/ac617c10.48550/arXiv.2111.01945}

\bibitem[{Annala {et~al.}(2018)Annala, Gorda, Kurkela, \& Vuorinen}]{AnEe2018}
Annala, E., Gorda, T., Kurkela, A., \& Vuorinen, A. 2018, Phys. Rev. Lett.,
  120, 172703, \dodoi{10.1103/PhysRevLett.120.172703}

\bibitem[{{Arcavi} {et~al.}(2017){Arcavi}, {Hosseinzadeh}, {Howell}, {McCully},
  {Poznanski}, {Kasen}, {Barnes}, {Zaltzman}, {Vasylyev}, {Maoz}, \&
  {Valenti}}]{Arcavi2017GW}
{Arcavi}, I., {Hosseinzadeh}, G., {Howell}, D.~A., {et~al.} 2017, \nat, 551,
  64, \dodoi{10.1038/nature24291}

\bibitem[{{Ascenzi} {et~al.}(2019){Ascenzi}, {Coughlin}, {Dietrich}, {Foley},
  {Ramirez-Ruiz}, {Piranomonte}, {Mockler}, {Murguia-Berthier}, {Fryer},
  {Lloyd-Ronning}, \& {Rosswog}}]{Ascenzi2019MNRAS}
{Ascenzi}, S., {Coughlin}, M.~W., {Dietrich}, T., {et~al.} 2019, \mnras, 486,
  672, \dodoi{10.1093/mnras/stz891}

\bibitem[{{Astropy Collaboration} {et~al.}(2013){Astropy Collaboration},
  {Robitaille}, {Tollerud}, {Greenfield}, {Droettboom}, {Bray}, {Aldcroft},
  {Davis}, {Ginsburg}, {Price-Whelan}, {Kerzendorf}, {Conley}, {Crighton},
  {Barbary}, {Muna}, {Ferguson}, {Grollier}, {Parikh}, {Nair}, {Unther},
  {Deil}, {Woillez}, {Conseil}, {Kramer}, {Turner}, {Singer}, {Fox}, {Weaver},
  {Zabalza}, {Edwards}, {Azalee Bostroem}, {Burke}, {Casey}, {Crawford},
  {Dencheva}, {Ely}, {Jenness}, {Labrie}, {Lim}, {Pierfederici}, {Pontzen},
  {Ptak}, {Refsdal}, {Servillat}, \& {Streicher}}]{2013AA...558A..33A}
{Astropy Collaboration}, {Robitaille}, T.~P., {Tollerud}, E.~J., {et~al.} 2013,
  \aap, 558, A33, \dodoi{10.1051/0004-6361/201322068}

\bibitem[{{Astropy Collaboration} {et~al.}(2018){Astropy Collaboration},
  {Price-Whelan}, {Sip{\H{o}}cz}, {G{\"u}nther}, {Lim}, {Crawford}, {Conseil},
  {Shupe}, {Craig}, {Dencheva}, {Ginsburg}, {VanderPlas}, {Bradley},
  {P{\'e}rez-Su{\'a}rez}, {de Val-Borro}, {Aldcroft}, {Cruz}, {Robitaille},
  {Tollerud}, {Ardelean}, {Babej}, {Bach}, {Bachetti}, {Bakanov}, {Bamford},
  {Barentsen}, {Barmby}, {Baumbach}, {Berry}, {Biscani}, {Boquien}, {Bostroem},
  {Bouma}, {Brammer}, {Bray}, {Breytenbach}, {Buddelmeijer}, {Burke},
  {Calderone}, {Cano Rodr{\'\i}guez}, {Cara}, {Cardoso}, {Cheedella}, {Copin},
  {Corrales}, {Crichton}, {D'Avella}, {Deil}, {Depagne}, {Dietrich}, {Donath},
  {Droettboom}, {Earl}, {Erben}, {Fabbro}, {Ferreira}, {Finethy}, {Fox},
  {Garrison}, {Gibbons}, {Goldstein}, {Gommers}, {Greco}, {Greenfield},
  {Groener}, {Grollier}, {Hagen}, {Hirst}, {Homeier}, {Horton}, {Hosseinzadeh},
  {Hu}, {Hunkeler}, {Ivezi{\'c}}, {Jain}, {Jenness}, {Kanarek}, {Kendrew},
  {Kern}, {Kerzendorf}, {Khvalko}, {King}, {Kirkby}, {Kulkarni}, {Kumar},
  {Lee}, {Lenz}, {Littlefair}, {Ma}, {Macleod}, {Mastropietro}, {McCully},
  {Montagnac}, {Morris}, {Mueller}, {Mumford}, {Muna}, {Murphy}, {Nelson},
  {Nguyen}, {Ninan}, {N{\"o}the}, {Ogaz}, {Oh}, {Parejko}, {Parley}, {Pascual},
  {Patil}, {Patil}, {Plunkett}, {Prochaska}, {Rastogi}, {Reddy Janga},
  {Sabater}, {Sakurikar}, {Seifert}, {Sherbert}, {Sherwood-Taylor}, {Shih},
  {Sick}, {Silbiger}, {Singanamalla}, {Singer}, {Sladen}, {Sooley},
  {Sornarajah}, {Streicher}, {Teuben}, {Thomas}, {Tremblay}, {Turner},
  {Terr{\'o}n}, {van Kerkwijk}, {de la Vega}, {Watkins}, {Weaver}, {Whitmore},
  {Woillez}, {Zabalza}, \& {Astropy Contributors}}]{2018AJ....156..123A}
{Astropy Collaboration}, {Price-Whelan}, A.~M., {Sip{\H{o}}cz}, B.~M., {et~al.}
  2018, \aj, 156, 123,
  \dodoi{10.3847/1538-3881/aabc4f10.48550/arXiv.1801.02634}

\bibitem[{{Bauswein} {et~al.}(2017){Bauswein}, {Just}, {Janka}, \&
  {Stergioulas}}]{BaJu2017}
{Bauswein}, A., {Just}, O., {Janka}, H.-T., \& {Stergioulas}, N. 2017, \apjl,
  850, L34, \dodoi{10.3847/2041-8213/aa9994}

\bibitem[{{Bellm} {et~al.}(2019){Bellm}, {Kulkarni}, {Graham}, {Dekany},
  {Smith}, {Riddle}, {Masci}, {Helou}, {Prince}, {Adams}, {Barbarino},
  {Barlow}, {Bauer}, {Beck}, {Belicki}, {Biswas}, {Blagorodnova}, {Bodewits},
  {Bolin}, {Brinnel}, {Brooke}, {Bue}, {Bulla}, {Burruss}, {Cenko}, {Chang},
  {Connolly}, {Coughlin}, {Cromer}, {Cunningham}, {De}, {Delacroix}, {Desai},
  {Duev}, {Eadie}, {Farnham}, {Feeney}, {Feindt}, {Flynn}, {Franckowiak},
  {Frederick}, {Fremling}, {Gal-Yam}, {Gezari}, {Giomi}, {Goldstein},
  {Golkhou}, {Goobar}, {Groom}, {Hacopians}, {Hale}, {Henning}, {Ho}, {Hover},
  {Howell}, {Hung}, {Huppenkothen}, {Imel}, {Ip}, {Ivezi{\'c}}, {Jackson},
  {Jones}, {Juric}, {Kasliwal}, {Kaspi}, {Kaye}, {Kelley}, {Kowalski},
  {Kramer}, {Kupfer}, {Landry}, {Laher}, {Lee}, {Lin}, {Lin}, {Lunnan},
  {Giomi}, {Mahabal}, {Mao}, {Miller}, {Monkewitz}, {Murphy}, {Ngeow},
  {Nordin}, {Nugent}, {Ofek}, {Patterson}, {Penprase}, {Porter}, {Rauch},
  {Rebbapragada}, {Reiley}, {Rigault}, {Rodriguez}, {van Roestel}, {Rusholme},
  {van Santen}, {Schulze}, {Shupe}, {Singer}, {Soumagnac}, {Stein}, {Surace},
  {Sollerman}, {Szkody}, {Taddia}, {Terek}, {Van Sistine}, {van Velzen},
  {Vestrand}, {Walters}, {Ward}, {Ye}, {Yu}, {Yan}, \&
  {Zolkower}}]{Bellm2019PASP}
{Bellm}, E.~C., {Kulkarni}, S.~R., {Graham}, M.~J., {et~al.} 2019, \pasp, 131,
  018002, \dodoi{10.1088/1538-3873/aaecbe}

\bibitem[{{Bertin} \& {Arnouts}(1996)}]{Bertin1996}
{Bertin}, E., \& {Arnouts}, S. 1996, \aaps, 117, 393,
  \dodoi{10.1051/aas:1996164}

\bibitem[{{Bianco} {et~al.}(2019){Bianco}, {Drout}, {Graham}, {Pritchard},
  {Biswas}, {Narayan}, {Andreoni}, {Cowperthwaite}, {Ribeiro}, {LSST
  Transient}, \& {Variable Stars Collaboration}}]{Bianco2019}
{Bianco}, F.~B., {Drout}, M.~R., {Graham}, M.~L., {et~al.} 2019, \pasp, 131,
  068002, \dodoi{10.1088/1538-3873/ab121a}

\bibitem[{Biscoveanu {et~al.}(2023)Biscoveanu, Landry, \&
  Vitale}]{biscoveanu_population_2023}
Biscoveanu, S., Landry, P., \& Vitale, S. 2023, Monthly Notices of the Royal
  Astronomical Society, 518, 5298, \dodoi{10.1093/mnras/stac3052}

\bibitem[{Broekgaarden {et~al.}(2021)Broekgaarden, Berger, Neijssel,
  {Vigna-G{\'o}mez}, Chattopadhyay, Stevenson, Chruslinska, Justham, {de~Mink},
  \& Mandel}]{broekgaarden_impact_2021}
Broekgaarden, F.~S., Berger, E., Neijssel, C.~J., {et~al.} 2021, Monthly
  Notices of the Royal Astronomical Society, 508, 5028,
  \dodoi{10.1093/mnras/stab2716}

\bibitem[{{Bulla}(2019)}]{Bulla2019}
{Bulla}, M. 2019, \mnras, 489, 5037, \dodoi{10.1093/mnras/stz2495}

\bibitem[{{Bulla}(2023)}]{Bulla2023MNRAS}
---. 2023, \mnras, 520, 2558, \dodoi{10.1093/mnras/stad232}

\bibitem[{{Bulla} {et~al.}(2022){Bulla}, {Coughlin}, {Dhawan}, \&
  {Dietrich}}]{Bulla2022Univ}
{Bulla}, M., {Coughlin}, M.~W., {Dhawan}, S., \& {Dietrich}, T. 2022, Universe,
  8, 289, \dodoi{10.3390/universe8050289}

\bibitem[{{Camilletti} {et~al.}(2022){Camilletti}, {Chiesa}, {Ricigliano},
  {Perego}, {Lippold}, {Padamata}, {Bernuzzi}, {Radice}, {Logoteta}, \&
  {Guercilena}}]{Camilletti2022}
{Camilletti}, A., {Chiesa}, L., {Ricigliano}, G., {et~al.} 2022, \mnras, 516,
  4760, \dodoi{10.1093/mnras/stac2333}

\bibitem[{{Chase} {et~al.}(2022){Chase}, {O'Connor}, {Fryer}, {Troja},
  {Korobkin}, {Wollaeger}, {Ristic}, {Fontes}, {Hungerford}, \&
  {Herring}}]{Chase2022}
{Chase}, E.~A., {O'Connor}, B., {Fryer}, C.~L., {et~al.} 2022, \apj, 927, 163,
  \dodoi{10.3847/1538-4357/ac3d25}

\bibitem[{{Chornock et al.}(2017)}]{ChBe2017}
{Chornock et al.} 2017, The Astrophysical Journal Letters, 848, L19.
\newblock \url{http://stacks.iop.org/2041-8205/848/i=2/a=L19}

\bibitem[{Coughlin {et~al.}(2019{\natexlab{a}})Coughlin, Dietrich, Margalit, \&
  Metzger}]{CoDi2018b}
Coughlin, M.~W., Dietrich, T., Margalit, B., \& Metzger, B.~D.
  2019{\natexlab{a}}, Monthly Notices of the Royal Astronomical Society:
  Letters, 489, L91, \dodoi{10.1093/mnrasl/slz133}

\bibitem[{Coughlin {et~al.}(2022)Coughlin, Farah, Emily, Abigail, Singer, \&
  Weizmann}]{coughlin_ligo/virgo/kagra_2022}
Coughlin, M.~W., Farah, A., Emily, {et~al.} 2022, {{LIGO}}/{{Virgo}}/{{KAGRA
  Observing Capabilities}}: {{Simulated Detections}} and {{Localization}} for
  {{O4}} and {{O5}} ({{October}} 2022 Edition),  {Zenodo},
  \dodoi{10.5281/ZENODO.7026209}

\bibitem[{Coughlin {et~al.}(2018)Coughlin, Dietrich, Doctor, Kasen, Coughlin,
  Jerkstrand, Leloudas, McBrien, Metzger, O’Shaughnessy, \&
  Smartt}]{CoDi2018}
Coughlin, M.~W., Dietrich, T., Doctor, Z., {et~al.} 2018, Monthly Notices of
  the Royal Astronomical Society, 480, 3871, \dodoi{10.1093/mnras/sty2174}

\bibitem[{Coughlin {et~al.}(2019{\natexlab{b}})Coughlin, Dietrich, Antier,
  Bulla, Foucart, Hotokezaka, Raaijmakers, Hinderer, \& Nissanke}]{CoDi2019b}
Coughlin, M.~W., Dietrich, T., Antier, S., {et~al.} 2019{\natexlab{b}}, Monthly
  Notices of the Royal Astronomical Society, 492, 863,
  \dodoi{10.1093/mnras/stz3457}

\bibitem[{Coughlin {et~al.}(2020{\natexlab{a}})Coughlin, Dietrich, Heinzel,
  Khetan, Antier, Bulla, Christensen, Coulter, \& Foley}]{CoDi2019}
Coughlin, M.~W., Dietrich, T., Heinzel, J., {et~al.} 2020{\natexlab{a}}, Phys.
  Rev. Research, 2, 022006, \dodoi{10.1103/PhysRevResearch.2.022006}

\bibitem[{Coughlin {et~al.}(2020{\natexlab{b}})Coughlin, Antier, Dietrich,
  Foley, Heinzel, Bulla, Christensen, Coulter, Issa, \& Khetan}]{CoAn2020}
Coughlin, M.~W., Antier, S., Dietrich, T., {et~al.} 2020{\natexlab{b}}, Nature
  Communications, 11, 4129, \dodoi{10.1038/s41467-020-17998-5}

\bibitem[{Coughlin {et~al.}(2020{\natexlab{c}})Coughlin, Dietrich, Antier,
  Bulla, Foucart, Hotokezaka, Raaijmakers, Hinderer, \&
  Nissanke}]{coughlin_implications_2020a}
Coughlin, M.~W., Dietrich, T., Antier, S., {et~al.} 2020{\natexlab{c}}, Monthly
  Notices of the Royal Astronomical Society, 492, 863,
  \dodoi{10.1093/mnras/stz3457}

\bibitem[{{Coulter} {et~al.}(2017){Coulter}, {Foley}, {Kilpatrick}, {Drout},
  {Piro}, {Shappee}, {Siebert}, {Simon}, {Ulloa}, {Kasen}, {Madore},
  {Murguia-Berthier}, {Pan}, {Prochaska}, {Ramirez-Ruiz}, {Rest}, \&
  {Rojas-Bravo}}]{Coulter17}
{Coulter}, D.~A., {Foley}, R.~J., {Kilpatrick}, C.~D., {et~al.} 2017, Science,
  358, 1556, \dodoi{10.1126/science.aap9811}

\bibitem[{{Cowperthwaite} {et~al.}(2017){Cowperthwaite}, {Berger}, {Villar},
  {Metzger}, {Nicholl}, {Chornock}, {Blanchard}, {Fong}, {Margutti},
  {Soares-Santos}, {Alexander}, {Allam}, {Annis}, {Brout}, {Brown}, {Butler},
  {Chen}, {Diehl}, {Doctor}, {Drout}, {Eftekhari}, {Farr}, {Finley}, {Foley},
  {Frieman}, {Fryer}, {Garc{\'{\i}}a-Bellido}, {Gill}, {Guillochon}, {Herner},
  {Holz}, {Kasen}, {Kessler}, {Marriner}, {Matheson}, {Neilsen}, {Quataert},
  {Palmese}, {Rest}, {Sako}, {Scolnic}, {Smith}, {Tucker}, {Williams},
  {Balbinot}, {Carlin}, {Cook}, {Durret}, {Li}, {Lopes}, {Louren{\c c}o},
  {Marshall}, {Medina}, {Muir}, {Mu{\~n}oz}, {Sauseda}, {Schlegel}, {Secco},
  {Vivas}, {Wester}, {Zenteno}, {Zhang}, {Abbott}, {Banerji}, {Bechtol},
  {Benoit-L{\'e}vy}, {Bertin}, {Buckley-Geer}, {Burke}, {Capozzi}, {Carnero
  Rosell}, {Carrasco Kind}, {Castander}, {Crocce}, {Cunha}, {D'Andrea}, {da
  Costa}, {Davis}, {DePoy}, {Desai}, {Dietrich}, {Drlica-Wagner}, {Eifler},
  {Evrard}, {Fernandez}, {Flaugher}, {Fosalba}, {Gaztanaga}, {Gerdes},
  {Giannantonio}, {Goldstein}, {Gruen}, {Gruendl}, {Gutierrez}, {Honscheid},
  {Jain}, {James}, {Jeltema}, {Johnson}, {Johnson}, {Kent}, {Krause}, {Kron},
  {Kuehn}, {Nuropatkin}, {Lahav}, {Lima}, {Lin}, {Maia}, {March}, {Martini},
  {McMahon}, {Menanteau}, {Miller}, {Miquel}, {Mohr}, {Neilsen}, {Nichol},
  {Ogando}, {Plazas}, {Roe}, {Romer}, {Roodman}, {Rykoff}, {Sanchez},
  {Scarpine}, {Schindler}, {Schubnell}, {Sevilla-Noarbe}, {Smith}, {Smith},
  {Sobreira}, {Suchyta}, {Swanson}, {Tarle}, {Thomas}, {Thomas}, {Troxel},
  {Vikram}, {Walker}, {Wechsler}, {Weller}, {Yanny}, \&
  {Zuntz}}]{Cowperthwaite17}
{Cowperthwaite}, P.~S., {Berger}, E., {Villar}, V.~A., {et~al.} 2017, \apjl,
  848, L17, \dodoi{10.3847/2041-8213/aa8fc7}

\bibitem[{Cowperthwaite {et~al.}(2017)}]{Cowperthwaite:2017dyu}
Cowperthwaite, P.~S., {et~al.} 2017, Astrophys. J., 848, L17,
  \dodoi{10.3847/2041-8213/aa8fc7}

\bibitem[{Dietrich {et~al.}(2020)Dietrich, Coughlin, Pang, Bulla, Heinzel,
  Issa, Tews, \& Antier}]{DiCo2020}
Dietrich, T., Coughlin, M.~W., Pang, P. T.~H., {et~al.} 2020, Science, 370,
  1450, \dodoi{10.1126/science.abb4317}

\bibitem[{{Doctor} {et~al.}(2017){Doctor}, {Kessler}, {Chen}, {Farr}, {Finley},
  {Foley}, {Goldstein}, {Holz}, {Kim}, {Morganson}, {Sako}, {Scolnic}, {Smith},
  {Soares-Santos}, {Spinka}, {Abbott}, {Abdalla}, {Allam}, {Annis}, {Bechtol},
  {Benoit-L{\'e}vy}, {Bertin}, {Brooks}, {Buckley-Geer}, {Burke}, {Carnero
  Rosell}, {Carrasco Kind}, {Carretero}, {Cunha}, {D'Andrea}, {da Costa},
  {DePoy}, {Desai}, {Diehl}, {Drlica-Wagner}, {Eifler}, {Frieman},
  {Garc{\'\i}a-Bellido}, {Gaztanaga}, {Gerdes}, {Gruendl}, {Gschwend},
  {Gutierrez}, {James}, {Krause}, {Kuehn}, {Kuropatkin}, {Lahav}, {Li}, {Lima},
  {Maia}, {March}, {Marshall}, {Menanteau}, {Miquel}, {Neilsen}, {Nichol},
  {Nord}, {Plazas}, {Romer}, {Sanchez}, {Scarpine}, {Schubnell},
  {Sevilla-Noarbe}, {Smith}, {Sobreira}, {Suchyta}, {Swanson}, {Tarle},
  {Walker}, {Wester}, \& {DES Collaboration}}]{Doctor2017}
{Doctor}, Z., {Kessler}, R., {Chen}, H.~Y., {et~al.} 2017, \apj, 837, 57,
  \dodoi{10.3847/1538-4357/aa5d09}

\bibitem[{Drozda {et~al.}(2022)Drozda, Belczynski, O'Shaughnessy, Bulik, \&
  Fryer}]{drozda_black_2022}
Drozda, P., Belczynski, K., O'Shaughnessy, R., Bulik, T., \& Fryer, C.~L. 2022,
  Astronomy and Astrophysics, 667, A126, \dodoi{10.1051/0004-6361/202039418}

\bibitem[{Essick {et~al.}(2023)Essick, Farr, Fishbach, Holz, \&
  Katsavounidis}]{essick_anisotropy_2023}
Essick, R., Farr, W.~M., Fishbach, M., Holz, D.~E., \& Katsavounidis, E. 2023,
  Physical Review D, 107, 043016, \dodoi{10.1103/PhysRevD.107.043016}

\bibitem[{{Foley} {et~al.}(2020){Foley}, {Coulter}, {Kilpatrick}, {Piro},
  {Ramirez-Ruiz}, \& {Schwab}}]{Foley2020}
{Foley}, R.~J., {Coulter}, D.~A., {Kilpatrick}, C.~D., {et~al.} 2020, \mnras,
  494, 190, \dodoi{10.1093/mnras/staa725}

\bibitem[{Foucart {et~al.}(2013)Foucart, Deaton, Duez, Kidder, MacDonald,
  {et~al.}}]{Foucart:2012vn}
Foucart, F., Deaton, M.~B., Duez, M.~D., {et~al.} 2013, Phys.Rev., D87, 084006,
  \dodoi{10.1103/PhysRevD.87.084006}

\bibitem[{Foucart {et~al.}(2018)Foucart, Hinderer, \&
  Nissanke}]{foucart_remnant_2018}
Foucart, F., Hinderer, T., \& Nissanke, S. 2018, Physical Review D, 98, 081501,
  \dodoi{10.1103/PhysRevD.98.081501}

\bibitem[{Fragione(2021)}]{fragione_black-hole-neutron-star_2021}
Fragione, G. 2021, The Astrophysical Journal Letters, 923, L2,
  \dodoi{10.3847/2041-8213/ac3bcd}

\bibitem[{{Ghirlanda} {et~al.}(2019){Ghirlanda}, {Salafia}, {Paragi},
  {Giroletti}, {Yang}, {Marcote}, {Blanchard}, {Agudo}, {An}, {Bernardini},
  {Beswick}, {Branchesi}, {Campana}, {Casadio}, {Chassande-Mottin}, {Colpi},
  {Covino}, {D'Avanzo}, {D'Elia}, {Frey}, {Gawronski}, {Ghisellini}, {Gurvits},
  {Jonker}, {van Langevelde}, {Melandri}, {Moldon}, {Nava}, {Perego},
  {Perez-Torres}, {Reynolds}, {Salvaterra}, {Tagliaferri}, {Venturi},
  {Vergani}, \& {Zhang}}]{Ghirlanda2019Sci}
{Ghirlanda}, G., {Salafia}, O.~S., {Paragi}, Z., {et~al.} 2019, Science, 363,
  968, \dodoi{10.1126/science.aau8815}

\bibitem[{{Goldstein} {et~al.}(2017){Goldstein}, {Veres}, {Burns}, {Briggs},
  {Hamburg}, {Kocevski}, {Wilson-Hodge}, {Preece}, {Poolakkil}, {Roberts},
  {Hui}, {Connaughton}, {Racusin}, {von Kienlin}, {Dal Canton}, {Christensen},
  {Littenberg}, {Siellez}, {Blackburn}, {Broida}, {Bissaldi}, {Cleveland},
  {Gibby}, {Giles}, {Kippen}, {McBreen}, {McEnery}, {Meegan}, {Paciesas}, \&
  {Stanbro}}]{Goldstein2017}
{Goldstein}, A., {Veres}, P., {Burns}, E., {et~al.} 2017, \apjl, 848, L14,
  \dodoi{10.3847/2041-8213/aa8f41}

\bibitem[{{Gompertz} {et~al.}(2018){Gompertz}, {Levan}, {Tanvir}, {Hjorth},
  {Covino}, {Evans}, {Fruchter}, {Gonz{\'a}lez-Fern{\'a}ndez}, {Jin}, {Lyman},
  {Oates}, {O'Brien}, \& {Wiersema}}]{Gompertz2018}
{Gompertz}, B.~P., {Levan}, A.~J., {Tanvir}, N.~R., {et~al.} 2018, \apj, 860,
  62, \dodoi{10.3847/1538-4357/aac206}

\bibitem[{{Gottlieb} {et~al.}(2018){Gottlieb}, {Nakar}, {Piran}, \&
  {Hotokezaka}}]{Gottlieb2018}
{Gottlieb}, O., {Nakar}, E., {Piran}, T., \& {Hotokezaka}, K. 2018, \mnras,
  479, 588, \dodoi{10.1093/mnras/sty1462}

\bibitem[{{Graham} {et~al.}(2019){Graham}, {Kulkarni}, {Bellm}, {Adams},
  {Barbarino}, {Blagorodnova}, {Bodewits}, {Bolin}, {Brady}, {Cenko}, {Chang},
  {Coughlin}, {De}, {Eadie}, {Farnham}, {Feindt}, {Franckowiak}, {Fremling},
  {Gezari}, {Ghosh}, {Goldstein}, {Golkhou}, {Goobar}, {Ho}, {Huppenkothen},
  {Ivezi{\'c}}, {Jones}, {Juric}, {Kaplan}, {Kasliwal}, {Kelley}, {Kupfer},
  {Lee}, {Lin}, {Lunnan}, {Mahabal}, {Miller}, {Ngeow}, {Nugent}, {Ofek},
  {Prince}, {Rauch}, {van Roestel}, {Schulze}, {Singer}, {Sollerman}, {Taddia},
  {Yan}, {Ye}, {Yu}, {Barlow}, {Bauer}, {Beck}, {Belicki}, {Biswas}, {Brinnel},
  {Brooke}, {Bue}, {Bulla}, {Burruss}, {Connolly}, {Cromer}, {Cunningham},
  {Dekany}, {Delacroix}, {Desai}, {Duev}, {Feeney}, {Flynn}, {Frederick},
  {Gal-Yam}, {Giomi}, {Groom}, {Hacopians}, {Hale}, {Helou}, {Henning},
  {Hover}, {Hillenbrand}, {Howell}, {Hung}, {Imel}, {Ip}, {Jackson}, {Kaspi},
  {Kaye}, {Kowalski}, {Kramer}, {Kuhn}, {Landry}, {Laher}, {Mao}, {Masci},
  {Monkewitz}, {Murphy}, {Nordin}, {Patterson}, {Penprase}, {Porter},
  {Rebbapragada}, {Reiley}, {Riddle}, {Rigault}, {Rodriguez}, {Rusholme}, {van
  Santen}, {Shupe}, {Smith}, {Soumagnac}, {Stein}, {Surace}, {Szkody}, {Terek},
  {Van Sistine}, {van Velzen}, {Vestrand}, {Walters}, {Ward}, {Zhang}, \&
  {Zolkower}}]{Graham2019PASP}
{Graham}, M.~J., {Kulkarni}, S.~R., {Bellm}, E.~C., {et~al.} 2019, \pasp, 131,
  078001, \dodoi{10.1088/1538-3873/ab006c}

\bibitem[{{Guidorzi} {et~al.}(2017){Guidorzi}, {Margutti}, {Brout}, {Scolnic},
  {Fong}, {Alexander}, {Cowperthwaite}, {Annis}, {Berger}, {Blanchard},
  {Chornock}, {Coppejans}, {Eftekhari}, {Frieman}, {Huterer}, {Nicholl},
  {Soares-Santos}, {Terreran}, {Villar}, \& {Williams}}]{Guidorzi2017}
{Guidorzi}, C., {Margutti}, R., {Brout}, D., {et~al.} 2017, \apjl, 851, L36,
  \dodoi{10.3847/2041-8213/aaa009}

\bibitem[{{Hallinan} {et~al.}(2017){Hallinan}, {Corsi}, {Mooley}, {Hotokezaka},
  {Nakar}, {Kasliwal}, {Kaplan}, {Frail}, {Myers}, {Murphy}, {De}, {Dobie},
  {Allison}, {Bannister}, {Bhalerao}, {Chandra}, {Clarke}, {Giacintucci}, {Ho},
  {Horesh}, {Kassim}, {Kulkarni}, {Lenc}, {Lockman}, {Lynch}, {Nichols},
  {Nissanke}, {Pallifyaguru}, {Peters}, {Piran}, {Rana}, {Sadler}, \&
  {Singer}}]{Hallinan:2017woc}
{Hallinan}, G., {Corsi}, A., {Mooley}, K.~P., {et~al.} 2017, Science, 358,
  1579, \dodoi{10.1126/science.aap9855}

\bibitem[{{Hjorth} {et~al.}(2017){Hjorth}, {Levan}, {Tanvir}, {Lyman},
  {Wojtak}, {Schr{\o}der}, {Mandel}, {Gall}, \& {Bruun}}]{Hjorth17}
{Hjorth}, J., {Levan}, A.~J., {Tanvir}, N.~R., {et~al.} 2017, \apjl, 848, L31,
  \dodoi{10.3847/2041-8213/aa9110}

\bibitem[{{Hotokezaka} \& {Nakar}(2020)}]{Hotokezaka2020}
{Hotokezaka}, K., \& {Nakar}, E. 2020, \apj, 891, 152,
  \dodoi{10.3847/1538-4357/ab6a98}

\bibitem[{{Hotokezaka} {et~al.}(2019){Hotokezaka}, {Nakar}, {Gottlieb},
  {Nissanke}, {Masuda}, {Hallinan}, {Mooley}, \&
  {Deller}}]{Hotokezaka2019NatAs}
{Hotokezaka}, K., {Nakar}, E., {Gottlieb}, O., {et~al.} 2019, Nature Astronomy,
  3, 940, \dodoi{10.1038/s41550-019-0820-1}

\bibitem[{{Hotokezaka} {et~al.}(2021){Hotokezaka}, {Tanaka}, {Kato}, \&
  {Gaigalas}}]{Hotokezaka2021}
{Hotokezaka}, K., {Tanaka}, M., {Kato}, D., \& {Gaigalas}, G. 2021, \mnras,
  506, 5863, \dodoi{10.1093/mnras/stab1975}

\bibitem[{{Hounsell} {et~al.}(2018){Hounsell}, {Scolnic}, {Foley}, {Kessler},
  {Miranda}, {Avelino}, {Bohlin}, {Filippenko}, {Frieman}, {Jha}, {Kelly},
  {Kirshner}, {Mandel}, {Rest}, {Riess}, {Rodney}, \&
  {Strolger}}]{Hounsell2018}
{Hounsell}, R., {Scolnic}, D., {Foley}, R.~J., {et~al.} 2018, \apj, 867, 23,
  \dodoi{10.3847/1538-4357/aac08b}

\bibitem[{Hunter(2007)}]{Hunter:2007}
Hunter, J.~D. 2007, Computing In Science \& Engineering, 9, 90

\bibitem[{{Ivezi{\'c}} {et~al.}(2019){Ivezi{\'c}}, {Kahn}, {Tyson}, {Abel},
  {Acosta}, {Allsman}, {Alonso}, {AlSayyad}, {Anderson}, {Andrew}, {Angel},
  {Angeli}, {Ansari}, {Antilogus}, {Araujo}, {Armstrong}, {Arndt}, {Astier},
  {Aubourg}, {Auza}, {Axelrod}, {Bard}, {Barr}, {Barrau}, {Bartlett}, {Bauer},
  {Bauman}, {Baumont}, {Bechtol}, {Bechtol}, {Becker}, {Becla}, {Beldica},
  {Bellavia}, {Bianco}, {Biswas}, {Blanc}, {Blazek}, {Bland ford}, {Bloom},
  {Bogart}, {Bond}, {Booth}, {Borgland}, {Borne}, {Bosch}, {Boutigny},
  {Brackett}, {Bradshaw}, {Brand t}, {Brown}, {Bullock}, {Burchat}, {Burke},
  {Cagnoli}, {Calabrese}, {Callahan}, {Callen}, {Carlin}, {Carlson}, {Chand
  rasekharan}, {Charles-Emerson}, {Chesley}, {Cheu}, {Chiang}, {Chiang},
  {Chirino}, {Chow}, {Ciardi}, {Claver}, {Cohen-Tanugi}, {Cockrum}, {Coles},
  {Connolly}, {Cook}, {Cooray}, {Covey}, {Cribbs}, {Cui}, {Cutri}, {Daly},
  {Daniel}, {Daruich}, {Daubard}, {Daues}, {Dawson}, {Delgado}, {Dellapenna},
  {de Peyster}, {de Val-Borro}, {Digel}, {Doherty}, {Dubois},
  {Dubois-Felsmann}, {Durech}, {Economou}, {Eifler}, {Eracleous}, {Emmons},
  {Fausti Neto}, {Ferguson}, {Figueroa}, {Fisher-Levine}, {Focke}, {Foss},
  {Frank}, {Freemon}, {Gangler}, {Gawiser}, {Geary}, {Gee}, {Geha}, {Gessner},
  {Gibson}, {Gilmore}, {Glanzman}, {Glick}, {Goldina}, {Goldstein}, {Goodenow},
  {Graham}, {Gressler}, {Gris}, {Guy}, {Guyonnet}, {Haller}, {Harris},
  {Hascall}, {Haupt}, {Hernand ez}, {Herrmann}, {Hileman}, {Hoblitt},
  {Hodgson}, {Hogan}, {Howard}, {Huang}, {Huffer}, {Ingraham}, {Innes},
  {Jacoby}, {Jain}, {Jammes}, {Jee}, {Jenness}, {Jernigan}, {Jevremovi{\'c}},
  {Johns}, {Johnson}, {Johnson}, {Jones}, {Juramy-Gilles}, {Juri{\'c}},
  {Kalirai}, {Kallivayalil}, {Kalmbach}, {Kantor}, {Karst}, {Kasliwal},
  {Kelly}, {Kessler}, {Kinnison}, {Kirkby}, {Knox}, {Kotov}, {Krabbendam},
  {Krughoff}, {Kub{\'a}nek}, {Kuczewski}, {Kulkarni}, {Ku}, {Kurita}, {Lage},
  {Lambert}, {Lange}, {Langton}, {Le Guillou}, {Levine}, {Liang}, {Lim},
  {Lintott}, {Long}, {Lopez}, {Lotz}, {Lupton}, {Lust}, {MacArthur}, {Mahabal},
  {Mand elbaum}, {Markiewicz}, {Marsh}, {Marshall}, {Marshall}, {May},
  {McKercher}, {McQueen}, {Meyers}, {Migliore}, {Miller}, {Mills}, {Miraval},
  {Moeyens}, {Moolekamp}, {Monet}, {Moniez}, {Monkewitz}, {Montgomery},
  {Morrison}, {Mueller}, {Muller}, {Mu{\~n}oz Arancibia}, {Neill}, {Newbry},
  {Nief}, {Nomerotski}, {Nordby}, {O'Connor}, {Oliver}, {Olivier}, {Olsen},
  {O'Mullane}, {Ortiz}, {Osier}, {Owen}, {Pain}, {Palecek}, {Parejko},
  {Parsons}, {Pease}, {Peterson}, {Peterson}, {Petravick}, {Libby Petrick},
  {Petry}, {Pierfederici}, {Pietrowicz}, {Pike}, {Pinto}, {Plante}, {Plate},
  {Plutchak}, {Price}, {Prouza}, {Radeka}, {Rajagopal}, {Rasmussen},
  {Regnault}, {Reil}, {Reiss}, {Reuter}, {Ridgway}, {Riot}, {Ritz}, {Robinson},
  {Roby}, {Roodman}, {Rosing}, {Roucelle}, {Rumore}, {Russo}, {Saha},
  {Sassolas}, {Schalk}, {Schellart}, {Schindler}, {Schmidt}, {Schneider},
  {Schneider}, {Schoening}, {Schumacher}, {Schwamb}, {Sebag}, {Selvy},
  {Sembroski}, {Seppala}, {Serio}, {Serrano}, {Shaw}, {Shipsey}, {Sick},
  {Silvestri}, {Slater}, {Smith}, {Smith}, {Sobhani}, {Soldahl},
  {Storrie-Lombardi}, {Stover}, {Strauss}, {Street}, {Stubbs}, {Sullivan},
  {Sweeney}, {Swinbank}, {Szalay}, {Takacs}, {Tether}, {Thaler}, {Thayer},
  {Thomas}, {Thornton}, {Thukral}, {Tice}, {Trilling}, {Turri}, {Van Berg},
  {Vanden Berk}, {Vetter}, {Virieux}, {Vucina}, {Wahl}, {Walkowicz}, {Walsh},
  {Walter}, {Wang}, {Wang}, {Warner}, {Wiecha}, {Willman}, {Winters},
  {Wittman}, {Wolff}, {Wood-Vasey}, {Wu}, {Xin}, {Yoachim}, \&
  {Zhan}}]{Ivezic2019}
{Ivezi{\'c}}, {\v{Z}}., {Kahn}, S.~M., {Tyson}, J.~A., {et~al.} 2019, \apj,
  873, 111, \dodoi{10.3847/1538-4357/ab042c}

\bibitem[{{KAGRA Collaboration}(2019)}]{AkEA2019}
{KAGRA Collaboration}. 2019, Nat. Astron., 3, 35,
  \dodoi{10.1038/s41550-018-0658-y}

\bibitem[{{Kasen} {et~al.}(2017){Kasen}, {Metzger}, {Barnes}, {Quataert}, \&
  {Ramirez-Ruiz}}]{Kasen17}
{Kasen}, D., {Metzger}, B., {Barnes}, J., {Quataert}, E., \& {Ramirez-Ruiz}, E.
  2017, \nat, 551, 80, \dodoi{10.1038/nature24453}

\bibitem[{{Kasliwal} {et~al.}(2017){Kasliwal}, {Nakar}, {Singer}, {Kaplan},
  {Cook}, {Van Sistine}, {Lau}, {Fremling}, {Gottlieb}, {Jencson}, {Adams},
  {Feindt}, {Hotokezaka}, {Ghosh}, {Perley}, {Yu}, {Piran}, {Allison},
  {Anupama}, {Balasubramanian}, {Bannister}, {Bally}, {Barnes}, {Barway},
  {Bellm}, {Bhalerao}, {Bhattacharya}, {Blagorodnova}, {Bloom}, {Brady},
  {Cannella}, {Chatterjee}, {Cenko}, {Cobb}, {Copperwheat}, {Corsi}, {De},
  {Dobie}, {Emery}, {Evans}, {Fox}, {Frail}, {Frohmaier}, {Goobar}, {Hallinan},
  {Harrison}, {Helou}, {Hinderer}, {Ho}, {Horesh}, {Ip}, {Itoh}, {Kasen},
  {Kim}, {Kuin}, {Kupfer}, {Lynch}, {Madsen}, {Mazzali}, {Miller}, {Mooley},
  {Murphy}, {Ngeow}, {Nichols}, {Nissanke}, {Nugent}, {Ofek}, {Qi}, {Quimby},
  {Rosswog}, {Rusu}, {Sadler}, {Schmidt}, {Sollerman}, {Steele}, {Williamson},
  {Xu}, {Yan}, {Yatsu}, {Zhang}, \& {Zhao}}]{Kasliwal17}
{Kasliwal}, M.~M., {Nakar}, E., {Singer}, L.~P., {et~al.} 2017, Science, 358,
  1559, \dodoi{10.1126/science.aap9455}

\bibitem[{Kasliwal {et~al.}(2019)Kasliwal, Kasen, Lau, Perley, Rosswog, Ofek,
  Hotokezaka, Chary, Sollerman, Goobar, \& Kaplan}]{KaKa2019}
Kasliwal, M.~M., Kasen, D., Lau, R.~M., {et~al.} 2019, Monthly Notices of the
  Royal Astronomical Society: Letters, \dodoi{10.1093/mnrasl/slz007}

\bibitem[{{Kasliwal} {et~al.}(2020){Kasliwal}, {Anand}, {Ahumada}, {Stein},
  {Carracedo}, {Andreoni}, {Coughlin}, {Singer}, {Kool}, {De}, {Kumar},
  {AlMualla}, {Yao}, {Bulla}, {Dobie}, {Reusch}, {Perley}, {Cenko}, {Bhalerao},
  {Kaplan}, {Sollerman}, {Goobar}, {Copperwheat}, {Bellm}, {Anupama}, {Corsi},
  {Nissanke}, {Agudo}, {Bagdasaryan}, {Barway}, {Belicki}, {Bloom}, {Bolin},
  {Buckley}, {Burdge}, {Burruss}, {Caballero-Garc{\'\i}a}, {Cannella},
  {Castro-Tirado}, {Cook}, {Cooke}, {Cunningham}, {Dahiwale}, {Deshmukh},
  {Dichiara}, {Duev}, {Dutta}, {Feeney}, {Franckowiak}, {Frederick},
  {Fremling}, {Gal-Yam}, {Gatkine}, {Ghosh}, {Goldstein}, {Golkhou}, {Graham},
  {Graham}, {Hankins}, {Helou}, {Hu}, {Ip}, {Jaodand}, {Karambelkar}, {Kong},
  {Kowalski}, {Khandagale}, {Kulkarni}, {Kumar}, {Laher}, {Li}, {Mahabal},
  {Masci}, {Miller}, {Mogotsi}, {Mohite}, {Mooley}, {Mroz}, {Newman}, {Ngeow},
  {Oates}, {Patil}, {Pandey}, {Pavana}, {Pian}, {Riddle},
  {S{\'a}nchez-Ram{\'\i}rez}, {Sharma}, {Singh}, {Smith}, {Soumagnac},
  {Taggart}, {Tan}, {Tzanidakis}, {Troja}, {Valeev}, {Walters}, {Waratkar},
  {Webb}, {Yu}, {Zhang}, {Zhou}, \& {Zolkower}}]{Kasliwal2020}
{Kasliwal}, M.~M., {Anand}, S., {Ahumada}, T., {et~al.} 2020, \apj, 905, 145,
  \dodoi{10.3847/1538-4357/abc335}

\bibitem[{{Kasliwal} {et~al.}(2022){Kasliwal}, {Kasen}, {Lau}, {Perley},
  {Rosswog}, {Ofek}, {Hotokezaka}, {Chary}, {Sollerman}, {Goobar}, \&
  {Kaplan}}]{Kasliwal2022}
{Kasliwal}, M.~M., {Kasen}, D., {Lau}, R.~M., {et~al.} 2022, \mnras, 510, L7,
  \dodoi{10.1093/mnrasl/slz007}

\bibitem[{{Kawaguchi} {et~al.}(2016){Kawaguchi}, {Kyutoku}, {Shibata}, \&
  {Tanaka}}]{Kawaguchi2016}
{Kawaguchi}, K., {Kyutoku}, K., {Shibata}, M., \& {Tanaka}, M. 2016, \apj, 825,
  52, \dodoi{10.3847/0004-637X/825/1/52}

\bibitem[{Kiendrebeogo {et~al.}(2023)Kiendrebeogo, Farah, Foley, Gray, Kunert,
  Puecher, Toivonen, VandenBerg, Anand, Ahumada, Karambelkar, Coughlin,
  Dietrich, Kam, Pang, Singer, \& Sravan}]{kiendrebeogo_updated_2023}
Kiendrebeogo, R.~W., Farah, A.~M., Foley, E.~M., {et~al.} 2023, Updated
  Observing Scenarios and Multi-Messenger Implications for the {{International
  Gravitational-wave Network}}'s {{O4}} and {{O5}},  {arXiv},
  \dodoi{10.48550/arXiv.2306.09234}

\bibitem[{{Kilpatrick} {et~al.}(2017){Kilpatrick}, {Foley}, {Kasen},
  {Murguia-Berthier}, {Ramirez-Ruiz}, {Coulter}, {Drout}, {Piro}, {Shappee},
  {Boutsia}, {Contreras}, {Di Mille}, {Madore}, {Morrell}, {Pan}, {Prochaska},
  {Rest}, {Rojas-Bravo}, {Siebert}, {Simon}, \& {Ulloa}}]{Kilpatrick2017}
{Kilpatrick}, C.~D., {Foley}, R.~J., {Kasen}, D., {et~al.} 2017, Science, 358,
  1583, \dodoi{10.1126/science.aaq0073}

\bibitem[{Lai {et~al.}(2019)Lai, Zhou, \& Xu}]{Lai2019}
Lai, X., Zhou, E., \& Xu, R. 2019, The European Physical Journal A, 55, 60,
  \dodoi{10.1140/epja/i2019-12720-8}

\bibitem[{{Lipunov} {et~al.}(2017){Lipunov}, {Gorbovskoy}, {Kornilov}, {.
  Tyurina}, {Balanutsa}, {Kuznetsov}, {Vlasenko}, {Kuvshinov}, {Gorbunov},
  {Buckley}, {Krylov}, {Podesta}, {Lopez}, {Podesta}, {Levato}, {Saffe},
  {Mallamachi}, {Potter}, {Budnev}, {Gress}, {Ishmuhametova}, {Vladimirov},
  {Zimnukhov}, {Yurkov}, {Sergienko}, {Gabovich}, {Rebolo}, {Serra-Ricart},
  {Israelyan}, {Chazov}, {Wang}, {Tlatov}, \& {Panchenko}}]{Lipunov2017}
{Lipunov}, V.~M., {Gorbovskoy}, E., {Kornilov}, V.~G., {et~al.} 2017, \apjl,
  850, L1, \dodoi{10.3847/2041-8213/aa92c0}

\bibitem[{{Ma} {et~al.}(2023){Ma}, {Lu}, {Guo}, {Zhang}, \& {Chu}}]{Ma2023}
{Ma}, H., {Lu}, Y., {Guo}, X., {Zhang}, S., \& {Chu}, Q. 2023, \mnras, 518,
  6183, \dodoi{10.1093/mnras/stac3418}

\bibitem[{{Margalit} \& {Metzger}(2017)}]{MaMe2017}
{Margalit}, B., \& {Metzger}, B. 2017, The Astrophysical Journal Letters, 850,
  \dodoi{10.3847/2041-8213/aa991c}

\bibitem[{{Margutti} \& {Chornock}(2020)}]{Margutti21}
{Margutti}, R., \& {Chornock}, R. 2020, arXiv e-prints, arXiv:2012.04810.
\newblock \doarXiv{2012.04810}

\bibitem[{{Margutti} {et~al.}(2017){Margutti}, {Berger}, {Fong}, {Guidorzi},
  {Alexander}, {Metzger}, {Blanchard}, {Cowperthwaite}, {Chornock},
  {Eftekhari}, {Nicholl}, {Villar}, {Williams}, {Annis}, {Brown}, {Chen},
  {Doctor}, {Frieman}, {Holz}, {Sako}, \& {Soares-Santos}}]{Margutti:2017cjl}
{Margutti}, R., {Berger}, E., {Fong}, W., {et~al.} 2017, \apjl, 848, L20,
  \dodoi{10.3847/2041-8213/aa9057}

\bibitem[{{McBrien} {et~al.}(2021){McBrien}, {Smartt}, {Huber}, {Rest},
  {Chambers}, {Barbieri}, {Bulla}, {Jha}, {Gromadzki}, {Srivastav}, {Smith},
  {Young}, {McLaughlin}, {Inserra}, {Nicholl}, {Fraser}, {Maguire}, {Chen},
  {Wevers}, {Anderson}, {M{\"u}ller-Bravo}, {Olivares E.}, {Kankare},
  {Gal-Yam}, \& {Waters}}]{McBrien2021MNRAS}
{McBrien}, O.~R., {Smartt}, S.~J., {Huber}, M.~E., {et~al.} 2021, \mnras, 500,
  4213, \dodoi{10.1093/mnras/staa3361}

\bibitem[{Metzger(2020)}]{Metzger:2019zeh}
Metzger, B.~D. 2020, Living Rev. Rel., 23, 1, \dodoi{10.1007/s41114-019-0024-0}

\bibitem[{{Mooley} {et~al.}(2022){Mooley}, {Anderson}, \& {Lu}}]{Mooley2022}
{Mooley}, K.~P., {Anderson}, J., \& {Lu}, W. 2022, \nat, 610, 273,
  \dodoi{10.1038/s41586-022-05145-7}

\bibitem[{{Mooley} {et~al.}(2018){Mooley}, {Deller}, {Gottlieb}, {Nakar},
  {Hallinan}, {Bourke}, {Frail}, {Horesh}, {Corsi}, \&
  {Hotokezaka}}]{Mooley2018Nat}
{Mooley}, K.~P., {Deller}, A.~T., {Gottlieb}, O., {et~al.} 2018, \nat, 561,
  355, \dodoi{10.1038/s41586-018-0486-3}

\bibitem[{Most {et~al.}(2018)Most, Weih, Rezzolla, \&
  Schaffner-Bielich}]{MoWe2018}
Most, E.~R., Weih, L.~R., Rezzolla, L., \& Schaffner-Bielich, J. 2018, Phys.
  Rev. Lett., 120, 261103, \dodoi{10.1103/PhysRevLett.120.261103}

\bibitem[{{Nakar}(2020)}]{Nakar2020PhR}
{Nakar}, E. 2020, \physrep, 886, 1, \dodoi{10.1016/j.physrep.2020.08.008}

\bibitem[{{Nativi} {et~al.}(2021{\natexlab{a}}){Nativi}, {Bulla}, {Rosswog},
  {Lundman}, {Kowal}, {Gizzi}, {Lamb}, \& {Perego}}]{Nativi2021a}
{Nativi}, L., {Bulla}, M., {Rosswog}, S., {et~al.} 2021{\natexlab{a}}, \mnras,
  500, 1772, \dodoi{10.1093/mnras/staa3337}

\bibitem[{{Nativi} {et~al.}(2021{\natexlab{b}}){Nativi}, {Lamb}, {Rosswog},
  {Lundman}, \& {Kowal}}]{Nativi2021b}
{Nativi}, L., {Lamb}, G.~P., {Rosswog}, S., {Lundman}, C., \& {Kowal}, G.
  2021{\natexlab{b}}, \mnras, \dodoi{10.1093/mnras/stab2982}

\bibitem[{{Nicholl} {et~al.}(2021){Nicholl}, {Margalit}, {Schmidt}, {Smith},
  {Ridley}, \& {Nuttall}}]{Nicholl2021}
{Nicholl}, M., {Margalit}, B., {Schmidt}, P., {et~al.} 2021, \mnras, 505, 3016,
  \dodoi{10.1093/mnras/stab1523}

\bibitem[{{Palmese} {et~al.}(2023){Palmese}, {Kaur}, {Hajela}, {Margutti},
  {McDowell}, \& {MacFadyen}}]{Palmese2023H0arXiv}
{Palmese}, A., {Kaur}, R., {Hajela}, A., {et~al.} 2023, arXiv e-prints,
  arXiv:2305.19914, \dodoi{10.48550/arXiv.2305.19914}

\bibitem[{Petrov {et~al.}(2022)Petrov, Singer, Coughlin, Kumar, Almualla,
  Anand, Bulla, Dietrich, Foucart, \& Guessoum}]{petrov_data-driven_2022a}
Petrov, P., Singer, L.~P., Coughlin, M.~W., {et~al.} 2022, The Astrophysical
  Journal, 924, 54, \dodoi{10.3847/1538-4357/ac366d}

\bibitem[{{Pian et al.}(2017)}]{PiDa2017}
{Pian et al.} 2017, Nature, 551, 67 EP .
\newblock \url{http://dx.doi.org/10.1038/nature24298}

\bibitem[{{Planck Collaboration} {et~al.}(2020){Planck Collaboration},
  {Aghanim}, {Akrami}, {Ashdown}, {Aumont}, {Baccigalupi}, {Ballardini},
  {Banday}, {Barreiro}, {Bartolo}, {Basak}, {Battye}, {Benabed}, {Bernard},
  {Bersanelli}, {Bielewicz}, {Bock}, {Bond}, {Borrill}, {Bouchet}, {Boulanger},
  {Bucher}, {Burigana}, {Butler}, {Calabrese}, {Cardoso}, {Carron},
  {Challinor}, {Chiang}, {Chluba}, {Colombo}, {Combet}, {Contreras}, {Crill},
  {Cuttaia}, {de Bernardis}, {de Zotti}, {Delabrouille}, {Delouis}, {Di
  Valentino}, {Diego}, {Dor{\'e}}, {Douspis}, {Ducout}, {Dupac}, {Dusini},
  {Efstathiou}, {Elsner}, {En{\ss}lin}, {Eriksen}, {Fantaye}, {Farhang},
  {Fergusson}, {Fernandez-Cobos}, {Finelli}, {Forastieri}, {Frailis},
  {Fraisse}, {Franceschi}, {Frolov}, {Galeotta}, {Galli}, {Ganga},
  {G{\'e}nova-Santos}, {Gerbino}, {Ghosh}, {Gonz{\'a}lez-Nuevo}, {G{\'o}rski},
  {Gratton}, {Gruppuso}, {Gudmundsson}, {Hamann}, {Handley}, {Hansen},
  {Herranz}, {Hildebrandt}, {Hivon}, {Huang}, {Jaffe}, {Jones}, {Karakci},
  {Keih{\"a}nen}, {Keskitalo}, {Kiiveri}, {Kim}, {Kisner}, {Knox},
  {Krachmalnicoff}, {Kunz}, {Kurki-Suonio}, {Lagache}, {Lamarre}, {Lasenby},
  {Lattanzi}, {Lawrence}, {Le Jeune}, {Lemos}, {Lesgourgues}, {Levrier},
  {Lewis}, {Liguori}, {Lilje}, {Lilley}, {Lindholm}, {L{\'o}pez-Caniego},
  {Lubin}, {Ma}, {Mac{\'\i}as-P{\'e}rez}, {Maggio}, {Maino}, {Mandolesi},
  {Mangilli}, {Marcos-Caballero}, {Maris}, {Martin}, {Martinelli},
  {Mart{\'\i}nez-Gonz{\'a}lez}, {Matarrese}, {Mauri}, {McEwen}, {Meinhold},
  {Melchiorri}, {Mennella}, {Migliaccio}, {Millea}, {Mitra},
  {Miville-Desch{\^e}nes}, {Molinari}, {Montier}, {Morgante}, {Moss}, {Natoli},
  {N{\o}rgaard-Nielsen}, {Pagano}, {Paoletti}, {Partridge}, {Patanchon},
  {Peiris}, {Perrotta}, {Pettorino}, {Piacentini}, {Polastri}, {Polenta},
  {Puget}, {Rachen}, {Reinecke}, {Remazeilles}, {Renzi}, {Rocha}, {Rosset},
  {Roudier}, {Rubi{\~n}o-Mart{\'\i}n}, {Ruiz-Granados}, {Salvati}, {Sandri},
  {Savelainen}, {Scott}, {Shellard}, {Sirignano}, {Sirri}, {Spencer},
  {Sunyaev}, {Suur-Uski}, {Tauber}, {Tavagnacco}, {Tenti}, {Toffolatti},
  {Tomasi}, {Trombetti}, {Valenziano}, {Valiviita}, {Van Tent}, {Vibert},
  {Vielva}, {Villa}, {Vittorio}, {Wandelt}, {Wehus}, {White}, {White},
  {Zacchei}, \& {Zonca}}]{Planck2020cosmo}
{Planck Collaboration}, {Aghanim}, N., {Akrami}, Y., {et~al.} 2020, \aap, 641,
  A6, \dodoi{10.1051/0004-6361/201833910}

\bibitem[{Radice {et~al.}(2018)Radice, Perego, Zappa, \& Bernuzzi}]{RaPe2018}
Radice, D., Perego, A., Zappa, F., \& Bernuzzi, S. 2018, The Astrophysical
  Journal Letters, 852, L29.
\newblock \url{http://stacks.iop.org/2041-8205/852/i=2/a=L29}

\bibitem[{{Rom{\'a}n-Garza} {et~al.}(2021){Rom{\'a}n-Garza}, Bavera, Fragos,
  Zapartas, Misra, Andrews, Coughlin, Dotter, Kovlakas, Serra, Qin, Rocha, \&
  Tran}]{roman-garza_the_2021}
{Rom{\'a}n-Garza}, J., Bavera, S.~S., Fragos, T., {et~al.} 2021, The
  Astrophysical Journal Letters, 912, L23, \dodoi{10.3847/2041-8213/abf42c}

\bibitem[{{Rossi} {et~al.}(2020){Rossi}, {Stratta}, {Maiorano}, {Spighi},
  {Masetti}, {Palazzi}, {Gardini}, {Melandri}, {Nicastro}, {Pian}, {Branchesi},
  {Dadina}, {Testa}, {Brocato}, {Benetti}, {Ciolfi}, {Covino}, {D'Elia},
  {Grado}, {Izzo}, {Perego}, {Piranomonte}, {Salvaterra}, {Selsing},
  {Tomasella}, {Yang}, {Vergani}, {Amati}, \& {Stephen}}]{Rossi2020}
{Rossi}, A., {Stratta}, G., {Maiorano}, E., {et~al.} 2020, \mnras, 493, 3379,
  \dodoi{10.1093/mnras/staa479}

\bibitem[{Rosswog {et~al.}(2017)Rosswog, Feindt, Korobkin, {et~al.}}]{RoFe2017}
Rosswog, S., Feindt, U., Korobkin, O., {et~al.} 2017, Class. Quant. Grav., 34,
  104001, \dodoi{10.1088/1361-6382/aa68a9}

\bibitem[{{Rosswog} {et~al.}(2018){Rosswog}, {Sollerman}, {Feindt}, {Goobar},
  {Korobkin}, {Wollaeger}, {Fremling}, \& {Kasliwal}}]{Rosswog2018}
{Rosswog}, S., {Sollerman}, J., {Feindt}, U., {et~al.} 2018, \aap, 615, A132,
  \dodoi{10.1051/0004-6361/201732117}

\bibitem[{{Salafia} {et~al.}(2019){Salafia}, {Ghirlanda}, {Ascenzi}, \&
  {Ghisellini}}]{Salafia2019}
{Salafia}, O.~S., {Ghirlanda}, G., {Ascenzi}, S., \& {Ghisellini}, G. 2019,
  \aap, 628, A18, \dodoi{10.1051/0004-6361/201935831}

\bibitem[{{Salafia} \& {Giacomazzo}(2021)}]{Salafia2021}
{Salafia}, O.~S., \& {Giacomazzo}, B. 2021, \aap, 645, A93,
  \dodoi{10.1051/0004-6361/202038590}

\bibitem[{{Scolnic} {et~al.}(2018){Scolnic}, {Kessler}, {Brout},
  {Cowperthwaite}, {Soares-Santos}, {Annis}, {Herner}, {Chen}, {Sako},
  {Doctor}, {Butler}, {Palmese}, {Diehl}, {Frieman}, {Holz}, {Berger},
  {Chornock}, {Villar}, {Nicholl}, {Biswas}, {Hounsell}, {Foley}, {Metzger},
  {Rest}, {Garc{\'\i}a-Bellido}, {M{\"o}ller}, {Nugent}, {Abbott}, {Abdalla},
  {Allam}, {Bechtol}, {Benoit-L{\'e}vy}, {Bertin}, {Brooks}, {Buckley-Geer},
  {Carnero Rosell}, {Carrasco Kind}, {Carretero}, {Castander}, {Cunha},
  {D'Andrea}, {da Costa}, {Davis}, {Doel}, {Drlica-Wagner}, {Eifler},
  {Flaugher}, {Fosalba}, {Gaztanaga}, {Gerdes}, {Gruen}, {Gruendl}, {Gschwend},
  {Gutierrez}, {Hartley}, {Honscheid}, {James}, {Johnson}, {Johnson}, {Krause},
  {Kuehn}, {Kuhlmann}, {Lahav}, {Li}, {Lima}, {Maia}, {March}, {Marshall},
  {Menanteau}, {Miquel}, {Neilsen}, {Plazas}, {Sanchez}, {Scarpine},
  {Schubnell}, {Sevilla-Noarbe}, {Smith}, {Smith}, {Sobreira}, {Suchyta},
  {Swanson}, {Tarle}, {Thomas}, {Tucker}, {Walker}, \& {DES
  Collaboration}}]{Scolnic2018}
{Scolnic}, D., {Kessler}, R., {Brout}, D., {et~al.} 2018, \apjl, 852, L3,
  \dodoi{10.3847/2041-8213/aa9d82}

\bibitem[{{Setzer} {et~al.}(2023){Setzer}, {Peiris}, {Korobkin}, \&
  {Rosswog}}]{Setzer2023}
{Setzer}, C.~N., {Peiris}, H.~V., {Korobkin}, O., \& {Rosswog}, S. 2023,
  \mnras, 520, 2829, \dodoi{10.1093/mnras/stad257}

\bibitem[{Singer {et~al.}(2022)Singer, Kiendr{\'e}b{\'e}ogo, \&
  Tnarikawa}]{singer_lpsinger/observing-scenarios-simulations:_2022}
Singer, L., Kiendr{\'e}b{\'e}ogo, W., \& Tnarikawa. 2022,
  Lpsinger/Observing-Scenarios-Simulations: {{Version}} 2, Zenodo,
  \dodoi{10.5281/ZENODO.5206852}

\bibitem[{{Smartt} {et~al.}(2017){Smartt}, {Chen}, {Jerkstrand}, {Coughlin},
  {Kankare}, {Sim}, {Fraser}, {Inserra}, {Maguire}, {Chambers}, {Huber},
  {Kr{\"u}hler}, {Leloudas}, {Magee}, {Shingles}, {Smith}, {Young}, {Tonry},
  {Kotak}, {Gal-Yam}, {Lyman}, {Homan}, {Agliozzo}, {Anderson}, {Angus},
  {Ashall}, {Barbarino}, {Bauer}, {Berton}, {Botticella}, {Bulla}, {Bulger},
  {Cannizzaro}, {Cano}, {Cartier}, {Cikota}, {Clark}, {De Cia}, {Della Valle},
  {Denneau}, {Dennefeld}, {Dessart}, {Dimitriadis}, {Elias-Rosa}, {Firth},
  {Flewelling}, {Fl{\"o}rs}, {Franckowiak}, {Frohmaier}, {Galbany},
  {Gonz{\'a}lez-Gait{\'a}n}, {Greiner}, {Gromadzki}, {Guelbenzu},
  {Guti{\'e}rrez}, {Hamanowicz}, {Hanlon}, {Harmanen}, {Heintz}, {Heinze},
  {Hernandez}, {Hodgkin}, {Hook}, {Izzo}, {James}, {Jonker}, {Kerzendorf},
  {Klose}, {Kostrzewa-Rutkowska}, {Kowalski}, {Kromer}, {Kuncarayakti},
  {Lawrence}, {Lowe}, {Magnier}, {Manulis}, {Martin-Carrillo}, {Mattila},
  {McBrien}, {M{\"u}ller}, {Nordin}, {O'Neill}, {Onori}, {Palmerio},
  {Pastorello}, {Patat}, {Pignata}, {Podsiadlowski}, {Pumo}, {Prentice}, {Rau},
  {Razza}, {Rest}, {Reynolds}, {Roy}, {Ruiter}, {Rybicki}, {Salmon}, {Schady},
  {Schultz}, {Schweyer}, {Seitenzahl}, {Smith}, {Sollerman}, {Stalder},
  {Stubbs}, {Sullivan}, {Szegedi}, {Taddia}, {Taubenberger}, {Terreran}, {van
  Soelen}, {Vos}, {Wainscoat}, {Walton}, {Waters}, {Weiland}, {Willman},
  {Wiseman}, {Wright}, {Wyrzykowski}, \& {Yaron}}]{Smartt17}
{Smartt}, S.~J., {Chen}, T.-W., {Jerkstrand}, A., {et~al.} 2017, \nat, 551, 75,
  \dodoi{10.1038/nature24303}

\bibitem[{{Soares-Santos} {et~al.}(2017){Soares-Santos}, {Holz}, {Annis},
  {Chornock}, {Herner}, {Berger}, {Brout}, {Chen}, {Kessler}, {Sako}, {Allam},
  {Tucker}, {Butler}, {Palmese}, {Doctor}, {Diehl}, {Frieman}, {Yanny}, {Lin},
  {Scolnic}, {Cowperthwaite}, {Neilsen}, {Marriner}, {Kuropatkin}, {Hartley},
  {Paz-Chinch{\'o}n}, {Alexander}, {Balbinot}, {Blanchard}, {Brown}, {Carlin},
  {Conselice}, {Cook}, {Drlica-Wagner}, {Drout}, {Durret}, {Eftekhari}, {Farr},
  {Finley}, {Foley}, {Fong}, {Fryer}, {Garc{\'\i}a-Bellido}, {Gill}, {Gruendl},
  {Hanna}, {Kasen}, {Li}, {Lopes}, {Louren{\c{c}}o}, {Margutti}, {Marshall},
  {Matheson}, {Medina}, {Metzger}, {Mu{\~n}oz}, {Muir}, {Nicholl}, {Quataert},
  {Rest}, {Sauseda}, {Schlegel}, {Secco}, {Sobreira}, {Stebbins}, {Villar},
  {Vivas}, {Walker}, {Wester}, {Williams}, {Zenteno}, {Zhang}, {Abbott},
  {Abdalla}, {Banerji}, {Bechtol}, {Benoit-L{\'e}vy}, {Bertin}, {Brooks},
  {Buckley-Geer}, {Burke}, {Carnero Rosell}, {Carrasco Kind}, {Carretero},
  {Castander}, {Crocce}, {Cunha}, {D'Andrea}, {da Costa}, {Davis}, {Desai},
  {Dietrich}, {Doel}, {Eifler}, {Fernandez}, {Flaugher}, {Fosalba},
  {Gaztanaga}, {Gerdes}, {Giannantonio}, {Goldstein}, {Gruen}, {Gschwend},
  {Gutierrez}, {Honscheid}, {Jain}, {James}, {Jeltema}, {Johnson}, {Johnson},
  {Kent}, {Krause}, {Kron}, {Kuehn}, {Kuhlmann}, {Lahav}, {Lima}, {Maia},
  {March}, {McMahon}, {Menanteau}, {Miquel}, {Mohr}, {Nichol}, {Nord},
  {Ogando}, {Petravick}, {Plazas}, {Romer}, {Roodman}, {Rykoff}, {Sanchez},
  {Scarpine}, {Schubnell}, {Sevilla-Noarbe}, {Smith}, {Smith}, {Suchyta},
  {Swanson}, {Tarle}, {Thomas}, {Thomas}, {Troxel}, {Vikram}, {Wechsler},
  {Weller}, {Dark Energy Survey}, \& {Dark Energy Camera GW-EM
  Collaboration}}]{Soares-Santos2017}
{Soares-Santos}, M., {Holz}, D.~E., {Annis}, J., {et~al.} 2017, \apjl, 848,
  L16, \dodoi{10.3847/2041-8213/aa9059}

\bibitem[{Tanaka {et~al.}(2014)Tanaka, Hotokezaka, Kyutoku, Wanajo, Kiuchi,
  Sekiguchi, \& Shibata}]{Tanaka:2013ixa}
Tanaka, M., Hotokezaka, K., Kyutoku, K., {et~al.} 2014, Astrophys. J., 780, 31,
  \dodoi{10.1088/0004-637X/780/1/31}

\bibitem[{{Tanvir} {et~al.}(2017){Tanvir}, {Levan},
  {Gonz{\'a}lez-Fern{\'a}ndez}, {Korobkin}, {Mandel}, {Rosswog}, {Hjorth},
  {D'Avanzo}, {Fruchter}, {Fryer}, {Kangas}, {Milvang-Jensen}, {Rosetti},
  {Steeghs}, {Wollaeger}, {Cano}, {Copperwheat}, {Covino}, {D'Elia}, {de Ugarte
  Postigo}, {Evans}, {Even}, {Fairhurst}, {Figuera Jaimes}, {Fontes}, {Fujii},
  {Fynbo}, {Gompertz}, {Greiner}, {Hodosan}, {Irwin}, {Jakobsson},
  {J{\o}rgensen}, {Kann}, {Lyman}, {Malesani}, {McMahon}, {Melandri},
  {O'Brien}, {Osborne}, {Palazzi}, {Perley}, {Pian}, {Piranomonte}, {Rabus},
  {Rol}, {Rowlinson}, {Schulze}, {Sutton}, {Th{\"o}ne}, {Ulaczyk}, {Watson},
  {Wiersema}, \& {Wijers}}]{Tanvir17}
{Tanvir}, N.~R., {Levan}, A.~J., {Gonz{\'a}lez-Fern{\'a}ndez}, C., {et~al.}
  2017, \apjl, 848, L27, \dodoi{10.3847/2041-8213/aa90b6}

\bibitem[{{The LIGO-Virgo-KAGRA
  Collaboration}(2023)}]{theligo-virgo-kagracollaboration_observing_2023}
{The LIGO-Virgo-KAGRA Collaboration}. 2023, Observing {{Capabilities}} -
  {{IGWN}} | {{Public Alerts User Guide}},
  https://emfollow.docs.ligo.org/userguide/capabilities.html

\bibitem[{{Troja} {et~al.}(2017){Troja}, {Piro}, {van Eerten}, {Wollaeger},
  {Im}, {Fox}, {Butler}, {Cenko}, {Sakamoto}, {Fryer}, {Ricci}, {Lien}, {Ryan},
  {Korobkin}, {Lee}, {Burgess}, {Lee}, {Watson}, {Choi}, {Covino}, {D'Avanzo},
  {Fontes}, {Gonz{\'a}lez}, {Khandrika}, {Kim}, {Kim}, {Lee}, {Lee}, {Kutyrev},
  {Lim}, {S{\'a}nchez-Ram{\'\i}rez}, {Veilleux}, {Wieringa}, \&
  {Yoon}}]{Troja2017}
{Troja}, E., {Piro}, L., {van Eerten}, H., {et~al.} 2017, \nat, 551, 71,
  \dodoi{10.1038/nature24290}

\bibitem[{Valenti {et~al.}(2017)Valenti, Sand, Yang, Cappellaro, Tartaglia,
  Corsi, Jha, Reichart, Haislip, \& Kouprianov}]{Valenti:2017ngx}
Valenti, S., Sand, D.~J., Yang, S., {et~al.} 2017, Astrophys. J., 848, L24,
  \dodoi{10.3847/2041-8213/aa8edf}

\bibitem[{{Villar} {et~al.}(2017){Villar}, {Guillochon}, {Berger}, {Metzger},
  {Cowperthwaite}, {Nicholl}, {Alexander}, {Blanchard}, {Chornock},
  {Eftekhari}, {Fong}, {Margutti}, \& {Williams}}]{Villar17}
{Villar}, V.~A., {Guillochon}, J., {Berger}, E., {et~al.} 2017, \apjl, 851,
  L21, \dodoi{10.3847/2041-8213/aa9c84}

\bibitem[{{Wang} \& {Giannios}(2021)}]{Wang2021H0}
{Wang}, H., \& {Giannios}, D. 2021, \apj, 908, 200,
  \dodoi{10.3847/1538-4357/abd39c}

\bibitem[{Watson {et~al.}(2019)Watson, Hansen, Selsing, Koch, Malesani,
  Andersen, Fynbo, Arcones, Bauswein, Covino, Grado, Heintz, Hunt, Kouveliotou,
  Leloudas, Levan, Mazzali, \& Pian}]{WaHa2019}
Watson, D., Hansen, C.~J., Selsing, J., {et~al.} 2019, Nature, 574, 497,
  \dodoi{10.1038/s41586-019-1676-3}

\end{thebibliography}



\end{document}